\pdfoutput=1
\documentclass[a4paper,12pt]{article}

\usepackage{amsfonts}
\usepackage{mathrsfs}
\usepackage{amsmath}
\usepackage{amssymb}
\usepackage{framed}

\usepackage[medium]{titlesec}
\usepackage{bm}
\usepackage{cite}

\usepackage[normalem]{ulem}
\usepackage{extarrows}
\usepackage{slashed}
\usepackage{isodateo}
\usepackage{graphicx}
\usepackage[dvipsnames]{xcolor}
\usepackage[bookmarksnumbered=true,bookmarksopen=true]{hyperref}
 \hypersetup{colorlinks,%
             linkcolor=NavyBlue, %
             citecolor=PineGreen, %
             urlcolor=PineGreen}
\usepackage[hmargin=.7in,vmargin=1.1in]{geometry}
\usepackage{indentfirst}
\usepackage{booktabs}

\usepackage{bbm}

\newcommand{\FR}[2]{\displaystyle\frac{\,{#1}\,}{#2}}
\newcommand{\fr}[2]{\mbox{$\frac{\,{#1}\,}{#2}$}}
\newcommand{\n}{\nonumber}
\renewcommand{\rm}{\mathrm}

\graphicspath{{fig/}}

\def\bge{\begin{equation}}
\def\ede{\end{equation}}
\def\bga{\begin{aligned}}
\def\eda{\end{aligned}}
\def\bgb{\begin{bmatrix}}
\def\edb{\end{bmatrix}}
\def\bgp{\begin{pmatrix}}
\def\edp{\end{pmatrix}}
\def\bgm{\begin{matrix}}
\def\edm{\end{matrix}}
\def\bgs{\begin{subequations}}
\def\eds{\end{subequations}}
\newcommand{\order}[1]{\mathcal{O}({#1})}
\def\di{{\mathrm{d}}}

\def\mb{\mathbf}

\def\pd{\partial}
\def\ld{{\mathscr{L}}}

\def\la{\langle}\def\ra{\rangle}

\def\to{\rightarrow}
\def\To{\Rightarrow}
\def\ii{\mathrm{i}}

\def\ga{\gamma}
\def\de{\delta}
\def\ep{\epsilon}
\def\ka{\kappa}
\def\lam{\lambda}

\def\si{\sigma}

\def\aa{\mathsf{a}}
\def\bb{\mathsf{b}}
\def\cc{\mathsf{c}}

\def\2F1{{}_2\mathrm{F}_1}
\def\3F2{{}_3\mathrm{F}_2}
\def\nn{\wt{\nu}}

\usepackage{mdframed} 
\usepackage{tikz}

\newmdenv[skipabove=0pt,%
          skipbelow=5pt,%
          leftmargin=0pt,%
          rightmargin=0pt,%
          innertopmargin=-5pt,%
          innerbottommargin=7pt,%
          innerleftmargin=2pt,%
          innerrightmargin=2pt,%
          splittopskip=0pt,%
          splitbottomskip=0pt,%
          linewidth=0pt,%
          nobreak=true]%
          {keyeqn2}

\newmdenv[backgroundcolor=gray!15,%
          skipabove=0pt,%
          skipbelow=5pt,%
          leftmargin=0pt,%
          rightmargin=0pt,%
          innertopmargin=-5pt,%
          innerbottommargin=7pt,%
          innerleftmargin=2pt,%
          innerrightmargin=2pt,%
          splittopskip=0pt,%
          splitbottomskip=0pt,%
          linewidth=0pt,%
          nobreak=true]%
          {keyeqn}

\usepackage{titlesec}          
\titleformat{\section}
{\normalfont\fontsize{15}{20}\bfseries}{\thesection}{1em}{}

\newcommand{\ob}[1]{\mkern 2mu \overline{\mkern -2mu #1 \mkern -2mu}\mkern 2mu}
\newcommand{\wt}[1]{\mkern 2mu \widetilde{\mkern -2mu #1 \mkern -2mu}\mkern 2mu}

\newcommand{\fnemail}[1]{\footnote{Email: \href{mailto:#1}{\nolinkurl{#1}}}}

\begin{document}

\title{\Large\textbf{Dispersive Bootstrap of Massive Inflation Correlators\\[2mm]}}

\author{Haoyuan Liu$^{\,a\,}$\fnemail{liuhy23@mails.tsinghua.edu.cn},~~~~~~Zhehan Qin$^{\,a\,}$\fnemail{qzh21@mails.tsinghua.edu.cn},~~~~~~Zhong-Zhi Xianyu$^{\,a,b\,}$\fnemail{zxianyu@tsinghua.edu.cn}\\[5mm]
$^a\,$\normalsize{\emph{Department of Physics, Tsinghua University, Beijing 100084, China} }\\
$^b\,$\normalsize{\emph{Peng Huanwu Center for Fundamental Theory, Hefei, Anhui 230026, China }}
}

\date{}
\maketitle

\begin{tikzpicture}[overlay]
\node[minimum width=40mm,minimum height=15mm] (b) at (16,9){\small\verb+USTC-ICTS/PCFT-24-23+};
\end{tikzpicture}

\vspace{20mm}

\begin{abstract}
\vspace{10mm}

Inflation correlators with massive exchanges are central observables of cosmological collider physics, and are also important theoretical data for us to better understand quantum field theories in dS. However, they are difficult to compute directly due to many technical complications of the Schwinger-Keldysh integral. In this work, we initiate a new bootstrap program for massive inflation correlators with dispersion relations on complex momentum planes. We classify kinematic variables of a correlator into vertex energies and line energies, and develop two distinct types of dispersion relations for both of them, respectively called vertex dispersion and line dispersion relations. These dispersion methods allow us to obtain full analytical results of massive correlators from a knowledge of their oscillatory signals alone, while the oscillatory signal at the tree level can be related to simpler subgraphs via the cutting rule. We further apply this method to massive loop correlators, and obtain new analytical expressions for loop diagrams much simpler than existing results from spectral decomposition. In particular, we show that the analyticity demands the existence of an ``irreducible background'' in the loop correlator, which is unambiguously defined, free of UV divergence, and independent of renormalization schemes.

\end{abstract}

\newpage
\tableofcontents

\newpage
\section{Introduction}\label{sec_intro}

There have been active and ongoing efforts in the study of $n$-point correlation functions of primordial curvature fluctuations in recent years \cite{Baumann:2022jpr,Maldacena:2011nz,Assassi:2012zq,Arkani-Hamed:2017fdk,Baumann:2017jvh,Arkani-Hamed:2018bjr,Arkani-Hamed:2018kmz,Baumann:2019oyu,Baumann:2020dch,Sleight:2019mgd,Sleight:2019hfp,Sleight:2020obc,Sleight:2021iix,Sleight:2021plv,Hillman:2019wgh,Pajer:2020wnj,Pajer:2020wxk,Goodhew:2020hob,Jazayeri:2021fvk,Melville:2021lst,Goodhew:2021oqg,Baumann:2021fxj,Gomez:2021qfd,Gomez:2021ujt,Bonifacio:2021azc,Meltzer:2021zin,Hogervorst:2021uvp,DiPietro:2021sjt,Cabass:2021fnw,Wang:2021qez,Premkumar:2021mlz,Hillman:2021bnk,Tong:2021wai,Heckelbacher:2022hbq,Pimentel:2022fsc,Wang:2022eop,Qin:2022lva,Jazayeri:2022kjy,Qin:2022fbv,Cabass:2022rhr,Cabass:2022oap,Xianyu:2022jwk,Bonifacio:2022vwa,Salcedo:2022aal,Lee:2022fgr,Qin:2023ejc,Werth:2023pfl,Pinol:2023oux,Qin:2023bjk,Qin:2023nhv,Lee:2023jby,Loparco:2023rug,AguiSalcedo:2023nds,De:2023xue,Stefanyszyn:2023qov,Xianyu:2023ytd,Green:2023ids,DuasoPueyo:2023kyh,Arkani-Hamed:2023bsv,Arkani-Hamed:2023kig,Chen:2023iix,Benincasa:2024leu,Benincasa:2024lxe,Werth:2024aui,Donath:2024utn,Du:2024hol,Fan:2024iek,Grimm:2024mbw,Melville:2024ove,Cohen:2024anu,Stefanyszyn:2024msm}. These functions are, on the one hand, observables extracted from cosmic microwave background (CMB) or large-scale structure (LSS) data, and, on the other hand, generated by quantum process of particle productions and interactions during the cosmic inflation. Therefore, these correlation functions, subsequently called \emph{inflation correlators}, are the central object that bridge the observational data with quantum field theory in inflationary spacetime. 

A particular class of correlation functions mediated by massive particles have attracted many attentions in recent years \cite{Chen:2009we,Chen:2009zp,Baumann:2011nk,Chen:2012ge,Pi:2012gf,Noumi:2012vr,Gong:2013sma,Arkani-Hamed:2015bza,Chen:2015lza,Chen:2016nrs,Chen:2016uwp,Chen:2016hrz,Lee:2016vti,Chen:2017ryl,An:2017hlx,An:2017rwo,Iyer:2017qzw,Kumar:2017ecc,Tong:2018tqf,Chen:2018sce,Saito:2018omt,Chen:2018xck,Chen:2018cgg,Chua:2018dqh,Kumar:2018jxz,Wu:2018lmx,Li:2019ves,Alexander:2019vtb,Lu:2019tjj,Hook:2019zxa,Hook:2019vcn,Kumar:2019ebj,Liu:2019fag,Wang:2019gbi,Wang:2019gok,Wang:2020uic,Li:2020xwr,Wang:2020ioa,Fan:2020xgh,Bodas:2020yho,Aoki:2020zbj,Maru:2021ezc,Lu:2021gso,Sou:2021juh,Lu:2021wxu,Pinol:2021aun,Cui:2021iie,Tong:2022cdz,Reece:2022soh,Chen:2022vzh,Niu:2022quw,Niu:2022fki,Aoki:2023tjm,Chen:2023txq,Tong:2023krn,Jazayeri:2023xcj,Jazayeri:2023kji,Chakraborty:2023qbp,Chakraborty:2023eoq,Aoki:2023wdc,Craig:2024qgy,McCulloch:2024hiz,Wu:2024wti,Aoki:2024uyi}. A propagating massive particle during inflation could impact the inflaton fluctuations through a resonant process, and leaves a distinct pattern in the inflation correlators as logarithmic oscillations in  momentum ratios. The logarithmic nature is a consequence of exponential expansion of the inflating universe \cite{Chen:2014joa,Chen:2015lza,Chen:2018cgg,Quintin:2024boj}, while the oscillations encode rich physical information about the massive particles. For these reasons, the logarithmic oscillations have been dubbed ``clock signals'' and ``cosmological collider (CC) signals.'' 

The phenomenological studies of CC physics have identified many scenarios producing large CC signals \cite{Chen:2009zp,An:2017hlx,Chen:2018xck,Liu:2019fag,Hook:2019zxa,Hook:2019vcn,Kumar:2018jxz,Kumar:2019ebj,Wang:2019gbi,Wang:2020ioa,Bodas:2020yho}, which are promising targets for the current and upcoming CMB and LSS observations \cite{Meerburg:2016zdz,MoradinezhadDizgah:2017szk,MoradinezhadDizgah:2018ssw,Kogai:2020vzz,Green:2023uyz,Cabass:2024wob,Sohn:2024xzd,Philcox:2024jpd,Goldstein:2024bky}. To connect theory predictions to observational data, it is crucial to perform efficient and accurate computations of inflation correlators. It's not surprising that progress from analytical studies can facilitate this process. Theory-wise, inflation correlators encode important data of  quantum field theories in the bulk de Sitter (dS), and are interesting objects in their own rights. Given the great success of amplitude program in other spacetime backgrounds such as Minkowski and AdS, we are now increasingly motivated in developing amplitude techniques in dS, which are more relevant to our very own universe.

Many progresses have been made recently in the study of dS correlators or cosmological correlators in general. Relevant to this work is the analytical structure of massive inflation correlators in momentum space, which have been explored in recent years from different angles, e.g., \cite{Arkani-Hamed:2018kmz,Baumann:2021fxj,Goodhew:2021oqg,Meltzer:2021zin,DiPietro:2021sjt,Tong:2021wai,Salcedo:2022aal,Qin:2023bjk,Qin:2023nhv,Grimm:2024mbw,Arkani-Hamed:2015bza}. To explain this analytical structure, it is convenient to start from a soft limit where the momentum $K$ of a bulk massive propagator goes to zero \cite{Arkani-Hamed:2015bza,Arkani-Hamed:2018kmz,Tong:2021wai,Qin:2023bjk,Qin:2023nhv}. As will be detailed below, a general graph in this limit can be separated into three pieces: a \emph{nonlocal signal} which is in nonanalytic in the soft momentum $K$ in the form of a branch cut; a \emph{local signal} which is analytic in $K$, but nonanalytic in the energy ratios also in the form of a branch cut; and finally, a \emph{background} which is analytic in both momentum $K$ and other energy variables. 

Although we use the analytical property to classify the signals and the background, this classification has a practical consequence when doing real computations. To explain this point, we note that a bulk computation of a given graph involves a time integral at each bulk vertex and a momentum integral for each independent loop \cite{Chen:2017ryl}. In particular, the bulk propagators contain a part that depends on the ordering of its two time variables, and this makes the bulk time integral heavily nested. Therefore, a direct integration is typically difficult.\footnote{See, however, a recently proposed method to compute arbitrary nested time integrals \cite{Xianyu:2023ytd,Fan:2024iek}.} 
However, a curious observation is that the computation of signals (both nonlocal and local) is generally simpler than the background. The reason is that, to get the signals, one can execute appropriate cuts of the graph to remove certain nested time integrals. The simplicity of signals also shows up in final results: Typically, both the signal and the background are (generalized) hypergeometric functions of momentum ratios, but the background is of higher ``transcendental weight''\footnote{Here we are using the term ``transcendental weight'' to characterize the complexity of hypergeometric series arising in inflation correlators. Very loosely, an irreducible hypergeometric function of $n$-variables can be thought of as having weight $n$. This meaning can be made precise by the family-chain decomposition, as explained in \cite{Fan:2024iek}.} than the signal \cite{Qin:2023ejc,Xianyu:2023ytd,Fan:2024iek,Aoki:2024uyi}. 

In addition, a closer inspection shows that the computation of nonlocal signal is simpler than that of local signal. To get the nonlocal signal, one can take a simpler \emph{nonlocal cut} of the graph, which replaces the cut propagator by its real part \cite{Qin:2023bjk}. The nonlocal signal also obeys the on-shell factorization at arbitrary loop orders \cite{Qin:2023nhv}. In comparison, the computation of local signals requires a subtle and asymmetric cut, which depends on external kinematics and also retains the imaginary part of the propagator \cite{Tong:2021wai}. Besides, it remains challenging to identify local signals at arbitrary loop orders although some progress is ongoing. 

To recapitulate, our past experience shows that there is a ``hierarchy'' in the complexity and also the difficulty of computing the three parts of a given graph: In descending order, we have background $>$ local signal $>$ nonlocal signal. Thus, it is tempting to ask if we can bootstrap the full result of a given graph starting from its signal part alone, or better, if we can bootstrap the full shape with the knowledge of the nonlocal signal only. 

\begin{figure}
\centering 
\includegraphics[width=0.98\textwidth]{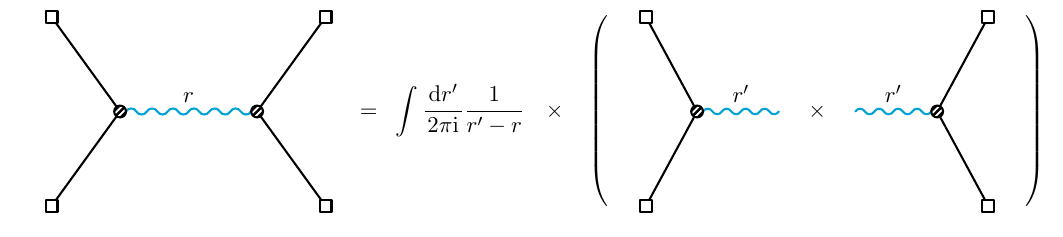} 
\caption{A schematic illustration of the dispersive bootstrap of inflation correlators with a massive exchange. On the left hand side, we have an 4-point boundary correlator of inflatons (black lines) mediated by a massive bulk propagator (cyan wiggly line) characterized by a (properly rescaled) momentum variable $r$. On the right hand side, this correlator is expressed as a dispersion integral along its branch cut, whose discontinuity can be obtained by cutting the graph open and computing the two simpler subgraphs.}
  \label{fig_disp}
\end{figure}

To answer these questions, in this work, we initiate a ``dispersive'' bootstrap program for massive inflation correlators, with the dispersion relation as a key ingredient. The dispersion relation is a very well studied technique, tailored to recover the full function from knowledge of its discontinuities. As the first step, we apply the dispersion relations and get full analytical expressions for a range of massive inflation correlators at both tree and 1-loop levels. The ingredient of the dispersion integral can be either the full signal (both local and nonlocal) or the nonlocal signal alone. Technically, these ingredients can be obtained by computing factorized time integrals, which correspond to simpler subgraphs at the tree level. The essential idea of this method is schematically illustrated in Fig.\ \ref{fig_disp}.

The dispersion relation is an old tool. It has played a central role in the flat-space S-matrix bootstrap program \cite{Eden:1966dnq,Mandelstam:1958xc,Toll:1956cya,Paulos:2016but,Danilkin:2014cra,Hoferichter:2018kwz,Garcia-Martin:2011nna,Bjorken:1965zz,Goldberger:1955zza,Goldberger:1955zz}. There have also been many studies on the cutting rule and dispersion relations in CFT \cite{Caron-Huot:2017vep,Alday:2017vkk,Carmi:2019cub,Penedones:2019tng,Bissi:2022mrs}. Given many types of cutting rules for inflation correlators proposed recently \cite{Baumann:2021fxj,Melville:2021lst,Goodhew:2021oqg,Tong:2021wai,Qin:2023bjk,Qin:2023nhv}, it is a natural next step to try to ``glue'' those cut subgraphs back together. While there are many discussions on dispersion relations at a conceptual level, we are not aware of any previous study using dispersion relations to explicitly bootstrap massive inflation correlators. We fill this gap by providing explicit calculations with dispersion relations for a few typical examples.

Our results at the tree level are not new; All the tree correlators considered in this work have been worked out using other methods, and our method here is by no means ``simpler'' than existing methods such as cosmological bootstrap \cite{Arkani-Hamed:2018kmz,Baumann:2019oyu,Qin:2023ejc} or partial Mellin-Barnes representation \cite{Qin:2022lva,Qin:2022fbv}. Rather, we use these known examples as tests of principle for the dispersive bootstrap method. We expect that one can use this method to ``glue'' more subgraphs and get full results for more complicated graphs, either analytically or numerically, where other methods may not be immediately applicable.

On the other hand, at the 1-loop level, we do obtain new analytical expressions for a class of 1-loop 3-point functions. Our expressions are substantially simpler than known results obtained with spectral decomposition \cite{Xianyu:2022jwk}, and are far easier to implement numerically. This result shows that the dispersive bootstrap can be a promising way to compute inflation correlators with massive loops, which we will further develop in a future study.

An appealing feature of our dispersion technique at the 1-loop level is that it is insensitive to the renormalization ambiguities, because the UV sensitive part of the 1-loop correlator can always be subtracted by a local counterterm and thus is local and analytic. In a sense, the background part of the 1-loop diagram obtained by the dispersion relation can be viewed as an ``irreducible'' companion of the signals, whose existence is enforced by the correct analytical behavior of the full correlator.

\paragraph{Outline of this work} At the heart of our dispersive bootstrap is a detailed understanding of the analytical structure of a specific graph contribution to an inflation correlator. In general, after properly removing all tensor structures, a tree-graph contribution to the inflation correlator is a scalar function of two types of kinematic variables: the \emph{vertex energies} and the \emph{line energies}. The vertex energy is the magnitude sum of momenta of all external lines at a vertex, while a line energy is the magnitude of the momentum flowing in an internal line. 

For physically reachable kinematical configurations (henceforth \emph{physical regions}), vertex and line energies are necessarily positive real. However, to develop dispersion relations, we need to study a graph as a function of complex energies. Our strategy is to consider only one variable being complex at a time, with all other variables staying in their physical regions. We can complexify either a vertex energy or a line energy. In both cases, a massive inflaton correlator develops branch points on the corresponding complex plane, connected by branch cuts. With these branch cuts, we can build corresponding dispersion integrals which compute the full correlator. Thus, we have two distinct types of dispersion relations: the \emph{vertex dispersion relation} built on a vertex energy complex plane, and the \emph{line dispersion relation} built on a line energy complex plane. As we shall see, for a four-point correlator with single massive exchange, the vertex dispersion relation computes the whole graph from its signal, both local and nonlocal. On the other hand, the line dispersion relation computes the whole graph from its nonlocal signal only.

While the vertex and line dispersion relations can be constructed for very general tree graphs, in this work, for definiteness, we will focus on 4-point correlators with $s$-channel massive exchange (Fig.\ \ref{fd_4pt_tree}) and the related 3-point single-exchange correlators (Fig.\ \ref{fd_3pt_tree}), the only exception being the 3-point 1-loop bubble graph (Fig.\ \ref{fd_3pt_1loop}), which is related to tree graphs via spectral decomposition. 

In Sec.\ \ref{sec_vertex}, we begin with a brief review of inflation correlators and the dispersion relation. In particular, we introduce the four-point seed integral $\mathcal{I}^{p_1p_2}(k_{12},k_{34},k_s)$ in (\ref{seed_tree_4pt_full}) which is the central object to be studied in this work. Here $k_i\equiv|\bm k_i|$ ($i=1,\cdots,4,s$) are magnitudes of momenta (also called energies) shown in Fig.\ \ref{fd_4pt_tree}, and $k_{ij}\equiv k_i+k_j$. A very important technical step is the analytical continuation of inflation correlators on the complex energy plane. Thus, in Sec.\ \ref{sec_contour_deform}, we use a few toy examples to explain how to take analytical continuation by contour deformation of an integral expression as a function of its (unintegrated) parameters. Then, we put this method in use in Sec.\ \ref{subsection_vertex_disp} and identify the branch cut of the seed integral $\mathcal{I}^{p_1p_2}(k_{12},k_{34},k_s)$ on the complex $k_{12}$ plane. With this method, we can compute the discontinuity of the seed integral across this branch cut without computing the integral itself, as summarized in  (\ref{disc_I_4pt_k}), which is the main result of this section. 

Then, in Sec.\ \ref{sec_vertex_disp}, we use the vertex dispersion relation to bootstrap a few 3-point and 4-point correlators. For the 3-point correlator, we also consider a one-loop example, where we make use of the loop signal computed via spectral decomposition and dispersively bootstrap the full loop correlator. While our computation of tree graphs recovers previously known results, we get a new analytical expression for the 3-point 1-loop correlator substantially simpler than the existing result. 

In Sec.\ \ref{sec_line}, we switch to a different perspective and consider the seed integral $\mathcal{I}^{p_1p_2}(k_{12},k_{34},k_s)$ on the complex $k_s$ plane. We show that the seed integral also possesses a few branch points on $k_s$ plane which are connected by branch cuts. The discontinuities of these branch cuts are again computable. Remarkably, all the discontinuities in this case can be related to the discontinuity of the nonlocal signal alone, as shown in (\ref{eq_discI_ks}). So, we can build up a line dispersion relation connecting the whole seed integral with its nonlocal signal. 

Then, in Sec.\ \ref{sec_linedisp}, we use the line dispersion to recover the full seed integral from the nonlocal signal. This calculation has the advantage that it uses a minimal set of data to bootstrap the full shape, but the drawback that the computation is complicated. It is nevertheless a useful proof of concept and points to possibilities of (analytical or semi-analytical) computation of more complicated correlators from their readily available nonlocal signal alone. We provide further discussions and outlooks in Sec.\ \ref{sec_conclusion}. In the first two appendices, we collect a few frequently used notations (App.\ \ref{appd_notation}) and special functions, together with their useful properties (App.\ \ref{appd_function}). We collect the details of analytical evaluations of vertex and line dispersion integrals in App.\ \ref{appd_vertex} and App.\ \ref{appd_line}, respectively. Finally, in App.\ \ref{app_Mink}, we use a simple 1-loop correlator in Minkowski spacetime to demonstrate the relation between the dispersive method and a conventional calculation with dimensional regularization.

\paragraph{Comparison with previous works}

The dispersion relation is a topic with rich history. It is not surprising that this relation, together with several closely related concepts such as discontinuities, the optical theorem, cutting rules, has been explored in the context of cosmological correlators (and, relatedly, the wavefunction coefficients) from various different angles \cite{Sleight:2021plv,Melville:2021lst,Goodhew:2021oqg,Tong:2021wai,Meltzer:2021zin,Baumann:2021fxj,Qin:2023bjk,Qin:2023nhv,AguiSalcedo:2023nds}. There are a few similarities and differences between the discontinuities studied in the previous works and the current work, on which we very briefly comment here. 

In previous works such as \cite{Cespedes:2020xqq,Goodhew:2021oqg,Melville:2021lst}, the discontinuity of an amplitude (typically a wavefunction coefficient) is normally defined to be the difference between the amplitude and its complex conjugate with one or several energies' signs flipped. In this combination, one can replace one or a product of several propagators by the real part. (It was the imaginary part in \cite{Goodhew:2021oqg,Melville:2021lst} due to a different convention.) Since the real part of a bulk propagator is always factorized, the discontinuity of an amplitude defined in this way possesses a cutting rule. The nice thing about this definition is that it has a natural origin from the the unitarity of the theory, and therefore, one can use this discontinuity to formulate an optical theorem for cosmological amplitudes \cite{Goodhew:2020hob}. Generalized to the loop level, such a discontinuity can be expressed as momentum integrals of products of (discontinuity of) tree sub-diagrams \cite{AguiSalcedo:2023nds}. The dispersion relations for wavefunctions were used to construct wavefunction coefficients  with massless scalars in \cite{Meltzer:2021zin}. Similar dispersion relations in full Mellin space were discussed in \cite{Sleight:2021plv}. 

In comparison, the discontinuity we are going to use is defined with respect to a correlator alone, without invoking its complex conjugate. More importantly, for the dispersion relation to work as a bootstrap tool, we need to identify all branch cuts of a correlator on the entire complex plane of an energy, where the energy can take arbitrary unphysical value. To extract this information, it is essential to take analytical continuation of a correlator beyond its physical domain, which is not a trivial task as we shall show.

 Furthermore, our starting point is the correlators rather than the bulk propagators, so our dispersive bootstrap can be used to directly construct the full correlators, for both tree and loop diagrams, rather than the integrand as in \cite{AguiSalcedo:2023nds}.

With that said, there is certainly a connection between our definition of discontinuity of a correlator and the discontinuity defined in previous works. For instance, we find that the discontinuity of a tree diagram is also factorized, and expressible in terms of factorized part of propagators. Also, in the line dispersion relation introduced in this work, the discontinuity in the squeezed limit corresponds exactly to the nonlocal signal, so the discontinuity also obeys the nonlocal cutting rule and the factorization theorem as the nonlocal signal \cite{Qin:2023bjk,Qin:2023nhv}. It would be interesting to explore the deeper connections between this work and previous works such as \cite{Goodhew:2020hob,Goodhew:2021oqg,Baumann:2021fxj,AguiSalcedo:2023nds} where basic properties of amplitudes such as unitarity and locality are manifest. We leave this to future exploration.

\paragraph{Notations and conventions}
We work in the slow-roll limit of the inflation where the spacetime is described by the inflation patch of the dS spacetime, and the spacetime metric reads $\di s^2=a^2(\tau)(-\di \tau^2+\di \bm{x}^2)$. Here $\bm x\in \mathbb{R}^3$ is the spatial comoving coordinate, $\tau\in(-\infty,0)$ is the conformal time, and $a(\tau)=-1/(H\tau)$ is the scale factor with $H$ being the constant Hubble parameter. We take the energy unit $H=1$ throughout this work. 
We use bold italic letters such as $\bm k$ to denote 3-momenta and the corresponding italic letter $k\equiv|\bm k|$ to denote its magnitude, which is also called an energy. For sums of several indexed quantities, we use a shorthand notation such as $k_{12}\equiv k_1+k_2$. Other frequently used variables are collected in App.\ \ref{appd_notation}. 
Finally, we make heavy use of the discontinuity of a complex function across its branch cut and it is useful to fix our convention from the very beginning. In this work, the branch cut of a function $f(z)$ appears almost always on the real axis of $z$. Therefore, we define the \textit{discontinuity} of a function $f(z)$ for such a branch cut as:
\bge
 \mathop{\rm{Disc}}_{z}f(z)\equiv\lim_{\epsilon\to0^+}\Bigl[f(z+\ii\epsilon)-f(z-\ii\epsilon)\Bigr].~~~~(z\in \mathbb{R})
\ede

\section{Analytical Structure on a Complex Vertex-Energy Plane}
\label{sec_vertex}

\subsection{Inflation correlators}\label{subsection_correlator}

In this subsection, we set the stage by reviewing the basic kinematic structure of the correlation functions to be studied in this work. We consider generic boundary correlators of a massless or conformal scalar field, with arbitrary massive bulk exchanges. Apart from a three-point example in the next section, we will mostly consider tree-level diagrams. Also, we assume all bulk fields are directly coupled, i.e., without derivatives acting on them. Generalizations to derivative couplings or spinning exchanges are straightforward by including appropriate tensor structures. 

\paragraph{Vertex energies and line energies}
Using the standard diagrammatic rule in the Schwinger-Keldysh (SK) formalism \cite{Chen:2017ryl}, it is straightforward to write down an integral expression for any tree-level correlation function. For definiteness, let us consider a scalar theory with a conformal scalar field $\phi_c$ and a collection of $N_F$ massive fields $\si_A$ $(A=1,\cdots, N_F)$. In dS, a conformal scalar field $\phi_c$ has an effective mass $m^2=2$, while the masses of $\si_A$ can be arbitrary. We assume these fields are coupled directly via polynomial interactions with (possibly) power time dependences. Then, the SK integral for a generic tree-level correlator of $\phi_c$ takes the following form:
\begin{align}
\label{eq_GraphInt}
  \mathcal{G}(\bm k_1,\cdots,\bm k_N)=\sum_{\aa_1,\cdots,\aa_V=\pm} \int_{-\infty}^0\prod_{\ell=1}^V\Big[\di\tau_\ell(-\ii\aa_\ell)(-\tau_\ell)^{p_{\ell}}\Big]\prod_{i=1}^{N}C_{\aa_i}(k_i,\tau_i)\prod_{j=1}^{I}D_{\aa_j\bb_j}(K_j;\tau_j,\tau_j').
\end{align}
This is an integral of $V$ time variables $\tau_\ell$ for all $V$ vertices, with the integrand being products of time-dependent coupling factors $(-\tau_\ell)^{p_\ell}$ and two types of propagators. We assume the powers $p_\ell$ are not too negative such that the graph remains perturbative in the $\tau\to 0$ limit. The bulk-to-boundary propagator  $C_\aa(k;\tau)$ is constructed from a conformal scalar field $\phi_c$ with mass $m^2=2$: 
\bge
\label{eq_CSProp}
  C_\aa(k;\tau)=\FR{\tau\tau_f}{2k}e^{\aa\ii k\tau}.
\ede
Here $|\tau_f|\ll 1$ is a final time cutoff, and is introduced to characterize the leading fall-off behavior of a conformal scalar as $\tau\to 0$. In physical situations with external modes being massless scalars or tensors, this cutoff is unnecessary.\footnote{Also, the case of external massless mode can be conveniently obtained from the conformal scalar case here by acting appropriate differential operators of kinematic variables \cite{Arkani-Hamed:2015bza,Arkani-Hamed:2018kmz,Baumann:2019oyu}.} Moreover, $D_{\aa\bb}(k;\tau_1,\tau_2)$ is the bulk propagator for the massive scalar field $\si$ with mass $m$:
\begin{align}
\label{eq_Dmp}
  D_{-+} (k;\tau_1,\tau_2)
  =&~\FR{\pi}{4}e^{-\pi\wt\nu}(\tau_1\tau_2)^{3/2}\mathrm{H}_{\ii\wt\nu}^{(1)}(-k\tau_1)\mathrm{H}_{-\ii\wt\nu}^{(2)}(-k\tau_2),\\
\label{eq_Dpm}
  D_{+-} (k;\tau_1,\tau_2)
  =&~\FR{\pi}{4}e^{-\pi\wt\nu}(\tau_1\tau_2)^{3/2}\mathrm{H}_{-\ii\wt\nu}^{(2)}(-k\tau_1)\mathrm{H}_{\ii\wt\nu}^{(1)}(-k\tau_2),\\
\label{eq_Dpmpm}
  D_{\pm\pm} (k;\tau_1,\tau_2)=&~D_{\mp\pm}^{(\wt\nu)}(k;\tau_1,\tau_2)\theta(\tau_1-\tau_2)+D_{\pm\mp}^{(\wt\nu)}(k;\tau_1,\tau_2)\theta(\tau_2-\tau_1),
\end{align}
where $\rm{H}_{\nu}^{(j)}(z)$ $(j=1,2)$ is the Hankel function of $j$'th type. In this work, we choose $\si$ to be in the principal series, namely, $m>3/2$, so that the mass parameter $\wt\nu\equiv\sqrt{m^2-9/4}$ is positive, and we get oscillatory signals from $\si$. Generalization to complementary scalar with $0<m<3/2$ is completely straightforward. 

In (\ref{eq_GraphInt}), we have summations over all SK indices $\aa_\ell=\pm$ for all $V$ vertices. When doing so, we require each of the SK indices appearing in the subscript of propagators to be identified with the corresponding index on the vertex to which the propagator attach. 

It is trivial to see that the conformal scalar bulk-to-boundary propagator (\ref{eq_CSProp}) satisfies the relation $C_\aa(k_1;\tau)\cdots C_\aa(k_n;\tau)=C_\aa(k_1+\cdots+k_n;\tau)$ up to multiplications of prefactors $\tau_\ell\tau_f/(2k_\ell)$. As a result, the graph $\mathcal{G}(\{\bm k_i\})$ depends on all spatial vector momenta $\{\bm k_i\}$ only through two particular classes of scalar variables, the \emph{vertex energies} $E_\ell$ $(\ell=1,\cdots,V)$ and the \emph{line energies} $K_j$ ($j=1,\cdots,I$): A vertex energy is assigned to each vertex of the tree diagram, and equals to the \emph{magnitude sum} of the momenta of all \emph{external lines} (bulk-to-boundary propagators) attached to the vertex. A line energy, on the other hand, is assigned to each internal line (bulk propagator) of the tree diagram, and equals to the magnitude of the momentum flowing through this bulk line. Clearly, by momentum conservation, a line energy can always be expressed as the magnitude of a \emph{vector sum} of the momenta of all external lines at either side of the bulk line. 

Following the above analysis, we can always write the graph as:
\bge\label{mathcalG_of_energy}
  \mathcal{G}(\bm k_1,\cdots,\bm k_N)=\prod_{i=1}^N\Big(\FR{\tau_f}{2k_i}\Big)\times \wt{\mathcal{G}}(E_1,\cdots, E_V;K_1,\cdots,K_I).
\ede
We emphasize that this dependence works only for a particular diagram. Since we will develop dispersion relations at the diagrammatic level, this set of variables suit our purpose well. Explicitly, we have:
\begin{align}
  \wt{\mathcal{G}}\big(\{E_\ell\};\{K_j\}\big)=\sum_{\aa_1,\cdots,\aa_V=\pm}\int_{-\infty}^0\prod_{\ell=1}^V\Big[\di\tau_\ell(-\ii\aa_\ell)(-\tau_\ell)^{p_{\ell}}e^{\ii\aa_\ell E_\ell\tau_\ell}\Big]\prod_{j=1}^{I}D_{\aa_j\bb_j}(K_j;\tau_j,\tau_j').
\end{align}

The dispersion relations always involve analytical continuation of the correlator in the complex plane of some variables. Typically, we consider the complex plane of only one variable at a time, and keep all other variables fixed in their physical region. For the tree diagram $\wt{\mathcal{G}}(\{E_\ell\},\{K_j\})$, we can choose to analytically continue a vertex energy $E_\ell$ or a line energy $K_j$. With these two choices, we can respectively develop a \emph{vertex dispersion relation}, and a \emph{line dispersion relation}. Each of them has its own merits and drawbacks.

\begin{figure}[t]
  \centering
  \includegraphics[width=0.50\textwidth]{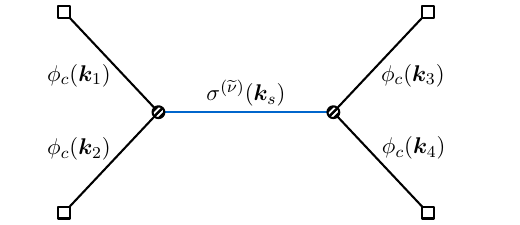}
  \caption{The 4-point correlator of conformal scalars $\phi_c$ with a single massive scalar exchange $\si$ in $s$-channel.}
  \label{fd_4pt_tree}
\end{figure}

\paragraph{Four-point seed integral} To be concrete, we will derive explicit dispersion relations for a tree-level four-point function of a conformal scalar $\phi_c$ with single exchange of a massive scalar $\si$ in the $s$-channel, shown in Fig.\ \ref{fd_4pt_tree}. Dispersion relations for more general correlation functions have similar structures and will be presented in a future work. Assuming a direct coupling $-\fr12\sqrt{-g}\lam\phi_c^2\si$, the integral expression for this graph reads:
\begin{align}
  \mathcal{G}_s(\bm k_1,\cdots,\bm k_4)
  =&-\lam^2\sum_{\aa,\bb=\pm}\aa\bb\int_{-\infty}^0\FR{\di\tau_1}{(-\tau_1)^4}\FR{\di\tau_2}{(-\tau_2)^4}\n\\
  &\times C_\aa(k_1;\tau_1)C_\aa(k_2;\tau_1)C_\bb(k_3;\tau_2)C_\bb(k_4;\tau_2)D_{\aa\bb}(k_s;\tau_1,\tau_2).
\end{align}
In light of the explicit expression for the conformal propagator (\ref{eq_CSProp}), it is useful to define the following dimensionless \emph{seed integral}, as introduced in \cite{Qin:2022fbv}, which enables direct generalization to arbitrary interactions and massless scalar/tensor external modes:\footnote{Note that our choice of arguments of the seed integral $\mathcal{I}^{p_1p_2}$ is different from previous papers including \cite{Qin:2022fbv}, where the seed integral is defined to be a function of two dimensionless momentum ratios, often chosen as $r_1=k_s/k_{12}$ and $r_2=k_s/k_{34}$.  Here, we prefer to explicitly retain the dependence on the three energies $k_{12}$, $k_{34}$, and $k_s$, since it is more transparent to consider the analytical property of the seed integral on the complex plane of an energy variable instead of a momentum ratio. }
\begin{align}\label{seed_tree_4pt}
    &\mathcal{I}^{p_1p_2}_{\mathsf{ab}}(k_{12},k_{34},k_s)
    =-\mathsf{ab}\ k_s^{5+p_{12}}\int_{-\infty}^{0}\di\tau_1\di\tau_2\,(-\tau_1)^{p_1}(-\tau_2)^{p_2}e^{\ii\mathsf{a}k_{12}\tau_1+\ii\mathsf{b}k_{34}\tau_2}D_{\mathsf{ab}} (k_s;\tau_1,\tau_2);\\
    \label{seed_tree_4pt_full}
    &\mathcal{I}^{p_1p_2}(k_{12},k_{34},k_s)\equiv \sum_{\aa,\bb=\pm}\mathcal{I}^{p_1p_2}_{\aa\bb}(k_{12},k_{34},k_s).
\end{align}
The introduction of arbitrary power factors $(-\tau_i)^{p_i}$ $(i=1,2)$ is to take account of various interaction types and external mode functions. The exponents $p_{1,2}$ can in general take complex values (as in models with resonant background). However, we will take $p_{1,2}$ to be real purely to reduce the complication of the analysis. The generalization to complex $p_{1,2}$ is straightforward. 

By construction, it is evident that the seed integral $\mathcal{I}^{p_1p_2}_{\mathsf{ab}}(k_{12},k_{34},k_s)$ is dimensionless, and thus can be expressed as a function of dimensionless momentum ratios. We will exploit this fact when doing explicit computations. 
Also, the graph ${\mathcal{G}}_s(\bm k_1,\cdots,\bm k_4)$ is expressible in terms of the seed integral as:
\begin{align}
\label{eq_TtoSI}
  \mathcal{G}_s(\bm k_1,\cdots,\bm k_4)= \FR{\lam^2\tau_f^4}{16k_1k_2k_3k_4 k_s} \mathcal{I}^{-2,-2}(k_{12},k_{34},k_s).
\end{align}
Thus, we have reduced the whole problem to an analysis of the seed integral. It is certainly possible to compute the entire seed integral by other methods such as partial Mellin-Barnes representation \cite{Qin:2022fbv} or bootstrap equations \cite{Arkani-Hamed:2018kmz,Qin:2022fbv}. However, to be in accordance with the spirit of the dispersive bootstrap, we avoid such a direct computation, but pay more attention to the analytical structure of the seed integral itself. Finally, it is worth noting that the physical regions of the energies $(k_{12},k_{34},k_s)$ are given by $0\leq k_s\leq k_{12}$ and $0\leq k_s\leq k_{34}$ due to the triangle inequalities from momentum conservation. 

\subsection{Dispersion relations}\label{subsection_intro_dispersion}

In the current and next subsections, we make some mathematical preparations for deriving the vertex dispersion relation in Sec.\ \ref{subsection_vertex_disp}. In this subsection, we very briefly explain what a dispersion relation is for nonexperts. Readers familiar with this topic are free to skip this entire subsection. 

\begin{figure}
\centering 
\parbox{0.38\textwidth}{\includegraphics[width=0.38\textwidth]{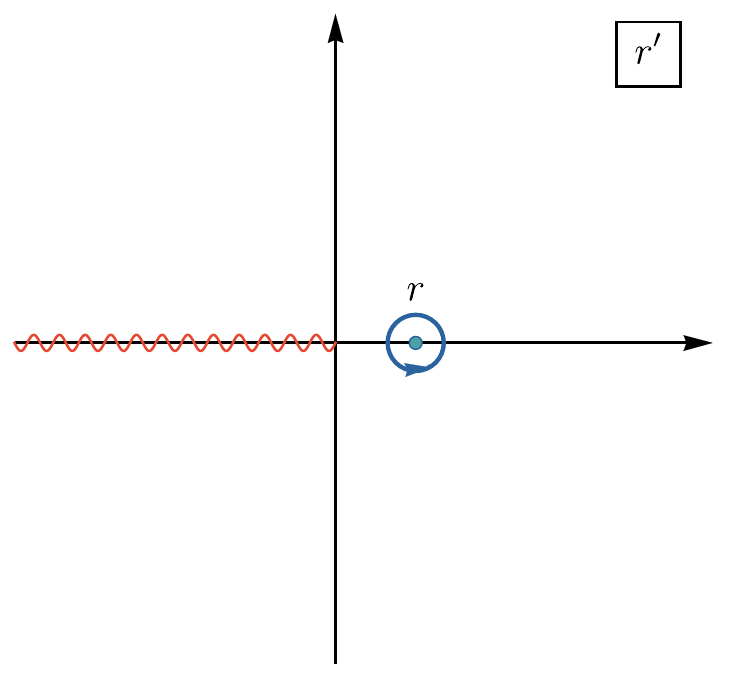}}
~~~~$\To$~~~~
\parbox{0.38\textwidth}{\includegraphics[width=0.38\textwidth]{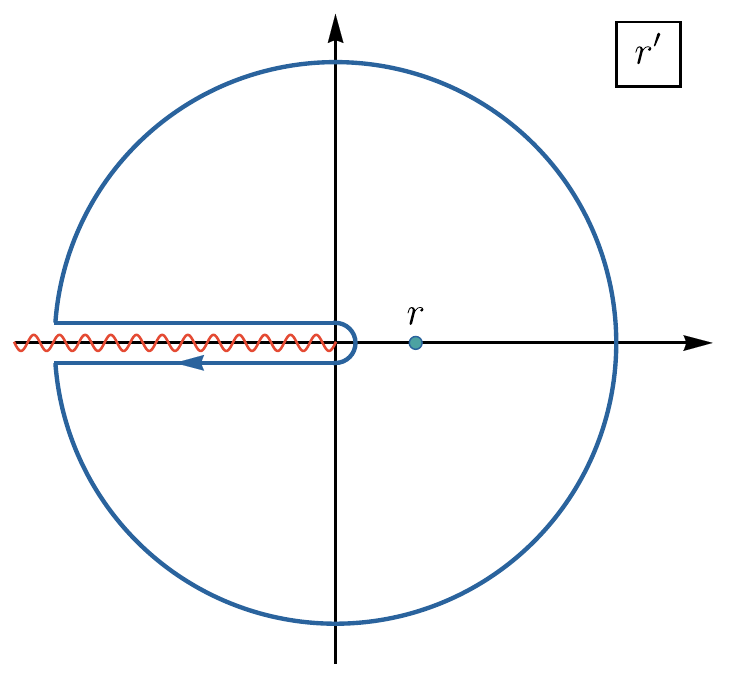}}
\caption{An illustration of the dispersion integral for the function $f(r)$ with a branch cut shown by the magenta wiggly line. (One can think of $f(r)$ as the correlator shown in Fig.\;\ref{fig_disp}.) So long as the function $f(r')$ in question decreases fast enough as $|r'|\to \infty$, one can deform the small circle $\mathcal{C}$ enclosing the point $r'=r$ to a larger contour $\mathcal{C}'$, and thus relate the value $f(r)$ with an integral along the branch cut.}
  \label{fig_dispcontour}
\end{figure}

 At the mathematical level, a dispersion relation is nothing but a clever manipulation of the contour integral on a complex plane. As a simple but very typical example, suppose we have $f(r)$ as a function of complex variable $r$, which possesses a branch cut along the negative real axis $r<0$, but is otherwise analytic everywhere. Furthermore, it is convenient (but not necessary) to assume that $f(r)$ decreases fast enough as $|r|\to\infty$. Suppose that all quantitative information we have about $f(r)$ is its \emph{discontinuity} along the branch cut:
\bge\label{def_disc}
  \mathop{\text{Disc}}_{r}f(r)\equiv \lim_{\ep\to 0^+}\Big[ f(r+\ii\ep)-f(r-\ii\ep)\Big].~~~~~(r\in\mathbb{R})
\ede
Then, a dispersion relation makes use of this quantitative information to recover the original function $f(r)$ for an arbitrary given point $r$ on the complex plane. As shown in the left panel of Fig.\;\ref{fig_dispcontour}, we enclose the given point $r$ by a small contour $\mathcal{C}$. Then, we have the following equality by virtue of the residue theorem:
\bge
  f(r)=\int_\mathcal{C}\FR{\di r'}{2\pi\ii}\FR{f(r')}{r'-r}.
\ede
Now, as shown in the right panel of Fig.\;\ref{fig_dispcontour}, we can deform the contour $\mathcal{C}$ to a big circle $\mathcal{C}'$ without changing the answer of the integration. The new contour $\mathcal{C}'$ is chosen with radius $|r'|\to \infty$ except on the negative real axis, to which the contour approaches from both sides. By our assumption of the analytical property of $f(r)$, the integration of $f(r')/(r-r')$ along the big circle at $|r'|\to \infty$ vanishes. Then, we get:
\bge
  f(r)=\int_{\mathcal{C}'}\FR{\di r'}{2\pi\ii}\FR{f(r')}{r'-r}=\int_{-\infty}^{+\infty}\FR{\di r'}{2\pi \ii}\FR{\mathop{\rm{Disc}}_{r'}f(r')}{r'-r}.
\ede
Thus, by performing an integration along the branch cut, we recover the value of $f(r)$ at any point $r$.

The requirement that $f(r')$ decreases faster enough when $|r'|\to \infty$ is to make sure that the integration over $f(r')/(r-r')$ vanishes along the large circle at infinity. This requirement can be loosen: So long as $f(r')$ is bounded by a power function of finite order, namely $|f(r')/r'^n|\to 0$ as $|r'|\to \infty$ for some $n\in\mathbb{Z}_+$, we can consider the following new function $g(r_1,\cdots,r_n;r')$:
\bge
 g(r_1,\cdots;r_n;r')\equiv \FR{f(r')}{(r'-r_1)\cdots (r'-r_n)} ,
\ede
where $r_1\cdots r_n$ are $n$ arbitrarily chosen points. Then it is clear that $g(r_1,\cdots,r_n;r')$ decreases fast enough at infinity. So, we can use $g$ in place of $f$ to do dispersion integral, at the expense that we need the values of $f(r')$ at $n$ discrete points $r'=r_1,\cdots, r_n$. This way of dealing with large-circle divergence is called \emph{subtraction}, and the number $n$ is called the \emph{order} of the subtraction.

It is worth mentioning that the study of dispersion relations has a long history in physics, with the Kramers-Krönig relation in classical electrodynamics as a notable early example \cite{kramers1927diffusion,deL.Kronig:26}. In the S-matrix bootstrap program for relativistic field theories, the dispersion relations played a central role  \cite{Eden:1966dnq,Mandelstam:1958xc,Toll:1956cya,Paulos:2016but,Danilkin:2014cra,Hoferichter:2018kwz,Garcia-Martin:2011nna,Bjorken:1965zz,Goldberger:1955zza,Goldberger:1955zz}. In these examples, the desired analytical property of the scattering amplitude is closely related to causality \cite{Eden:1966dnq,Mandelstam:1958xc,Toll:1956cya,Bjorken:1965zz,Goldberger:1955zz}. At the perturbative level, the analytical properties can also be diagnosed by methods such as Landau analysis \cite{Eden:1966dnq,Salcedo:2022aal,Bjorken:1965zz}. At a fixed order in perturbation theory, the dispersion relation relates loop amplitudes with tree amplitudes, and in well-situated cases, it allows one to reconstruct loop amplitudes from simpler tree amplitudes. More remarkably, one can exploit the dispersion relation beyond the perturbation theory \cite{Eden:1966dnq,Bjorken:1965zz}. This has been shown useful in the study of hadron physics, e.g., \cite{Goldberger:1955zza,Goldberger:1955zz}. Also, one can use the dispersion relation to connect UV and IR parts of a theory and derive nontrivial positivity bounds for low-energy effective theories \cite{Adams:2006sv,Nicolis:2009qm,deRham:2017avq,deRham:2017zjm}.

\subsection{Analytical continuation by contour deformation}
\label{sec_contour_deform}

To derive a dispersion relation for the seed integral in (\ref{seed_tree_4pt}), we need to understand its analytical property as a function of \emph{complex} energies. Now we face an obvious problem: While the original seed integral is well defined for energies taking physical values, it is not for arbitrary complex energies. Therefore, we need to redefine the seed integral so that it also applies to complex energies. We want to do it without evaluating the full integral. To see how this works, we demonstrate our method with three toy examples, before considering the full seed integral in the next subsection. 

\paragraph{One-fold integral} First, let us consider a complex function $I(z)$ for $z\in\mathbb{C}$ defined by the following integral:
\begin{align}
\label{eq_Iz}
    I(z)\equiv\int_{0}^{\infty}\di w\, e^{\ii zw},
\end{align}
where the integral contour is chosen to be the positive real axis. The integral is convergent when $z\neq 0$ and $\text{Arg}\, z\in(0,\pi)$, and is integrated to the following result:
\begin{align}
    I(z)=-\FR{1}{\ii z}.
\end{align}
Clearly, this expression is analytic everywhere in $z$ except when $z=0$. In particular, it is analytic for $\text{Arg}\,z\notin (0,\pi)$, where the original integral (\ref{eq_Iz}) is no longer well defined. So the question is how we can modify the original integral so that it is well defined for arbitrary $z\neq 0$. The answer for this example is simple enough. Indeed, let us consider the following integral:
\bge
  \tilde I_\de(z)\equiv  \int_0^{z^{-1}e^{\ii\de}\infty}\di w\,e^{\ii z w},
\ede 
where $\de$ is a small positive real number. That is, the contour is deformed to approach $w=\infty$ from the direction $\text{Arg}\,w=-\text{Arg}\,z+\de$. This integral is convergent for any $z\neq 0$, and, in the mean time, we have:
\bge
  \lim_{\de\to 0^+}\tilde I_\de(z)=I(z).~~~~(\text{Im}\,z>0)
\ede
Therefore, we can take $\tilde I_{\de\to 0^+}(z)$ as the analytical continuation of the original integral $I(z)$ to any $z\neq 0$. The lesson here is that, when $z$ takes a value at which the original integral $I(z)$ is not convergent (at infinity), we can deform the contour properly so that the integral is convergent again. 

\paragraph{One-fold integral with a branch cut}
Taking analytical continuation by deforming the integral contour may have obstructions when the integrand contains branch cuts. To see this, consider the following example:
\bge
\label{eq_Jtoy}
  J(z)\equiv\int_{0}^{\infty}\di w\, e^{\ii zw}\sqrt{w},
\ede
The situation is similar: The integral is well defined when $\text{Arg}\,z\in(0,\pi)$, but the integrated result is analytic in a larger region:
\bge
  J(z)=\FR{\sqrt{\pi}e^{3\ii\pi/4}}{2z^{3/2}}.~~~~~(0<\text{Arg}\,z<\pi)
\ede
This time we can consider the following integral:
\bge
  \tilde J_\de(z)= \int_0^{z^{-1}e^{\ii\de}\infty}\di w\,e^{\ii z w}\sqrt{w}.
\ede
We still have $\tilde J_{\de\to 0^+}(z)=J(z)$ when $\text{Im}\,z>0$. However, the new phenomenon here is that the integral possesses a branch point at $z=0$ due to the factor $\sqrt{w}$. 
For definiteness, we can take the branch cut to be along the negative real axis $w\in(-\infty,0)$. Then we see that this branch cut implies the existence of a branch cut of the integral $\tilde J_\delta(z)$ along $z\in (-\infty,0)$ when $\de\to0^+$. Let us compute the discontinuity of this branch cut:
\begin{align}
\label{eq_DiscJdelta}
  \mathop{\text{Disc}}_{z}\tilde J_\de(z)
 = &~\tilde J_\de(ze^{-\ii\ep})-\tilde J_\de(ze^{+\ii\ep})\n\\
 =&\int_0^{z^{-1}e^{+\ii\ep}e^{\ii\de}\infty}\di w\,e^{\ii z w}\sqrt{w}-\int_0^{z^{-1}e^{-\ii\ep}e^{\ii\de}\infty}\di w\,e^{\ii z w}\sqrt{w}\n\\
 =&-\int_0^{-\infty}\di u\,\mathop{\text{Disc}}_{u}\Big[ e^{-\ii|z|u e^{\ii\de}}\sqrt{u}\Big]
 = 2\ii\int_0^{\infty}\di u\, e^{+\ii|z|u e^{\ii\de}}\sqrt{u} .
\end{align}
Then we let $\de\to 0$, and get:
\begin{align}
  \mathop{\text{Disc}}_{z}J(z)
 =  2\ii\lim_{\de\to 0^+}\int_0^{\infty}\di u\, e^{+\ii|z|u e^{\ii\de}}\sqrt{u}
 =-\FR{\sqrt\pi e^{+\ii\pi/4}}{ |z|^{3/2}}.
\end{align}
Therefore, we have found a relation between the discontinuity of the integral $J(z)$ and the integrand. The lesson from this example is that deforming the contour to approach the branch cut of the integrand from two different directions will lead to a discontinuity of the integral itself, and this contour deformation procedure provides us a way to relate the discontinuities of the integral and the integrand.  

\paragraph{Two-fold integral}
We will have to deal with time orderings when studying the seed integral. So, our third example will be a two-fold time-ordered integral: 
\bge
\label{eq_Kz1z2}
  K(z_1,z_2)\equiv \int_{0}^\infty\di w_1\di w_2\,e^{\ii z_1w_1+\ii z_2w_2}\theta(w_1-w_2).
\ede
Again, when $\text{Im}\,z_1>0$ and $\text{Im}\,z_2>0$ hold at the same time, the integral is well defined, and can be directly integrated to:
\bge
  K(z_1,z_2)=-\FR{1}{z_1(z_1+z_2)}.~~~~~(\text{Im}\,z_1>0 \text{~and~}\text{Im}\,z_2>0)
\ede
Now we want to analyze the above integral for more general choice for $z_1$ and $z_2$. In particular, we assume that $z_2>0$ stays in the positive real axis, while $z_1\in\mathbb{C}$ can take arbitrary complex values. Then, we can first rewrite the original integral as an iterated integral:
\bge
  K(z_1,z_2)\equiv \int_{0}^\infty\di w_1\int_0^{w_1}\di w_2\,e^{\ii z_1w_1+\ii z_2w_2} =\FR{1}{\ii z_2}\int_{0}^\infty\di w_1 \,\Big[e^{\ii (z_1+z_2)w_1}-e^{\ii z_1w_1}\Big].
\ede
As shown above, the inner-layer integral is trivially convergent, and we only need to deal with the $w_1$-integral, which may be divergent. Now, we want to deform the contour of $w_1$-integral to make it convergent for any $z_1\neq 0$ and $z_1+z_2\neq 0$. As a consistent deformation of the original integral, we should use the same contour for both terms. Then, we need a judicious choice for the direction along which the contour goes to infinity. That is, we want to deform the integral contour in the following way:
\bge
  \FR{1}{\ii z_2}\bigg[\int_{0}^{(z_1+z_2)^{-1}e^{\ii\de_1}\infty}\di w_1 \,e^{\ii (z_1+z_2)w_1}-\int_{0}^{z_1^{-1}e^{\ii\de_2}\infty}\di w_1 \,e^{\ii z_1w_1}\bigg],
\ede
such that two conditions hold at the same time: 1) $0<\de_1,\de_2<\pi$ so that both integrals converge; 2) $\de_1-\de_2=\text{Arg}\,(z_1+z_2)-\text{Arg}\,z_2~\text{mod}\,2\pi$, so that both integrals share the same contour. Clearly, the two conditions can always be satisfied simultaneously, except when $\text{Arg}\,(z_1+z_2)-\text{Arg}\,z_1=\pi$, in which case no contour deformation works. For $z_2>0$, this corresponds to $z_1>-z_2$. So, we conclude that, the above contour deformation always works well for any $z_1\neq -z_2$ and $z_2>0$ so long as $z_1$ is away from an interval on the negative real axis $(-z_2,0)$. 

How to deal with this interval? The solution is to rewrite the original integral in a different way:
\bge
  K(z_1,z_2)=\int_{0}^\infty\di w_1\di w_2\,e^{\ii z_1w_1+\ii z_2w_2}\Big[1-\theta(w_2-w_1)\Big].
\ede  
Then, the first term is factorized and thus is trivial, and the second term is again a nested integral but with the role of $w_1$ and $w_2$ switched. Thus, all above analysis still applies to this nested integral, and the contour deformation trick applies for all $z_1\neq-z_2$ except in the interval $z_1\in(-\infty,-z_2)$.  

So the lesson is that, when we try to take the analytical continuation of a nested integral by deforming the contour, the two cases of $z_1<-z_2$ and $-z_2<z_1<0$ should be treated separately.

\subsection{Vertex dispersion relation of the seed integral}\label{subsection_vertex_disp}

Now we come back to the seed integral (\ref{seed_tree_4pt}). We observe that the integrands of $\mathcal{I}_{\mathsf{ab}}^{p_1p_2}$ contain exponential functions, power functions and Hankel functions. Moreover, the opposite-sign integrals $\mathcal{I}_{\pm\mp}^{p_1p_2}$ are \textit{factorized}, meaning that the integrands are of product form $f(\tau_1)g(\tau_2)$. On the contrary, the same-sign integrals $\mathcal{I}_{\pm\pm}^{p_1p_2}$ are \textit{nested}, due to the time-ordered factor $\theta(\tau_1-\tau_2)$ or $\theta(\tau_2-\tau_1)$. Although the seed integrals are much more complicated than the toy examples considered above, they share some common features. In particular, the integrand of a seed integral is regular along the integral path, so that any potential singular behavior must be from a diveregence in the early time limit. (The integral is always convergent in the late-time limit by our assumption of IR regularity.) Therefore, let us consider the asymptotic behavior of Hankel functions in the early time limit $\tau\to-\infty$:
\begin{align}\label{eq_Hankel_asymptotic}
    &\rm{H}^{(1)}_{\ii\nn}(-k\tau)\sim \FR{C_1}{\sqrt{-k\tau}}e^{-\ii k\tau},
    &&\rm{H}^{(2)}_{-\ii\nn}(-k\tau)\sim \FR{C_2}{\sqrt{-k\tau}}e^{+\ii k\tau},
\end{align}
where $C_{1}$ and $C_2$ are $k\tau$-independent constants. So, we see that, although the integrand of the seed integral is complicated, its behavior at the early time limit is simple, and is controlled by exponential functions, much like the toy examples considered above. Then, the previous discussion shows that, if we allow $\{k_{12},k_{34},k_s\}$ to be outside the physical region, the seed integral in its original form (\ref{seed_tree_4pt}) can not always be convergent. To make sense of the seed integral  for arbitrary $\{k_{12},k_{34},k_s\}$, we need analytical continuation. Below, we carry out this analytical continuation on the complex plane of the vertex energy $k_{12}$ while $k_{34}$ and $k_s$ are fixed within their physical regions. From this result, we will derive the vertex dispersion relation. 

\paragraph{Factorized integrals}
We start from the factorized integrals $\mathcal{I}_{\pm\mp}^{p_1p_2}$ which are free of time orderings and thus simpler.  Without loss of generality, we focus on $\mathcal{I}^{p_1p_2}_{+-}$, and the treatment for $\mathcal{I}^{p_1p_2}_{-+}$ is very similar. Below we rewrite $\mathcal{I}^{p_1p_2}_{+-}$ in an explicitly factorized form:
\begin{align}
    \label{eq_IpmtoJpJm}
    \mathcal{I}^{p_1p_2}_{\mathsf{+-}}(r_1,r_2)
    =&~\FR{\pi}{4}e^{-\pi\wt\nu}  \mathcal{U}^{p_1}_{+}(k_{12},k_s)\mathcal{U}^{p_2}_{-}(k_{34},k_s).\\
    \label{eq_Uplus}
    \mathcal{U}^{p_1}_{+}(k_{12},k_s)=&~k_s^{5/2+p_{1}}\int_{-\infty}^{0}\di\tau_1 \,(-\tau_1)^{3/2+p_1} e^{+\ii k_{12}\tau_1}\mathrm{H}_{-\ii\wt\nu}^{(2)}(-k_s\tau_1),\\
    \mathcal{U}^{p_2}_{-}(k_{34},k_s)=&~k_s^{5/2+p_{2}}\int_{-\infty}^{0} \di\tau_2\, (-\tau_2)^{3/2+p_2}e^{-\ii k_{34}\tau_2}\mathrm{H}_{\ii\wt\nu}^{(1)}(-k_s\tau_2).
\end{align}
To analyze the behavior of $\mathcal{I}^{p_1p_2}_{+-}$ on the complex $k_{12}$ plane, it suffices to consider $\mathcal{U}^{p_1}_{+}(k_{12},k_s)$ alone. Of course, the integrals  $\mathcal{U}^{p_1}_{\pm}(k_{12},k_s)$ are simple enough to be done directly. However, we prefer to analyze their analytical structure without really evaluating them. Thus we will defer the direct integration until next section, where one can find the explicit results of $\mathcal{U}^{p_1}_{\pm}(k_{12},k_s)$ in (\ref{eq_UplusResult}).

For convenience, let us fix $k_s$ in the physical region $k_s>0$. (We can also fix $k_{34}$ in the physical region $k_{34}>k_s$ although this is irrelevant for the analysis of $\mathcal{U}^{p_1}_{+}(k_{12},k_s)$.) From (\ref{eq_Hankel_asymptotic}), we know that, at the early time limit, the integrand of $\mathcal{U}^{p_1}_{+}(k_{12},k_s)$ is controlled by the exponential factor:
\begin{align}
    e^{\ii(k_{12}+k_s)\tau_1} .
\end{align}
Therefore, for fixed integral contour $\tau_1\in(-\infty,0)$, the phase of $k_{12}+k_s$ controls the convergence of $\mathcal{U}^{p_1}_+(k_{12},k_s)$ when $\tau_1\to-\infty$. For example, if $\rm{Im}[k_{12}+k_s]>0$, the integral in (\ref{eq_Uplus}) will diverge, although the function $\mathcal{U}^{p_1}_+(k_{12},k_s)$ can actually be analytically continued to this region. In order to make this analytical continuation, we improve the definition of $\mathcal{U}_+^{p_1}$ in (\ref{eq_Uplus}) by deforming the integration contour of $\mathcal{U}^{p_1}_+(k_{12},k_s)$ in the following way, similar to what we did for the first two toy examples before:
\begin{align}
\label{eq_Udef}
  {\mathcal{U}}_{+}^{p_1}(k_{12},k_s)\equiv k_s^{5/2+p_{1}}\int_{-(1-\ii 0^+)(k_{12}+k_s)^{-1}\infty}^{0}\di\tau_1 \,(-\tau_1)^{3/2+p_1} e^{\ii k_{12}\tau_1}\mathrm{H}_{-\ii\wt\nu}^{(2)}(-k_s\tau_1).
\end{align}
Clearly, this new definition agrees with (\ref{eq_Uplus}) for all $k_{12}$ and $k_s$ in the interior of the physical region. On the other hand, the change of integration path is continuous in $k_{12}$, so is the integral ${\mathcal{U}}_{+}^{p_1}(k_{12},k_s)$ for generic values of $k_{12}$. (One can see this point more explicitly by taking derivative of ${\mathcal{U}}_{+}^{p_1}(k_{12},k_s)$ with respect to $k_{12}$.) However, like the second toy example (\ref{eq_Jtoy}), the integrand of ${\mathcal{U}}_{+}^{p_1}(k_{12},k_s)$ contains a branch point at $\tau_1=0$ due to the Hankel function and the power factor. The branch point emanates a branch cut which we take to be on the positive real axis $\tau_1\in(0,+\infty)$, and this branch cut can be an obstacle for contour deformation. As a result, when the integration contour is brought to the vicinity of the branch cut of the integrand, a discontinuity may occur and lead to a branch cut for ${\mathcal{U}}_{+}^{p_1}(k_{12},k_s)$ with respect to $k_{12}$. 

Since the branch cut of the integrand in (\ref{eq_Udef}) is on the positive real axis of $\tau_1$, the integral contour of ${\mathcal{U}}_{+}^{p_1}(k_{12},k_s)$ has a chance to approach this branch cut if $(k_{12}+k_s)$ has a phase close to $\pm\pi$. For $k_s>0$, this corresponds to $k_{12}\in(-\infty,-k_s)$. Thus we conclude that the only possible branch cut of ${\mathcal{U}}_{+}^{p_1}(k_{12},k_s)$ on the complex $k_{12}$ plane for fixed $k_s>0$ is in the interval $k_{12}\in (-\infty,-k_s)$, with the two branch points $k_{12}=-\infty$ and $k_{12}=-k_s$. (In fact, the point $k_{12}=-k_s$ is often divergent, since the integral ${\mathcal{U}}_{+}^{p_1}(k_{12},k_s)$ at this point is typically divergent in the early time limit no matter how we deform the contour. This is called a partial energy singularity in the literature.) Apart from this integral as well as the two endpoints, the function ${\mathcal{U}}_{+}^{p_1}(k_{12},k_s)$ is analytical in $k_{12}$ everywhere else. 

Now let us determine the discontinuity across the branch cut of ${\mathcal{U}}_{+}^{p_1}(k_{12},k_s)$  at $k_{12}\in(-\infty, -k_s)$. Using the method identical to (\ref{eq_DiscJdelta}), we have:
\begin{align}
    \mathop{\rm{Disc}}_{k_{12}}{\mathcal{U}}_{+}^{p_1}(k_{12},k_s)
    = k_s^{5/2+p_1} \int_{\infty}^{0}\di\tau_1\,\mathop{\rm{Disc}}_{\tau_1}\Bigl[e^{+\ii k_{12}\tau_1}(-\tau_1)^{3/2+p_1}\rm{H}^{(2)}_{-\ii\nn}(-k_s\tau_1)\Bigr].~~~~(k_{12}<-k_s)
\end{align}
Then, using the known discontinuity of the power function and the Hankel function as given in (\ref{disc of hankel}) and (\ref{disc of power}), we get:
\begin{align}
   \mathop{\rm{Disc}}_{\tau_1}\Bigl[e^{+\ii k_{12}\tau_1}(-\tau_1)^{3/2+p_1}\rm{H}^{(2)}_{-\ii\nn}(-k_s\tau_1)\Bigr] 
    = 2\ii\cosh(\pi\nn)(-1)^{p_1}e^{+\ii k_{12}\tau_1}\tau_1^{3/2+p_1}\rm{H}^{(2)}_{-\ii\nn}(k_s\tau_1) \theta(\tau_1) . 
\end{align}
Therefore, 
\begin{align}
    \mathop{\rm{Disc}}_{k_{12}}\mathcal{U}_{+}^{p_1}(k_{12},k_s)=&~2\ii\cosh(\pi\nn)(-1)^{p_1} k_s^{5/2+p_1} \int_{\infty}^{0}\di\tau_1\ e^{+\ii k_{12}\tau_1}\tau_1^{3/2+p_1}\rm{H}^{(2)}_{-\ii\nn}(k_s\tau_1)\n\\
    =&~2\ii\cosh(\pi\nn)(-1)^{p_1+1} k_s^{5/2+p_1} \int_{-\infty}^{0}\di\tau_1\ e^{-\ii k_{12}\tau_1}(-\tau_1)^{3/2+p_1}\rm{H}^{(2)}_{-\ii\nn}(-k_s\tau_1)\n\\
    =&~2\ii\cosh(\pi\nn)(-1)^{p_1+1}\mathcal{U}_{+}^{p_1}(-k_{12},k_s).
\end{align}
Now it is trivial to put back all factors independent of $k_{12}$ in (\ref{eq_IpmtoJpJm}), and get the discontinuity for $\mathcal{I}_{+-}^{p_1p_2}$. The discontinuity of the other factorized seed integral $\mathcal{I}_{-+}^{p_1p_2}$ can be analyzed in the same way and the result is very similar. So we summarize the results for both factorized seed integrals together:
\begin{keyeqn}
\begin{align}
\label{eq_Disck12Ipmmp}
    \mathop{\rm{Disc}}_{k_{12}}\mathcal{I}_{\pm\mp}^{p_1p_2}(k_{12},k_{34},k_s)=2\ii\cosh(\pi\nn)(-1)^{p_1+1}\mathcal{I}_{\pm\mp}^{p_1p_2}(-k_{12},k_{34},k_s)\theta(-k_{12}-k_s). 
\end{align}
\end{keyeqn} 

\paragraph{Nested integrals}
Now we move on to the nested seed integrals $\mathcal{I}_{\pm\pm}^{p_1p_2}$. We will focus on $\mathcal{I}_{++}^{p_1p_2}$ and the treatment of $\mathcal{I}_{--}^{p_1p_2}$ is similar. Substituting the same-sign type propagators (\ref{eq_Dpmpm}) in (\ref{seed_tree_4pt}), we get an explicit expression for the integral $\mathcal{I}_{++}^{p_1p_2}$:
\begin{align}
\label{eq_IppFull}
  &\mathcal{I}_{++}^{p_1p_2}(k_{12},k_{34},k_s)
  =-\FR{\pi}4 e^{-\pi\wt\nu}k_s^{5+p_{12}}\int_{-\infty}^{0}\di\tau_1\di\tau_2\, (-\tau_1)^{3/2+p_1}(-\tau_2)^{3/2+p_2}e^{\ii k_{12}\tau_1+\ii k_{34}\tau_2}\n\\
    &\times\Big[\rm{H}^{(1)}_{\ii\nn}(-k_s\tau_1)\rm{H}^{(2)}_{-\ii\nn}(-k_s\tau_2)\theta(\tau_1-\tau_2)+\rm{H}^{(2)}_{-\ii\nn}(-k_s\tau_1)\rm{H}^{(1)}_{\ii\nn}(-k_s\tau_2)\theta(\tau_2-\tau_1)\Big].
\end{align}
As before, we fix $k_s$ and $k_{34}$ to be in the interior of the physical region, i.e., $k_{34}>k_s>0$, and analyze the integral $\mathcal{I}_{++}^{p_1p_2}(k_{12},k_{34},k_s)$ on the complex $k_{12}$ plane, where we need to perform analytical continuation by contour deformation.

The way to deform the contour has been indicated in the third toy example (\ref{eq_Kz1z2}). In particular, for fixed values of $k_{34}>k_s>0$ and for arbitrary real $k_{12}$, we need to consider separately two cases: $k_{12}<-k_{34}$ and $-k_{34}<k_{12}<k_s$, both in the unphysical region. In each case, we need to pick up a specific ordering for the two time variables. 

Let us first analyze the case of $k_{12}<-k_{34}$, for which we choose to rewrite the integral as:
\begin{align}
    \mathcal{I}_{++}^{p_1p_2}=&~\mathcal{I}_{++,\rm{F},>}^{p_1p_2}+\mathcal{I}_{++,\rm{N},>}^{p_1p_2},\\
    \label{Ipp_F_>}
    \mathcal{I}_{++,\rm{F},>}^{p_1p_2}\equiv&-\FR{\pi}{4}e^{-\pi\nn}k_s^{5+p_{12}}\int_{-\infty}^{0}\di\tau_1\di\tau_2\, e^{\ii k_{12}\tau_1+\ii k_{34}\tau_2}\n\\
    &~\times (-\tau_1)^{3/2+p_1}(-\tau_2)^{3/2+p_2}\rm{H}^{(1)}_{\ii\nn}(-k_s\tau_1)\rm{H}^{(2)}_{-\ii\nn}(-k_s\tau_2),\\
\label{Ipp_N_>}
    \mathcal{I}_{++,\rm{N},>}^{p_1p_2}\equiv&-\FR{\pi}{4}e^{-\pi\nn}k_s^{5+p_{12}}\int_{-\infty}^{0}\di\tau_1\di\tau_2\, e^{\ii k_{12}\tau_1+\ii k_{34}\tau_2}(-\tau_1)^{3/2+p_1}(-\tau_2)^{3/2+p_2}\n\\
    &\times\Big[\rm{H}^{(2)}_{-\ii\nn}(-k_s\tau_1)\rm{H}^{(1)}_{\ii\nn}(-k_s\tau_2)-\rm{H}^{(1)}_{\ii\nn}(-k_s\tau_1)\rm{H}^{(2)}_{-\ii\nn}(-k_s\tau_2)\Big]\theta(\tau_2-\tau_1).
\end{align}
The subscript $>$ means that we are working with the condition $|k_{12}|>|k_{34}|$. This notation is in line with the one taken in \cite{Qin:2022fbv}. The analysis for the factorized integral $\mathcal{I}_{++,\rm{F},>}^{p_1p_2}$ is identical to that for $\mathcal{I}_{\pm\mp}^{p_1p_2}$, and we have:
\begin{align}
   \mathop{\rm{Disc}}_{k_{12}}\mathcal{I}_{++,\rm{F},>}^{p_1p_2}(k_{12},k_{34},k_s)
    =
    &~2\ii\cosh(\pi\nn)(-1)^{p_1+1}\mathcal{I}_{++,\rm{F},>}^{p_1p_2}(-k_{12},k_{34},k_s)\theta(-k_{12}-k_{34}).
\end{align}
On the other hand, given the asymptotic behavior of the Hankel functions (\ref{eq_Hankel_asymptotic}), the analysis for the iterated integral $\mathcal{I}_{++,\rm{N},>}^{p_1p_2}$ is in parallel with the one for our third toy example (\ref{eq_Kz1z2}). In particular, when $\tau_1,\tau_2\to-\infty$, the integrand of $\mathcal{I}_{++,\rm{N},>}^{p_1p_2}$ behaves, up to unimportant power functions (denoted as $\#$), like:
\bge
  \# e^{\ii (k_{12}+k_s)\tau_1}e^{\ii (k_{34}-k_s)\tau_2}
  -\#  e^{\ii (k_{12}-k_s)\tau_1}e^{\ii (k_{34}+k_s)\tau_2}.
\ede
Thus, after finishing the inner-layer integral over $\tau_2$, we get four terms which behave in the $\tau_1\to-\infty$ limit like (again, up to unimportant power functions and constant coefficients):
\begin{align}
  &e^{\ii (k_{12}+k_{34})\tau_1},
  &&e^{\ii (k_{12}+k_s)\tau_1},
  &e^{\ii (k_{12}+k_{34})\tau_1} ,
  &&e^{\ii (k_{12}-k_s)\tau_1} .
\end{align}
Therefore, for fixed $k_{34}>k_s>0$, one can deform the integration contour on the $\tau_1$ plane to make all above four terms convergent, if $k_{12}$ is away from the interval $(-k_{34},0)$. This is exactly the condition $k_{12}<-k_{34}$ that we imposed from the very beginning. Then, we see that the integral $\mathcal{I}_{++,\rm{N},>}^{p_1p_2}$ is analytic everywhere in $k_{12}$ when $k_{12}$ is away from the negative real axis. On the negative real axis, the interval $(-k_{34},0)$ is not covered by the current case, while the interval $(-\infty,-k_{34})$ may contain a branch cut due to the potential discontinuities of the integrand of $\mathcal{I}_{++,\rm{N},>}^{p_1p_2}$.

However, by a direct computation, we can show that $\mathcal{I}_{++,\rm{N},>}^{p_1p_2}$ is in fact free of branch cut even in $(-\infty,-k_{34})$. Explicitly: 
\begin{align}\label{eq_DiscIppN}
  &\mathop{\rm{Disc}}_{k_{12}}\mathcal{I}_{++,\rm{N},>}^{p_1p_2}(k_{12},k_{34},k_s)\n\\
  =&-\FR{\pi}4 e^{-\pi\wt\nu}k_s^{5+p_{12}}\lim_{\epsilon\to0^+}\bigg\{\int_{\infty e^{\ii\epsilon}}^{0}\di\tau_1\int_{\tau_1}^{0}\di\tau_2\ e^{\ii(-k_{12}+\ii\epsilon)\tau_1+\ii k_{34}\tau_2}(-\tau_1)^{3/2+p_1}(-\tau_2)^{3/2+p_2} \n\\
  & \times\Big[\rm{H}^{(2)}_{-\ii\nn}(-k_s\tau_1)\rm{H}^{(1)}_{\ii\nn}(-k_s\tau_2)-\rm{H}^{(1)}_{\ii\nn}(-k_s\tau_1)\rm{H}^{(2)}_{-\ii\nn}(-k_s\tau_2)\Big]-(\ep\to-\ep)\bigg\}.
\end{align}
We can reparameterize the two time variables in (\ref{eq_DiscIppN}) so that the two nested integrals in the curly brackets can be combined: 
\begin{align}
\label{eq_DiscIppN2}
 &-\FR{\pi}{4}e^{-\pi\nn}k_s^{5+p_{12}}\int_{\infty}^{0}\di \tau_1\int_{\tau_1}^{0}\di\tau_2\bigg\{ e^{\ii(-k_{12}+\ii\ep)\tau_1^+}e^{\ii k_{34}\tau_2^+}(-\tau_1^+)^{3/2+p_1}(-\tau_2^+)^{3/2+p_2}\n\\
  &\times \Big[\rm{H}^{(2)}_{-\ii\nn}(-k_s\tau_1^+)\rm{H}^{(1)}_{\ii\nn}(-k_s\tau_2^+)-\rm{H}^{(1)}_{\ii\nn}(-k_s\tau_1^+)\rm{H}^{(2)}_{-\ii\nn}(-k_s\tau_2^+)\Big]-(\ep\to-\ep)\bigg\},
\end{align}
where $\tau_{1,2}^+\equiv \tau_{1,2}e^{\ii\ep}$. 
Thus, (\ref{eq_DiscIppN2}) says that we can find the discontinuity of the nested integral $\mathcal{I}_{++,\rm{N},>}^{p_1p_2}$ by computing a ``discontinuity'' of its integrand. Then, using the known discontinuities of the Hankel and power functions on their branch cuts, collected in (\ref{disc of hankel}) and (\ref{disc of power}), it is straightforward to show that the integrand of (\ref{eq_DiscIppN2}) actually vanishes. Thus we conclude that $\mathcal{I}_{++,\rm{N},>}^{p_1p_2}$ has no branch cut in $k_{12}$ when $k_{12}\in(-\infty,-k_{34})$.

The other case with $-k_{34}<k_{12}<k_s$ always satisfies $|k_{12}|<|k_{34}|$, and therefore we separate the integral $\mathcal{I}_{++}^{p_1p_2}$ in a different way:
\begin{align}
    \mathcal{I}_{++}^{p_1p_2}
    =&~\mathcal{I}_{++,\rm{F},<}^{p_1p_2}+\mathcal{I}_{++,\rm{N},<}^{p_1p_2},\\
    \label{Ipp_F_<}
    \mathcal{I}_{++,\rm{F},<}^{p_1p_2}
    \equiv&-\FR{\pi}{4}e^{-\pi\nn}k_s^{5+p_{12}}\int_{-\infty}^{0}\di\tau_1\di\tau_2\ e^{\ii k_{12}\tau_1}e^{\ii k_{34}\tau_2}\n\\
    &\times(-\tau_1)^{3/2+p_1}(-\tau_2)^{3/2+p_2}\rm{H}^{(2)}_{-\ii\nn}(-k_s\tau_1)\rm{H}^{(1)}_{\ii\nn}(-k_s\tau_2),\\
    \label{Ipp_N_<}
    \mathcal{I}_{++,\rm{N},<}^{p_1p_2}
    \equiv&-\FR{\pi}{4}e^{-\pi\nn}k_s^{5+p_{12}}\int_{-\infty}^{0}\di\tau_1\di\tau_2\ e^{\ii k_{12}\tau_1}e^{\ii k_{34}\tau_2}(-\tau_1)^{3/2+p_1}(-\tau_2)^{3/2+p_2}\n\\
    &\times\bigg[\rm{H}^{(1)}_{\ii\nn}(-k_s\tau_1)\rm{H}^{(2)}_{-\ii\nn}(-k_s\tau_2)-\rm{H}^{(2)}_{-\ii\nn}(-k_s\tau_1)\rm{H}^{(1)}_{\ii\nn}(-k_s\tau_2)\bigg]\theta(\tau_1-\tau_2) .
\end{align}
Like before, the subscript ``$<$'' here means that the way we split the integral works when $|k_{12}|<|k_{34}|$. Then, in complete parallel with the previous case, we can show that the the factorized part $\mathcal{I}_{++,\rm{F},<}^{p_1p_2}$ has a branch cut in the interval $-k_{34}<k_{12}<-k_s$, whose discontinuity is proportional to the factorized integral $\mathcal{I}_{++,\rm{F},<}^{p_1p_2}$ itself:
\begin{align}
  \mathop{\rm{Disc}}_{k_{12}}\mathcal{I}^{p_1p_2}_{++,\rm{F},<}(k_{12},k_{34},k_s) 
  &=2\ii\cosh(\pi\nn)(-1)^{p_1+1}\mathcal{I}^{p_1p_2}_{\pm\pm,\rm{F},<}(-k_{12},k_{34},k_s)\theta(k_{12}+k_{34})\theta(-k_{12}-k_s).
\end{align}
On the other hand, the nested part $\mathcal{I}_{++,\rm{N},<}^{p_1p_2}$ does not have any branch cut in the region where it is defined (namely, $|k_{12}|<|k_{34}|$). Therefore, the discontinuity in this case is also fully from the factorized integral. 

Above we have present a detailed analysis for the integral $\mathcal{I}_{++}^{p_1p_2}$. The treatment for $\mathcal{I}_{--}^{p_1p_2}$ is completely the same. In particular, one can separate $\mathcal{I}_{--}^{p_1p_2}$ into $\mathcal{I}_{{--},\text{F},\gtrless}^{p_1p_2}$ and $\mathcal{I}_{--,\text{N},\gtrless}^{p_1p_2}$  when $|k_{12}|\gtrless|k_{34}|$. So, we can summarize our result for both same-sign seed integrals as follows:
\begin{keyeqn}
\begin{align}
\label{eq_Disck12Ipmpm}
    \mathop{\rm{Disc}}_{k_{12}}\mathcal{I}_{\pm\pm}^{p_1p_2}(k_{12},k_{34},k_s)
        =2\ii\cosh(\pi\nn)(-1)^{p_1+1}\mathcal{I}_{\pm\pm,\text{F},>}^{p_1p_2}(-k_{12},k_{34},k_s)\theta(-k_{12}-k_{34})\n\\
    +2\ii\cosh(\pi\nn)(-1)^{p_1+1}\mathcal{I}^{p_1p_2}_{\pm\pm,\rm{F},<}(-k_{12},k_{34},k_s)\theta(k_{12}+k_{34})\theta(-k_{12}-k_s).
\end{align}
\end{keyeqn}

\begin{figure}
\centering 
\includegraphics[width=0.5\textwidth]{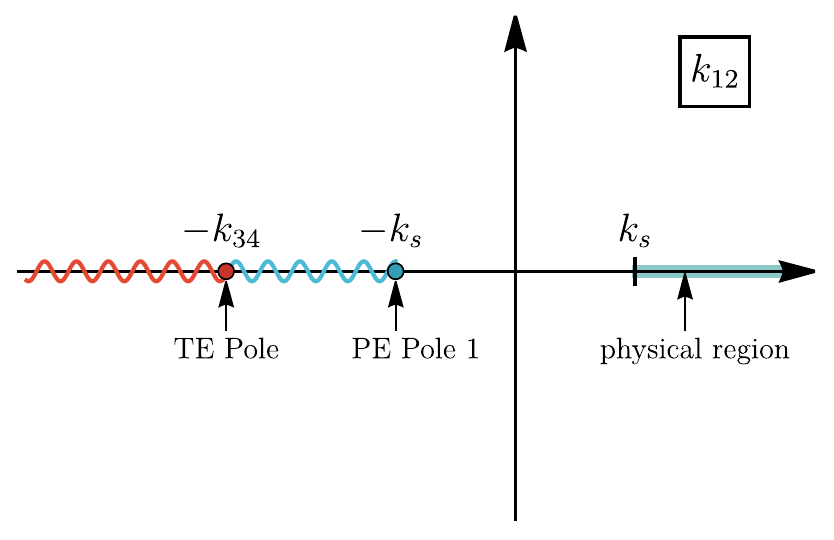} 
\caption{The analytical structure of the seed integral $\mathcal{I}^{p_1p_2}(k_{12},k_{34},k_s)$ in (\ref{seed_tree_4pt_full}) on the complex $k_{12}$ plane, with $k_{34}$ and $k_s$ staying in their physical regions $k_{34}>k_s>0$. The physical region for $k_{12}$ ($k_{12}\geq k_s$) is marked in green. The seed integral is analytic in $k_{12}$ everywhere except when $k_{12}\in (-\infty,-k_s]$. In this interval, we have three branch points: a partial energy (PE) pole at $k_{12}=-k_s$, the total energy (TE) pole at $k_{12}=-k_{34}$. The other partial energy pole $k_{34}=-k_s$ is never met for finite $k_{12}$, and can be thought of as sitting at $k_{12}=-\infty$ due to the scale invariance of the seed integral. The branch cut can be put in the entire interval $k_{12}\in(-\infty,-k_s]$, with the discontinuity itself being discontinuous at $k_{12}=-k_{34}$.  }
  \label{fig_k12}
\end{figure}

\paragraph{Summary} Now we have completed the analysis for the seed integral $\mathcal{I}^{p_1p_2}_{\aa\bb}$ on the complex $k_{12}$ plane, with $k_{34}$ and $k_s$ fixed in the interior of their physical region $k_{34}\geq k_s\geq 0$. The discontinuities of all four SK branches are given in (\ref{eq_Disck12Ipmmp}) and (\ref{eq_Disck12Ipmpm}), respectively. 

When performing the dispersion integrals, we don't have to separate the seed integral according to their SK branches. Therefore, it is useful to sum over SK indices $\aa,\bb=\pm$ and to get the analytical structure for the full seed integral $\mathcal{I}^{p_1p_2}$ in (\ref{seed_tree_4pt_full}):
\begin{keyeqn}
\begin{align}\label{disc_I_4pt_k}
  \mathop{\text{Disc}}_{k_{12}}\mathcal{I}^{p_1p_2}(k_{12},k_{34},k_s)=2\ii\cosh(\pi\nn)(-1)^{p_1+1}\mathcal{I}^{p_1p_2}_\text{S}(-k_{12},k_{34},k_s)\theta(-k_{12}-k_s).
\end{align}
\end{keyeqn}
In this expression, we have defined the \emph{signal} part of the seed integral as:
\begin{align}
\label{eq_IS}
  \mathcal{I}_\text{S}^{p_1p_2}(k_{12},k_{34},k_s)
  =&~\Bigl(\mathcal{I}_{++,\text{F},>}^{p_1p_2}(k_{12},k_{34},k_s)+\mathcal{I}_{--,\text{F},>}^{p_1p_2}(k_{12},k_{34},k_s)\Bigr)\theta\Big(|k_{12}|-|k_{34}|\Big)\n\\
  &+\Bigl(\mathcal{I}_{++,\text{F},<}^{p_1p_2}(k_{34},k_{12},k_s)+\mathcal{I}_{--,\text{F},<}^{p_1p_2}(k_{12},k_{34},k_s)\Bigr)\theta\Big(|k_{34}|-|k_{12}|\Big)\n\\
  &+\mathcal{I}_{+-}^{p_1p_2}(k_{12},k_{34},k_s)+\mathcal{I}_{-+}^{p_1p_2}(k_{12},k_{34},k_s).
\end{align}
Eqs.\;(\ref{disc_I_4pt_k}) and (\ref{eq_IS}) are the main results of this section. They form the basis for the vertex dispersion relation, detailed in the next section. 

Note that the ``signal'' defined in (\ref{eq_IS}) is nothing but all factorized pieces in (\ref{eq_Disck12Ipmmp}) and (\ref{eq_Disck12Ipmpm}) summed, and it is this signal piece that is responsible for all discontinuities of the seed integral on $k_{12}$ plane. On the other hand, it does agree with the signal defined in previous works through the analytical properties in $k_s/k_{12}$ and $k_s/k_{34}$ as $k_s\to 0$ \cite{Tong:2021wai,Qin:2022lva,Qin:2022fbv}. Thus the results (\ref{disc_I_4pt_k}) and (\ref{eq_IS}) make precise our intuition that the CC signal corresponds to the nonanalyticity of the correlator.

\section{Bootstrapping Correlators with Vertex Dispersion Relation}
\label{sec_vertex_disp}

In this section, we will put the vertex dispersion relation in use, to bootstrap a few 3-point and 4-point correlators with massive exchanges. We begin with the simplest example, the 3-point tree correlator with single massive exchange in Sec.\ \ref{subsec_3ptTree}. The dispersive bootstrap yields a closed-form analytical expression for this example, identical to the one found with improved bootstrap equation in \cite{Qin:2023ejc}. Then, in Sec.\ \ref{subsec_3ptLoop}, we bootstrap the 3-point correlator mediated by two massive fields via a bubble loop. We will show that, with an additional input of spectral decomposition explored in a previous work \cite{Xianyu:2022jwk}, the vertex dispersion relation can be generalized to loop processes, leading to analytical expressions much simpler than the one found with pure spectral method in \cite{Xianyu:2022jwk}. In particular, our one-loop result here features a neat separation of the renormalization-dependent local part and the renormalization-independent nonlocal part, thus allows for unambiguous extraction of on-shell effects from loop process.  Finally, in Sec.\ \ref{subsec_4ptTree_VB}, we bootstrap the 4-point correlator with a single massive exchange in the $s$-channel. This is a well studied example, and we use it to demonstrate the use of vertex dispersion relation for  kinematics more complicated than 3-point examples.

\subsection{Three-point single-exchange graph}
\label{subsec_3ptTree}

\begin{figure}[t]
    \centering
    \includegraphics[width=0.50\textwidth]{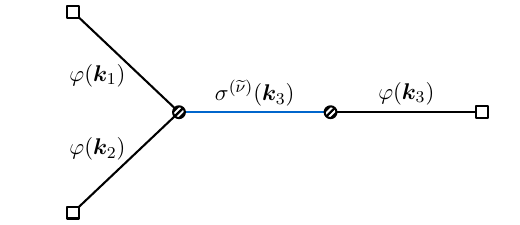}
    \caption{The 3-point correlator of a massless scalar $\varphi$ with a single massive scalar exchange $\si$. }
    \label{fd_3pt_tree}
\end{figure}

We begin with the simplest nontrivial example, namely a 3-point correlator with a single massive exchange. To be specific, we will consider the single massive exchange from the following interactions:
\begin{align}
\label{eq_Lag3pt}
  \Delta\ld=\lam_2 a^3\varphi'\si+\FR{1}{2}\lam_3a^2\varphi'^{\,2}\si,
\end{align}
where $\varphi$ is a massless scalar field (typically the inflaton fluctuation in the context of CC physics), and $\si$ is a real massive scalar field of mass $m$. For convenience, we take $m>3/2$ so that the mass parameter $\wt\nu$ is a positive real, although generalization to light mass $0<m<3/2$ is straightforward. Also, $\lam_3$ and $\lam_2$ in (\ref{eq_Lag3pt}) are coupling constants, and the powers of the scale factor $a=-1/\tau$ are inserted to ensure the scale invariance. Then, there is a single independent tree diagram, shown in Fig.\ \ref{fd_3pt_tree} that contributes to the 3-point correlator $\la\varphi\varphi\varphi\ra$ at the leading order $\order{\lam_2\lam_3}$, together with other two obtained by trivial momentum permutations. This process appears in a simple realization of the original quasi-single-field inflation with dim-6 inflaton-spectator coupling $(\pd_\mu\phi)^2\si^2$ \cite{Wang:2019gbi}, and turns out to be the leading signal in this model with comparable signal strength with double-massive-exchange and triple-massive-exchange graphs.

The time integral for the diagram in Fig.\ \ref{fd_3pt_tree} can be expressed in terms of the 4-point seed integral in (\ref{seed_tree_4pt_full}) as:
\bge
  \la\varphi_{\bm k_1}\varphi_{\bm k_2}\varphi_{\bm k_3}\ra'= \FR{\lam_2\lam_3}{k_1k_2k_3^4}\bigg[\mathcal{I}^{0,-2}(k_{12},k_3,k_3)+\text{2 perms}\bigg],
\ede
Therefore, technically, the 3-point function we are going to compute can be viewed as a limiting case of a 4-point correlator with $k_4\to 0^+$.

Then, the problem reduces to the computation of $\mathcal{I}^{0,-2}(k_{12},k_{3},k_3)$. For this particular integral, it turns out useful to use a new variable $u\equiv2k_{3}/k_{123}$. With this definition, the physical region $0\leq k_{3}\leq k_{12}$ can be written as $u\in[0,1]$. It is known that this variable is useful for obtaining closed-form analytical expressions for many 3-point functions \cite{Qin:2023ejc}. From the perspective of dispersion integral, the simplification can be observed from the fact that the partial-energy limit and the total-energy limit merge into a single limit $u\to -\infty$, while the branch point $k_{12}=\infty$ corresponds to $u=0$. Thus, the branch cut of the full 3-point function extends from $0$ to $-\infty$ on the entire negative real axis on the complex $u$ plane, which makes the dispersion integral simpler. 

To avoid potential confusions, we use a new notation $\mathcal{X}(u)$ to denote the dimensionless three-point seed integral as a function of $u=2k_3/k_{123}$:
\bge
\label{eq_XtoI}
  \mathcal{X}_{\aa\bb}(u=2k_3/k_{123})\equiv\mathcal{I}^{0,-2}_{\aa\bb}(k_{12},k_{3},k_3).
\ede
Then, we can rewrite the full 3-point seed integral as:
\begin{align}
  \mathcal{X}(u)\equiv\sum_{\aa,\bb=\pm}\mathcal{X}_{\aa\bb}(u)=\Big[\mathcal{X}_{++,\text{N}}(u)+\mathcal{X}_{++,\text{F}}(u)+\mathcal{X}_{+-}(u)\Big]+\text{c.c.},
\end{align}
where $\mathcal{X}_\text{++,\text{N}}$ and $\mathcal{X}_\text{++,\text{F}}$ are nested and factorized part of $\mathcal{X}_{++}$, defined from the corresponding seed integrals as in (\ref{eq_XtoI}); See (\ref{Ipp_F_>}) and (\ref{Ipp_N_>}). 

Our dispersive bootstrap of $\mathcal{X}(u)$ comes naturally with two steps following the result in (\ref{disc_I_4pt_k}): First, we will compute the signal part $\mathcal{X}(u)$ and its discontinuity across the branch cut. Second, we will perform the dispersion integral along the branch cut to get the full result. Below we carry out these two steps in turn.

\paragraph{Computing the signal}
From the analysis of the previous section, we know that the discontinuity of $\mathcal{X}(u)$ on the negative real axis is fully from $\mathcal{X}_{\pm\pm,\text{F}}(u)$ and $\mathcal{X}_{\pm\mp}(u)$, which can be combined together into the signal $\mathcal{X}_{\rm{S}}(u)$:
\begin{align}
\label{eq_XSdef}
  \mathcal{X}_\text{S}(u)
  \equiv&~ \mathcal{X}_{++,\text{F}}(u)+\mathcal{X}_{+-}(u)+\text{c.c.} \n\\
 =&~\mathcal{I}^{0,-2}_{++,\rm{F},>}(k_{12},k_3,k_3)+\mathcal{I}^{0,-2}_{+-}(k_{12},k_3,k_3)+\rm{c.c.}\n\\
  =&~\FR{\pi}4e^{-\pi\wt\nu}\Big[-{\mathcal{U}}_{-}^{0}(e^{\ii\pi}k_{12},k_3){\mathcal{U}}_{+}^{-2}(k_{3},k_3) +{\mathcal{U}}_{+}^{0}(k_{12},k_3){\mathcal{U}}_{-}^{-2}(k_{3},k_3)\Big]+\text{c.c.}.
\end{align}
In writing this expression, we have removed the $\theta$ functions in the original expression (\ref{eq_IS}), because the relation $|k_{12}|>|k_{34}|$ always holds true in the regions of our interest, including the region $u\in(-\infty,0)$ where the branch cut lies, and the physical region $u\in(0,1)$. Also, we note that, in the final expression, we have a factor ${\mathcal{U}}^{0}_{-}(e^{\ii\pi}k_{12},k_3)$, in which the first argument $k_{12}$ should be analytically continued by a rotation of $e^{\ii\pi}$. One can readily check that this way of analytical continuation brings ${\mathcal{U}}^{p_1}_{-}(k_{12},k_3)$ to the corresponding $\tau_1$-integral appears in $\mathcal{I}_{++,F,>}^{p_1p_2}$.

As mentioned before, the single-layer integrals $\mathcal{U}$ can be directly done, and the results are:
\begin{align}  
\label{eq_UplusResult} &{\mathcal{U}}_{\pm}^p(K_1,K_2)=\FR{e^{\mp\fr{\ii\pi}{4}(3+2p)+\fr{\pi\nn}{2}}}{\pi}\Bigl(\FR{K_2}{K_1}\Bigr)^{5/2+p}\biggl[\Bigl(\FR{K_2}{K_1}\Bigr)^{-\ii\nn}\mathbf{F}^p_{\nn}(K_2/K_1)+(\nn\to-\nn)\biggr],
\end{align}
where we have defined a function $\mb{F}_{\wt\nu}^p(z)$ for later convenience:
\begin{align}
\label{eq_mbFp}
  \mathbf{F}^{p}_{\nn}(z)\equiv-\ii\pi^{1/2}2^{3/2+p}\rm{csch}(\pi\nn){}_2\mathcal{F}_1\biggl[\bgm\fr{5}{4}+\fr{p}{2}-\fr{\ii\nn}{2},\fr{7}{4}+\fr{p}{2}-\fr{\ii\nn}{2}\\1-\ii\nn\edm\bigg|z^2\biggr].
\end{align}
Here ${}_2\mathcal{F}_1$ is the dressed version of Gauss's hypergeometric function, whose definition is collected in App. \ref{appd_function}.

It is straightforward to insert (\ref{eq_UplusResult}) into (\ref{eq_XSdef}) to get an expression for the signal $\mathcal{X}_\text{S}$. However, there are two small technical points worth mentioning. First, we want to write $\mathcal{X}_\text{S}$ as a function of $u=2k_3/k_{123}$, and this can be easily done by using the following identity of the hypergeometric function:
\begin{align}
  \2F1\biggl[\bgm a,b\\2b\edm\bigg|\FR{2r}{1+r}\biggr]=(1+r)^a\2F1\biggl[\bgm \fr{a}{2},\fr{a+1}{2}\\b+\frac{1}{2}\edm\bigg|r^2\biggr].
\end{align}
Second, in (\ref{eq_XSdef}) we have a factor ${\mathcal{U}}_{+}^{-2}(k_{3},k_3)$, which involves the cancellation of divergence between the two functions $\mathbf{F}^{-2}_{\pm\nn}(k_3/k_3)$:
\begin{align}
  \mathbf{F}^{-2}_{\nn}(k_3/k_3)+\mathbf{F}^{-2}_{-\nn}(k_3/k_3)=(2\pi^3)^{1/2}\rm{sech}(\pi\nn).
\end{align}
With these two points clarified, we obtain the signal part of 3-point correlator:
\begin{align}
\label{result_X_S}
  \mathcal{X}_\text{S}(u)= \FR{\pi\bigl(\ii+\sinh(\pi\nn)\bigr)}{4\sinh(2\pi\nn)}\,{}_2\mathcal{F}_1\biggl[\bgm\fr{1}{2}-\ii\nn,\fr{5}{2}-\ii\nn\\1-2\ii\nn\edm\bigg|u\biggr]u^{5/2-\ii\nn}+(\nn\to-\nn).
\end{align}

Now, we can quote our previous result  (\ref{disc_I_4pt_k}) to get the discontinuity of $\mathcal{X}(u)$. After switching to $u=2k_3/k_{123}$ as the argument, the result reads:
\begin{align}\label{disc_I_3pt_tree_u_relation}
  \mathop{\rm{Disc}}_{u}{\mathcal{X}}(u)&=-\mathop{\rm{Disc}}_{k_{12}}\mathcal{I}^{0,-2}(k_{12},k_3,k_3)\n\\
  &=2\ii\cosh(\pi\nn)\mathcal{I}^{0,-2}_{\rm{S}}(-k_{12},k_3,k_3)\theta(-k_{12}-k_s)\n\\
  &=2\ii\cosh(\pi\nn)\mathcal{X}_{\rm{S}}\Bigl(\FR{u}{u-1}\Bigr)\theta(-u),
\end{align}
in which the minus sign in the first line follows from the relation $u=2k_3/k_{123}$. Then, from (\ref{result_X_S}), we find the discontinuity of the full 3-point correlator as:
\begin{align}
\label{disc_tree_3pt_u_0-2}
    \mathop{\rm{Disc}}_u\mathcal{X}(u)=\FR{\pi}{4}\bigl(\ii-\rm{csch}(\pi\nn)\bigr)\,{}_2\mathcal{F}_1\biggl[\bgm\fr{1}{2}-\ii\nn,\fr{5}{2}-\ii\nn\\1-2\ii\nn\edm\bigg|u\biggr](-u)^{5/2-\ii\nn}+(\wt{\nu}\to-\wt{\nu}).
\end{align}
Here we have used the following identity to simplify the expression:
\begin{align}
  \2F1\biggl[\bgm a,b\\c\edm\bigg|u\biggr]=(1-u)^{-a}\2F1\biggl[\bgm a,c-b\\c\edm\bigg|\FR{u}{u-1}\biggr].
\end{align}

Of course, the discontinuity in (\ref{disc_tree_3pt_u_0-2}) can be directly read from the analytical expression for $\mathcal{X}_\text{S}$ in (\ref{result_X_S}) by using the known analytical properties of hypergeometric function ${}_2\mathcal{F}_1[\cdots|u]$ and the power function $u^{5/2-\ii\wt\nu}$. However, from (\ref{disc_I_3pt_tree_u_relation}), we can check that the discontinuity from the hypergeometric functions get canceled. The net discontinuity (\ref{disc_tree_3pt_u_0-2}) is fully from the power factor $u^{5/2-\ii\wt\nu}$, and this will be the key ingredient for our computation of dispersion integral in the next part.

\paragraph{Dispersion integral}
With the discontinuity of the function $\mathcal{X}(u)$ known, we are ready to form a dispersion integral, which computes the full correlator from its (factorized) discontinuity. 
\begin{align}
\label{eq_I3ptDispInt}
  \mathcal{X}(u)=\FR{u^3}{2\pi\ii}\int_{-\infty}^0\FR{\di u'}{u'^{\,3}(u'-u)}\mathop{\rm{Disc}}_{u'}\mathcal{X}(u').
\end{align} 
Here we have introduced a third-order subtraction ($u'^3$) to make sure that contour integral vanishes on the large circle. To understand this choice, we note that the large $u$ limit of $\mathcal{X}(u)$ corresponds to the total-energy limit $k_{123}\to 0$. By power counting of time, one can see that the seed integral behaves like $\mathcal{X}(u)\sim u^2$ as $|u|\to \infty$. Thus, a third-order subtraction suffices to make the dispersion integral well defined.

\begin{figure}[t]
  \centering
  \parbox{0.4\textwidth}{\includegraphics[width=0.4\textwidth]{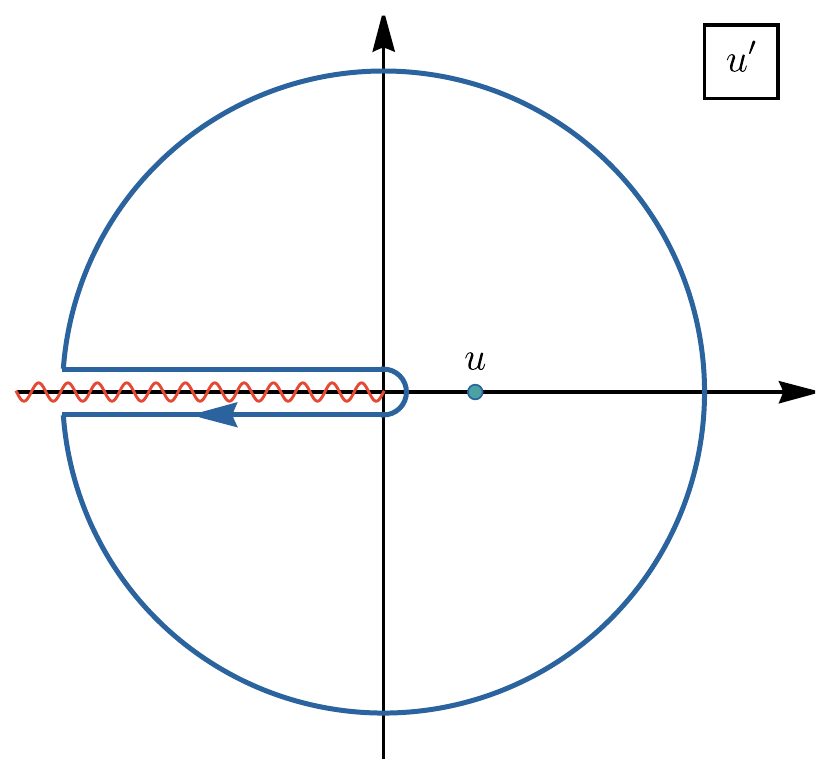}}
  ~~~$\To$~~~
  \parbox{0.45\textwidth}{\includegraphics[width=0.45\textwidth]{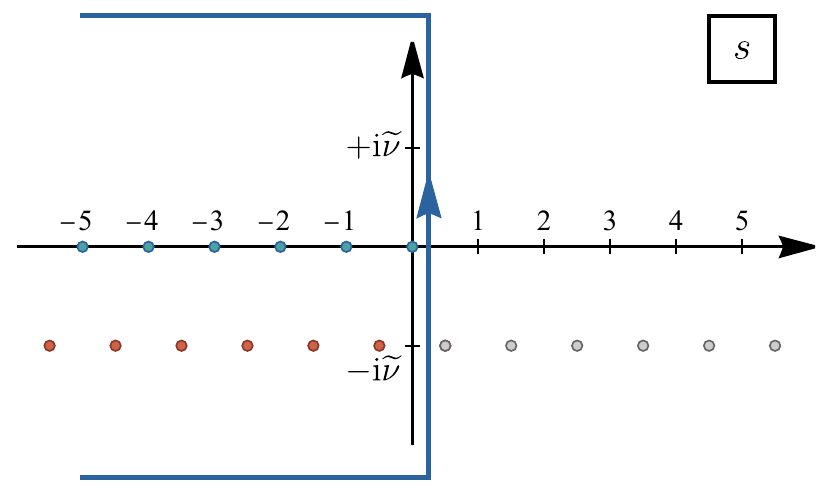}}
  \caption{The dispersion integrals for the 3-point correlator. The left panel shows the integral (\ref{eq_I3ptDispInt}) on the complex $u'$ plane, where we have a branch cut of the 3-point correlator in the unphysical region $u'\in(-\infty,0)$, and an inserted pole at the physical point $u'=u$ with $0<u<1$. On the right panel, we show the corresponding Mellin integral (\ref{eq_I3ptMellinInt}) on the complex plane of Mellin variable $s$. The blue and red poles on the left side of the integration contour contribute to the signal and the background of the full 3-point correlator, respectively. }
  \label{fig_3pt_u&s}
\end{figure}

(\ref{eq_I3ptDispInt}) is a well documented integral and can be directly done by \verb+Mathematica+. However, it is instructive to compute this integral more explicitly, by using the partial Mellin-Barnes (PMB) representation \cite{Qin:2022lva,Qin:2022fbv}. This method will be useful for more complicated integrals in the following sections where we do not have readily available integral formulae. Also, as we shall see, there is a nice correspondence between the pole structure of the Mellin integral and the analytical property of the final result. (See Fig.\ \ref{fig_3pt_u&s}).

To apply the PMB technique, we use the following MB representation for the hypergeometric function:\footnote{Generally, there is certain flexibility to deform the integral contour, so long as all poles coming from ``$\Gamma(+a s+\cdots)$ are to the left of the contour, and those poles from ``$\Gamma(-b s+\cdots)$'' are to the right. (Here $a,b\in \mathbb{R}_+$.) For convenience, here we just label the lower/upper bound of the integral as $\mp\ii\infty$.} 
\begin{align}\label{MB_rep_3pt_tree_u}
    {}_2\mathcal{F}_1\biggl[\bgm\fr{1}{2}\mp\ii\wt{\nu},\fr{5}{2}\mp\ii\wt{\nu}\\1\mp2\ii\wt{\nu}\edm\bigg|u'\biggr]=\int_{-\ii\infty}^{\ii\infty}\FR{\di s}{2\pi\ii}\ (-u')^{-s}\Gamma\biggl[\bgm s,\fr{1}{2}-s\mp\ii\wt{\nu},\fr{5}{2}-s\mp\ii\wt{\nu}\\1-s\mp2\ii\wt{\nu}\edm\biggr].
\end{align}
We see that the MB representation effectively turns a complicated dependence on $u'$ into a simple power function $(-u')^{-s}$. As a result, the dispersion integral over $u'$ in (\ref{eq_I3ptDispInt}) can now be trivially carried out as:
\begin{equation}
    \FR{u^3}{2\pi\ii}\int_{-\infty}^{0}\FR{\di u'\,(-u')^{5/2\mp\ii\wt{\nu}-s}}{u'^{\,3}(u'-u)}=\FR{\sec\big[\pi(s\pm\ii\wt{\nu})\big]}{2\ii}u^{5/2-s\mp\ii\wt{\nu}}.
\end{equation}
It then remains to finish the Mellin integral over $s$:
\begin{align}
\label{eq_I3ptMellinInt}
   \mathcal{X}(u)=&\ \int_{-\ii\infty}^{\ii\infty}\FR{\di s}{2\pi\ii}{u^{5/2-s-\ii\wt{\nu}}}\sec\big[{\color{BrickRed}\pi(s+\ii\wt{\nu})}\big]\Gamma\biggl[\bgm {\color{RoyalBlue}s},\fr{1}{2}-s-\ii\wt{\nu},\fr{5}{2}-s-\ii\wt{\nu}\\1-s-2\ii\wt{\nu}\edm\biggr]\n\\
   &\times\FR{\pi}{8}\bigl(1+\ii\ \rm{csch}(\pi\nn)\bigr)+(\nn\to-\nn).
\end{align}
Since we have taken the variable $u$ in the physical region, i.e., $u\in (0,1)$, we can perform the above Mellin integral by closing the contour from the left side. The integrand decreases fast enough when $s$ goes to infinity in the left plane, so that the integral over the large semi-circle on the left plane vanishes, and we can finish the integral by collecting the residues of all poles to the left side of the original integration contour. From (\ref{eq_I3ptMellinInt}), we see that there are two sets of left poles contributing to the final results, whose origins are highlighted in red and blue colors:\footnote{When computing integrals via PMB representation, if a Gamma function contributes poles, then all of its poles need to be collected. For example, here there are a set of poles from $\Gamma[s]$, then we need to pick up the whole set of these poles, i.e., $s=-n$ where $n=0,1,2,\cdots$. The case where poles come from $\sec[\pi(s+\ii\nn)]$ is a little subtler: If we change the upper bound of the integral over $u'$ from $0$ to $-\epsilon$ where $\epsilon$ is a small positive real number, one can find we will get $\Gamma[1/2+s+\ii\nn]$ instead of $\sec[\pi(s+\ii\nn)]$. It is only in the limit $\epsilon\to0$ that the Gamma function $\Gamma[1/2+s+\ii\nn]$ will meet another Gamma function $\Gamma[1/2-s-\ii\nn]$ and give rise to $\sec[\pi(s+\ii\nn)]$. This implies when considering poles from $\sec[\pi(s+\ii\nn)]$ we actually need to collect all poles from $\Gamma[1/2+s+\ii\nn]$, while poles from $\Gamma[1/2-s-\ii\nn]$ should be omitted. This analysis gives us another set of poles, i.e., $s=-1/2-n-\ii\nn$ where $n=0,1,2,\cdots$.}
\begin{equation}
\left\{
\begin{aligned}
    &{\color{BrickRed}s=-\fr{1}{2}-n-\ii\wt{\nu}};\\
    &{\color{RoyalBlue}s=-n}.~~~~(n=0,1,2,\cdots)
\end{aligned}
\right.
\end{equation} 
We also show these poles in the right panel of Fig.~\ref{fig_3pt_u&s}.
Clearly, from the factor $u^{5/2-s-\ii\wt\nu}$ in the integrand in (\ref{eq_I3ptMellinInt}), we see that the poles $s=-1/2-n-\ii\wt\nu$ correspond to the background, whose residues sum to:
\begin{align}
    \mathcal{X}_\text{B}(u)&=\sum_{n=0}^{\infty}\bigl(1+\ii\ \rm{csch}(\pi\nn)\bigr)\FR{(-1)^n}{8}u^{3+n}\Gamma\biggl[\bgm 1+n,3+n,-\frac{1}{2}-n-\ii\nn\\\frac{3}{2}+n-\ii\nn\edm\biggr]+(\nn\to-\nn)\n\\
    &=-\FR{2u^3}{1+4\nn^2}\3F2\biggl[\bgm1,1,3\\\frac{3}{2}+\ii\nn,\frac{3}{2}-\ii\nn\edm\bigg|u\biggr].
\end{align}
On the other hand, the poles at $s=-n$ in the integrand of (\ref{eq_I3ptMellinInt}) give rise to the signal:
\begin{align}
    \mathcal{X}_\text{S}(u)&=\sum_{n=0}^{\infty}\rm{sech}(\pi\nn)\bigl(1+\ii\ \rm{csch}(\pi\nn)\bigr)\FR{\pi}{8}u^{5/2+n-\ii\nn}\Gamma\biggl[\bgm\frac{1}{2}+n-\ii\nn,\frac{5}{2}+n-\ii\nn\\1+n,1+n-2\ii\nn\edm\biggr]+(\nn\to-\nn)\n\\
    &=\FR{\pi\bigl(\ii+\sinh(\pi\nn)\bigr)}{4\sinh(2\pi\nn)}\,{}_2\mathcal{F}_1\biggl[\bgm\fr{1}{2}-\ii\nn,\fr{5}{2}-\ii\nn\\1-2\ii\nn\edm\bigg|u\biggr]u^{5/2-\ii\nn}+(\nn\to-\nn).
\end{align}
Thus, the whole three-point correlator ${\mathcal{X}}(u)$ is neatly expressed as a sum of the signal and the background:
\begin{keyeqn}
\begin{align}
  \mathcal{X}(u) 
  =&~\bigg\{\FR{\pi\bigl(\ii+\sinh(\pi\nn)\bigr)}{4\sinh(2\pi\nn)}\,{}_2\mathcal{F}_1\biggl[\bgm\fr{1}{2}-\ii\nn,\fr{5}{2}-\ii\nn\\1-2\ii\nn\edm\bigg|u\biggr]u^{5/2-\ii\nn}+(\nn\to-\nn)\bigg\}\n\\
  &~-\FR{2u^3}{1+4\nn^2}\,\3F2\biggl[\bgm1,1,3\\\frac{3}{2}+\ii\nn,\frac{3}{2}-\ii\nn\edm\bigg|u\biggr].
\end{align}
\end{keyeqn}
This agrees with the results found previously using a different method \cite{Qin:2023ejc}.

To recapitulate our strategy, the PMB representation converts special functions into simple power functions, making the dispersion integrals easier to compute. Thereafter, the integration over Mellin variables can be directly computed via residue theorem. Therefore, the PMB representation provides a convenient way to calculate dispersion integrals analytically. For inflation correlators more complicated than the one considered here, the PMB representation remains useful, and will be shown below. 

\subsection{Three-point one-loop bubble graph}
\label{subsec_3ptLoop}

\begin{figure}[t]
  \centering
  \includegraphics[width=0.45\textwidth]{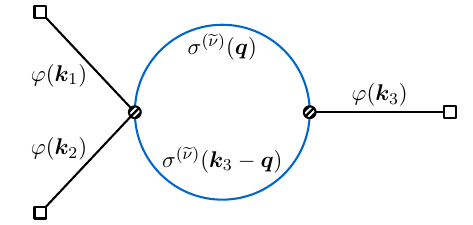}
  \caption{The 3-point correlator with 1-loop massive scalar exchange. For simplicity, we take the two bulk-to-bulk propagators to have the same mass parameter $\nn$. }
  \label{fd_3pt_1loop}
\end{figure}

Although most of the discussions of this work focus on tree-level processes, the dispersion technique can also be applied to loop processes. In this subsection, we will explore a simple 1-loop diagram with dispersion relations, with the help of the technique of spectral decomposition \cite{Xianyu:2022jwk}. Our example comes from the following  interactions between the massless scalar $\varphi$ and the principal scalar $\si$:
\begin{align}
  \Delta\ld=\FR{\ka_{4}}{4}a^2\varphi'^2\sigma^2+\FR{\ka_{3}}{2}a^3\varphi'\sigma^2.
\end{align}
Then, at $\order{\ka_4\ka_3}$, there is a unique diagram (up to trivial permutations) contributing to the 3-point correlator of $\varphi$ with a bubble loop formed by $\si$; See Fig. \ref{fd_3pt_1loop}.

Similar to the tree-level case, we can extract a dimensionless seed integral from the correlator:
\begin{align}
\label{eq_phi3loop}
  \la\varphi_{\bm k_1}\varphi_{\bm k_2}\varphi_{\bm k_3}\ra'_\text{1-loop}=\FR{\ka_3\ka_4}{8k_1k_2k_3^4}\bigg[ {\mathcal{J}}^{0,-2}\Big(\FR{2k_3}{k_{123}}\Big)+\text{2 perms}\bigg].
\end{align}
Here $\mathcal{J}^{p_1p_2}$ is the corresponding seed integral, defined as a function of the momentum ratio $u=2k_3/k_{123}$:
\begin{align}\label{seed_loop}
  \mathcal{J}^{p_1p_2}(u)\equiv-\FR{1}{2}\sum_{\mathsf{a,b}=\pm}\mathsf{ab}\ k_3^{5+p_{12}}\int_{-\infty}^0\di\tau_1\di\tau_2(-\tau_1)^{p_1}(-\tau_2)^{p_2}e^{\ii\mathsf{a}k_{12}\tau_1+\ii\mathsf{b}k_{3}\tau_2}\mathbf{Q}^{(\nn)}_{\mathsf{ab}}(k_3;\tau_1,\tau_2).
\end{align}
Here, $\mathbf{Q}^{(\nn)}_{\mathsf{ab}}$ denotes the 3-momentum loop integral:
\begin{align}\label{loop_propagator}
  \mathbf{Q}^{(\nn)}_{\mathsf{ab}}(k_3;\tau_1,\tau_2)\equiv\int\FR{\di^3\bm{q}}{(2\pi)^3}D^{(\nn)}_{\mathsf{ab}}\Bigl(q;\tau_1,\tau_2\Bigr)D^{(\nn)}_{\mathsf{ab}}\Bigl(|\bm k_3-\bm q|;\tau_1,\tau_2\Bigr).
\end{align}
Here we mark out the mass parameter $\nn$ of propagators as $\nn$ is important in the following analysis. 

As explained in previous works \cite{Marolf:2010zp,Xianyu:2022jwk,Loparco:2023rug}, the loop integral (\ref{loop_propagator}) can be recast as a (continuous) linear superposition of massive propagators ${D}^{(\nn')}_{\mathsf{ab}}$ with different values of $\nn'$, weighted by a spectral function $\rho^{\rm{dS}}_{\nn}(\nn')$:
\begin{align}\label{spectral_decomposition}
    \mathbf{Q}^{(\nn)}_{\mathsf{ab}}(k_3;\tau_1,\tau_2)=\int_{-\infty-\ii\epsilon}^{+\infty+\ii\epsilon}\di\nn'\FR{\nn'}{\pi\ii}\rho^{\rm{dS}}_{\nn}(\nn')D^{(\nn')}_{\mathsf{ab}}(k_3;\tau_1,\tau_2).
\end{align}
With the assumption that both the time integrals in (\ref{seed_loop}) and the spectral integral in (\ref{spectral_decomposition}) are convergent,\footnote{The convergence of the spectral integral (\ref{spectral_decomposition}) requires a proper regularization procedure, such as dimensional regularization, to make the spectral function $\rho^{\rm{dS}}_{\nn}(\nn')$ finite in the first place. However, as we will see below, our treatment of the loop process is completely independent of the regularization, and we can safely stay in $d=3$ throughout the discussion.} we can switch the order of two integrals, and write the loop correlator $\mathcal{J}^{p_1p_2}_{\wt\nu}$ as a spectral integral over tree correlator $\mathcal{I}^{p_1p_2}_{\wt\nu'}$:
\begin{align}\label{spectral_decomposition_seed}
    \mathcal{J}^{p_1p_2}_{\nn}(u)=\int_{-\infty-\ii\epsilon}^{+\infty+\ii\epsilon}\di\nn'\FR{\nn'}{2\pi\ii}\rho^{\rm{dS}}_{\nn}(\nn')\mathcal{I}^{p_1p_2}_{\nn'}(k_{12},k_3,k_3).
\end{align}

Now we specialize to the case of $(p_1,p_2)=(0,-2)$ as indicated in (\ref{eq_phi3loop}), and form a dispersion integral for $\mathcal{J}^{0,-2}$. Such a dispersion integral is possible, because all the 3-point tree-level correlators $\mathcal{I}_{\nn'}^{0,-2}$ with different mass parameters $\nn'$ satisfy the same dispersion relation (\ref{eq_I3ptDispInt}). Therefore, their linear superposition in (\ref{spectral_decomposition_seed}) should satisfies a similar dispersion integral. However, we should expect that the subtraction order for the loop correlator differs from the tree due to the different UV behavior. Therefore, let us write down the dispersion integral for the loop seed integral $\mathcal{J}^{0,-2}(u)$ on $u$ plane in the following way:
\begin{align}\label{dispersion_loop}
    {\mathcal{J}}^{0,-2}_{}(u)=\FR{u^m}{2\pi\ii}\int_{-\infty}^{0}\FR{\di u'}{u'^m(u'-u)}\mathop{\rm{Disc}}_{u'}{\mathcal{J}}_{\rm{S}}^{0,-2}(u').
\end{align}
Here we leave the subtraction order $m$ arbitrary, and we will determine it later. 

As mentioned above, the loop seed integral $\mathcal{J}^{p_1p_2}$ has been computed purely from spectral decomposition in \cite{Xianyu:2022jwk}. However, the result in \cite{Xianyu:2022jwk} shows a significant hierarchy in the degree of complication between the signal and the background: The signal part of the loop diagram is a discrete sum of tree signals weighted by a simple coefficients, which can be understood as  summing over all quasinormal modes of the loop. On the other hand, the background part is quite complicated, which, after regularization and renormalization, contains a highly intricate special function in the renormalized spectral function. Below, we shall exploit this hierarchy, using the signal computed via the spectral decomposition to bootstrap the full correlator, and thus bypassing any  complications of regularization and renormalization. 

Thus, our starting point will be the signal part of the loop seed integral computed via the spectral decomposition \cite{Xianyu:2022jwk}:
\begin{align}\label{1loop_sn_u}
    \mathcal{J}^{0,-2}_{\text{S}}(u)=&\ \FR{u^{4+2\ii\wt{\nu}}}{128\pi\sin(-2\pi\ii\wt{\nu})}\sum_{n=0}^{\infty}\FR{(3+4\ii\wt{\nu}+4n)(1+n)_{\frac{1}{2}}(1+2\ii\wt{\nu}+n)_{\frac{1}{2}}}{(\fr{1}{2}+\ii\wt{\nu}+n)_{\frac{1}{2}}(\fr{3}{2}+\ii\wt{\nu}+n)_{\frac{1}{2}}}\n\\
    &\times{}_2\mathcal{F}_1\biggl[\bgm2+2\ii\wt{\nu}+2n,4+2\ii\wt{\nu}+2n\\4+4\ii\wt{\nu}+4n\edm\bigg|u\biggr]u^{2n}+(\wt{\nu}\to-\wt{\nu}).
\end{align}
We then need to get the discontinuity of the signal along the branch cut. For the signal (\ref{1loop_sn_u}), its discontinuity along the branch cut $u\in(-\infty,0)$ is simply contributed by the $u^{\pm2\ii\nn}$ factor, which is similar to the tree-level case.
The result is
\begin{align}\label{disc_loop_3pt_u}
    \mathop{\rm{Disc}}_{u}{\mathcal{J}}^{0,-2}_{\rm{S}}(u)=&\ \FR{(-u)^{4+2\ii\wt{\nu}}}{64\pi\ii}\sum_{n=0}^{\infty}\FR{(3+4\ii\wt{\nu}+4n)(1+n)_{\frac{1}{2}}(1+2\ii\wt{\nu}+n)_{\frac{1}{2}}}{(\frac{1}{2}+\ii\wt{\nu}+n)_{\frac{1}{2}}(\frac{3}{2}+\ii\wt{\nu}+n)_{\frac{1}{2}}}\n\\
    &\times{}_2\mathcal{F}_1\biggl[\bgm2+2\ii\wt{\nu}+2n,4+2\ii\wt{\nu}+2n\\4+4\ii\wt{\nu}+4n\edm\bigg|u\biggr]u^{2n}\theta(-u)+(\wt{\nu}\to-\wt{\nu}).
\end{align}

Now we are ready to use (\ref{dispersion_loop}) and (\ref{disc_loop_3pt_u}) to compute the full correlator. However, at this point, we need to choose a subtraction (namely, to choose a value of $m$ in (\ref{dispersion_loop})) to make sure that the integral (\ref{dispersion_loop}) converges when $u\to0$ and $u\to-\infty$. Examining the behavior of the integrand in these two limits, we see that the convergence as $u\to 0$ requires $m\leq4$ while the convergence as $u\to -\infty$ prefers a large $m$. So, $m=4$ is an optimal choice.

Similar to the 3-point tree-level case, for every term in (\ref{disc_loop_3pt_u}), the dispersion integral can be done either by \verb|Mathematica| directly or by PMB representation. The final result for the loop seed integral $\mathcal{J}^{0,-2}(u)$ is again the sum of the signal $\mathcal{J}_\text{S}^{0,-2}(u)$ and the background $\mathcal{J}^{0,-2}_\text{BG}(u)$. The signal is already given in (\ref{1loop_sn_u}), and the background is given by:
\begin{align}\label{1loop_bg_dispersion_0-2_u^4}
    {\mathcal{J}}^{0,-2}_{\rm{BG}}(u)=\ &\FR{u^4}{128\pi\sin(2\pi\ii\wt{\nu})}\sum_{n=0}^{\infty}\FR{(3+4n+4\ii\wt{\nu})(1+n)_{\frac{1}{2}}(1+n+2\ii\wt{\nu})_{\frac{1}{2}}}{(\fr{1}{2}+n+\ii\wt{\nu})_{\frac{1}{2}}(\fr{3}{2}+n+\ii\wt{\nu})_{\frac{1}{2}}}\n\\
    &\times{}_3{\mathcal{F}}_2\biggl[\bgm1,2,4\\1-2n-2\ii\wt{\nu},4+2n+2\ii\wt{\nu}\edm\bigg|u\biggr]+(\wt{\nu}\to-\wt{\nu}).
\end{align}
Here  ${}_3\mathcal{F}_2$ is the dressed version of the generalized hypergeometric function, whose definition is collected in App. \ref{appd_function}. 

Some readers may find it mysterious that no UV divergence ever shows up in our calculation.  
The reason is in fact clear: The UV divergence in this 1-loop correlator can be fully subtracted by a local counterterm $\ld \supset \delta_\lambda a\varphi'^3$ with divergent coefficient $\de_\lam$. At the correlator level, this counterterm produces a contact diagram $\propto \delta_\lambda u^3$, and thus is analytical on the entire $u$ plane. If we follow the standard loop calculation, we would find a divergent part proportional to $u^3$, plus a finite part with more complicated $u$ dependence. Then we can use any convenient regularization method to remove the divergence, and use any proper renormalization scheme to determine the finite coefficient of the $u^3$ term. The arbitrariness of the coefficient of the $u^3$ term is an intrinsic uncertainty of the loop calculation. 

We think that this is an important lesson, especially for readers not very familiar with loop calculations, so let us reiterate it: When computing a superficially UV divergent loop correlator, the UV divergence is simply an artifact of our computation method and unphysical. Therefore, we may find a method so that UV divergences never appear and we never need to do UV regularization. Indeed, our dispersion method here is such an example where the regularization is never needed. 

On the contrary, when computing a 1-loop correlator with whatever methods, the result may contain a finite number of terms (in our case, the number is 1), whose kinematic dependence is totally fixed but coefficient undetermined. Indeed, the kinematic dependences of these terms are simply given by the corresponding tree graphs from the local counterterm in ordinary calculations, while the coefficients of these terms are never fixed by computation only; Instead, they should be determined by a renormalization condition, or, in a loose sense, by experimental data. Thus, to summarize: in a UV-divergent loop correlator, the UV divergence may be avoidable, but the renormalization ambiguity is not avoidable. 

For readers familiar with flat-space loop calculations with dimensional regularization, in App.\;\ref{app_Mink}, we provide a direct comparison between our dispersive calculation and the more conventional computation for a Minkowski 1-loop correlator, where one can see explicitly that the dispersion integral itself is free of any UV divergence or renormalization dependence, and that all renormalization-dependent information is fully encoded in the subtraction point. 

Back to our dispersion method, it is now clear that the renormalization ambiguity cannot be probed by the nonanalyticity of the correlator, and therefore we are not going to recover them from a dispersion integral. What we did recover in (\ref{1loop_bg_dispersion_0-2_u^4}), therefore, is a background free of any UV ambiguity, whose existence is demanded by analyticity of the correlator. For this reason, we call it the \emph{irreducible background}. 

The physical meaning of this irreducible background is clear: For the loop diagram in question, we can imagine to integrate out all loop modes and get infinitely many effective 3-point self-interaction vertices of the external mode, with increasing number of derivatives. These derivative couplings contribute to the 3-point function in the form of a Taylor expansion of $u$, starting from $u^3$. Except the renormalization-dependent term $\propto u^3$, all terms starting from $\order{u^4}$ are UV free and unambiguously determined by the loop computation. They can still be treated as from local (albeit derivative) interactions, but the coefficients of these interactions are unambiguous prediction of the model. Our result  for the background (\ref{1loop_bg_dispersion_0-2_u^4}) precisely recovers these terms.

With the above remark on renormalization ambiguity in mind, we can summarize our result for the loop seed integral as:
\begin{keyeqn}
\begin{align}
\label{eq_Jresult}
      &{\mathcal{J}}^{0,-2}(u)
      =Cu^3-\FR{u^4}{128\pi\sin(2\pi\ii\wt{\nu})}\sum_{n=0}^{\infty}\FR{(3+4\ii\wt{\nu}+4n)(1+n)_{\frac{1}{2}}(1+2\ii\wt{\nu}+n)_{\frac{1}{2}}}{(\fr{1}{2}+\ii\wt{\nu}+n)_{\frac{1}{2}}(\fr{3}{2}+\ii\wt{\nu}+n)_{\frac{1}{2}}}\n\\
    &\times\bigg\{{}_2\mathcal{F}_1\biggl[\bgm2+2\ii\wt{\nu}+2n,4+2\ii\wt{\nu}+2n\\4+4\ii\wt{\nu}+4n\edm\bigg|u\biggr]u^{2n+2\ii\wt{\nu}}-{}_3{\mathcal{F}}_2\biggl[\bgm1,2,4\\1-2n-2\ii\wt{\nu},4+2n+2\ii\wt{\nu}\edm\bigg|u\biggr]\bigg\}\n\\
    &+(\wt{\nu}\to-\wt{\nu}).
\end{align}
\end{keyeqn}
Here the first term $Cu^3$ is a local term, whose coefficient $C$ is to be determined by a renormalization scheme. The rest of terms, including the signal and the irreducible background, are free from renormalization ambiguities. They are both organized as an infinite summation over quasi-normal modes of the bubble loop. 

Although it is difficult to analytically compare our result (\ref{1loop_bg_dispersion_0-2_u^4}) with the known background obtained from the spectral decomposition in \cite{Xianyu:2022jwk} , we find that their numerical results only differ by a $u^3$-term, which is exactly the undetermined local part $Cu^3$ in (\ref{eq_Jresult}). Given the very complicated form of the background in \cite{Xianyu:2022jwk}, we consider this agreement a rather nontrivial check of both methods.

\subsection{Four-point single-exchange graph}\label{subsec_4ptTree_VB}

\begin{figure}[t]
    \centering
    \parbox{0.4\textwidth}{\includegraphics[width=0.45\textwidth]{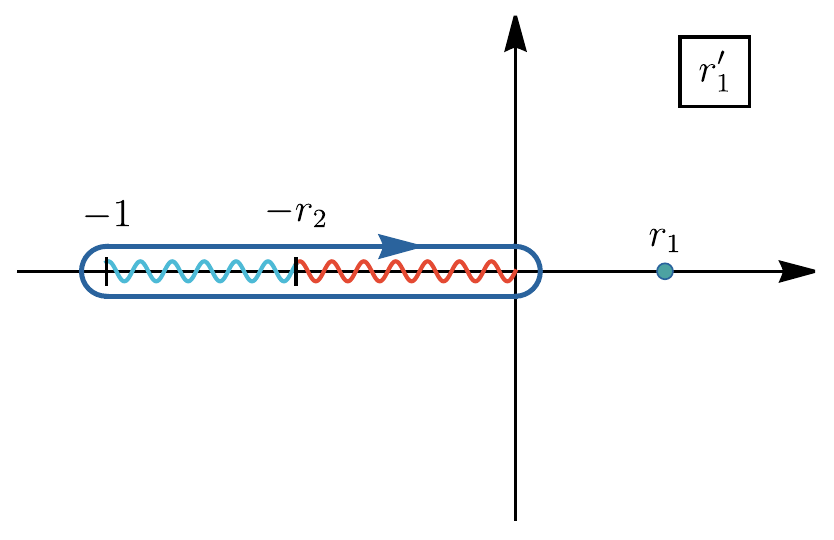}} 
    \caption{The analytical structure of the 4-point seed integral $\mathcal{Y}(r_1',r_2)$ on the complex $r_1'$ plane, with $r_2\in(0,1)$ fixed in the interior of its physical region. There are three branch points at $r_1'=0$ (signal branch point), $r_1'=-r_2$ (total-energy pole), and $r_1'=-1$ (partial-energy pole). with a branch cut connecting them in the interval $r_1'\in(-1,0)$. In this plot, we also show the insertion of a pole at $r_1'=r_1$ at which we compute the seed integral via the dispersion integral (\ref{dispersion_int_vertex_4pt}), and the blue circle surrounding the branch cut is the contour for the dispersion integral. }
    \label{fig_4pt_r1plane}
\end{figure}

As our last application of the vertex dispersion relation, we return to the 4-point seed integral (\ref{seed_tree_4pt_full}). Once again, we work with a particular choice of the exponents $(p_1,p_2)=(-2,-2)$. As explained in (\ref{eq_TtoSI}), this  corresponds to the case of nonderivative coupling $\phi_c^2\si$ between the conformal scalar $\phi_c$ in the external legs and a general principal massive scalar $\si$ in the bulk line. 

Similar to the previous 3-point examples, we want to exploit the scale invariance of the process, which implies that the the seed integral $\mathcal{I}^{-2,-2}(k_{12},k_{34},k_s)$ depends on three energy variables only through two independent momentum ratios. For the 4-point case, it is convenient to choose the following pair of ratios:
\begin{align}\label{def_r}
  r_1\equiv\FR{k_s}{k_{12}},&&r_2\equiv\FR{k_s}{k_{34}}.
\end{align}
The physical region $0\leq k_s\leq \text{min}\{k_{12},k_{34}\}$ then corresponds to $r_{1,2}\in[0,1]$. 
We then translate the analytical structure of the seed integral on the complex $k_{12}$ plane (Fig.\;\ref{fig_k12}) to the complex $r_1$ plane,  keeping $r_2\in(0,1)$ staying in the interior of the physical region. We show the result in Fig.\;\ref{fig_4pt_r1plane}, where the total-energy pole $k_{12}=-k_{34}$, the partial energy pole $k_{12}=-k_s$, and the signal branch point $k_{12}\to-\infty$ correspond to $r_1=-r_2$, $r_1=-1$, and $r_1=0$, respectively. Also, the branch cut is now entirely moved to the interval $r_1\in(-1,0)$.  

To highlight that we are working with  $r_1$ and $r_2$ as arguments of the seed integral, we use a new notation $\mathcal{Y}(r_1,r_2)$ for the 4-point seed integral:
\bge\label{eq_YtoI}
  \mathcal{Y}(r_1=k_s/k_{12},r_2=k_s/k_{34})\equiv \mathcal{I}^{-2,-2}(k_{12},k_{34},k_s).
\ede
Then, from (\ref{eq_IS}), we can read the signal $\mathcal{Y}_\text{S}(r_1,r_2)\equiv\mathcal{I}_\text{S}^{-2,-2}(k_{12},k_{34},k_s)$ of the seed integral, which is responsible for all the discontinuities:
\begin{align}
\label{eq_YSresult}
  \mathcal{Y}_{\text{S}}(r_1,r_2)
  =&~\mathcal{Y}_{\text{S},>}(r_1,r_2)\theta(r_2-r_1)+\mathcal{Y}_{\text{S},<}(r_1,r_2)\theta(r_1-r_2);\\
  \mathcal{Y}_{\text{S},>}(r_1,r_2)=&~\mathcal{I}^{-2,-2}_{+-}(k_{12},k_{34},k_s)+\mathcal{I}^{-2,-2}_{++,\rm{F},>}(k_{12},k_{34},k_s)+\rm{c.c.}\n\\
  =&~\FR{\pi}4e^{-\pi\wt\nu}\Big[{\mathcal{U}}_{+}^{-2}(k_{12},k_s){\mathcal{U}}_{-}^{-2}(k_{34},k_s) -{\mathcal{U}}_{-}^{-2}(e^{\ii\pi}k_{12},k_{s}){\mathcal{U}}_{+}^{-2}(k_{34},k_s) \Big]+\text{c.c.};\\
  \label{eq_YSless}
  \mathcal{Y}_{\rm{S},<}(r_1,r_2)=&~\mathcal{Y}_{\rm{S},>}(r_2,r_1).
\end{align}
Using the expressions for $\mathcal{U}^{p}_{\pm}$ in (\ref{eq_UplusResult}), we can find the explicit result for the signal:
\begin{align}\label{4pt_S_>_r}
    \mathcal{Y}_{\rm{S},>}(r_1,r_2)=\Bigl(\FR{1-\ii\sinh(\pi\nn)}{2\pi}r_1^{1/2-\ii\nn}\mathbf{F}_{\nn}^{-2}(r_1)+(\nn\to-\nn)\Bigr)\Bigl(r_2^{1/2-\ii\nn}\mathbf{F}_{\nn}^{-2}(r_2)+(\nn\to-\nn)\Bigr),
\end{align}
where $\mb{F}_{\wt\nu}^{p}$ is defined in (\ref{eq_mbFp}). Then, with $r_2\in(0,1)$ fixed in the interior of the physical region, the discontinuity of the seed function $\mathcal{Y}(r_1,r_2)$ on the real axis of $r_1$ is itself a piecewise function of $r_1$:
\begin{align}
\label{eq_DiscY}
    &\mathop{\rm{Disc}}_{r_1}\mathcal{Y}(r_1,r_2)
    =2\ii\cosh(\pi\nn)\mathcal{Y}_{\rm{S}}(-r_1,r_2)\theta(-r_1)\theta(r_1+1)\n\\
    &=2\ii\cosh(\pi\nn)\Bigl[\mathcal{Y}_{\rm{S},>}(-r_1,r_2)\theta(r_1+r_2)\theta(-r_1)+\mathcal{Y}_{\rm{S},<}(-r_1,r_2)\theta(r_1+1)\theta(-r_1-r_2)\Bigr].
\end{align}
This result is derived directly from (\ref{disc_I_4pt_k}), although there is a sign difference in $\mathop{\rm{Disc}}\limits_{r_1}\mathcal{Y}(r_1,r_2)$ and $\mathop{\rm{Disc}}\limits_{k_{12}}\mathcal{I}^{-2,-2}(k_{12},k_{34},k_s)$ due to the relation $r_1=k_s/k_{12}$.
Since the seed integral $\mathcal{Y}(r_1,r_2)$ is regular when $|r_1|\to\infty$ and $r_2$ fixed at a finite point,  we can directly construct a dispersion integral for $\mathcal{Y}(r_1,r_2)$ from (\ref{eq_DiscY}), with a first-order subtraction to ensure the vanishing integral along the large circle: 
\begin{align}\label{dispersion_int_vertex_4pt}
   \mathcal{Y}(r_1,r_2)= &~\FR{r_1}{2\pi\ii}\int_{-1}^{0}\FR{\di r}{r(r-r_1)}\mathop{\rm{Disc}}_r\mathcal{Y}(r,r_2)\n\\
    = &~\cosh(\pi\nn)\FR{r_1}{\pi}\bigg[\int_{-r_2}^{0}\di r\FR{\mathcal{Y}_{\rm{S},>}(-r,r_2)}{r(r-r_1)}+\int_{-1}^{-r_2}\di r\FR{\mathcal{Y}_{\rm{S},<}(-r,r_2)}{r(r-r_1)}\bigg],
\end{align}
With the explicit expressions for the signal $\mathcal{Y}_\text{S}$ in (\ref{4pt_S_>_r}) and (\ref{eq_YSless}), the dispersion integral (\ref{dispersion_int_vertex_4pt}) can be rewritten as:
\begin{align}
\label{eq_Yr1r2dispInt}
   & \mathcal{Y}(r_1,r_2)=\bigg[\FR{1-\ii\sinh(\pi\nn)}{2\pi^2}\cosh(\pi\nn){r_1}\mathbf{I}_{\wt\nu}^{(1)}(r_1,r_2)+(\nn\to-\nn)\bigg
    ]\bigg[r_2^{1/2-\ii\nn}\mathbf{F}_{\nn}^{-2}(r_2)+(\nn\to-\nn)\bigg]\n\\
    &+\bigg[{r_1}\mathbf{I}_{\wt\nu}^{(2)}(r_1,r_2)+(\nn\to-\nn)\bigg]\bigg[\FR{1-\ii\sinh(\pi\nn)}{2\pi^2}\cosh(\pi\nn)r_2^{1/2-\ii\nn}\mathbf{F}_{\nn}^{-2}(r_2)+(\nn\to-\nn)\bigg],
\end{align}
where $\mathbf{I}_{\wt\nu}^{(1)}$ and $\mathbf{I}_{\wt\nu}^{(2)}$ are the two integrals that are derived from the vertex dispersion relation:
\begin{align}
    \mathbf{I}_{\wt\nu}^{(1)}(r_1,r_2)\equiv&\int_{-r_2}^{0}\di r\FR{(-r)^{1/2-\ii\nn}\mathbf{F}_{\nn}^{-2}(-r)}{r(r-r_1)},\\
    \mathbf{I}_{\wt\nu}^{(2)}(r_1,r_2)\equiv&\int_{-1}^{-r_2}\di r\FR{(-r)^{1/2-\ii\nn}\mathbf{F}_{\nn}^{-2}(-r)}{r(r-r_1)}.
\end{align}
Unlike the 3-point case where the integrals extend from $-\infty$ to 0, the integrals here are defined on finite intervals $(-r_2,0)$ and $(-1,-r_2)$, making the calculation more involved. Still, we can get their analytical results by using the PMB representation, although the actual computation is quite lengthy. We collect the main steps and the final results for these two integrals in App.\ \ref{appd_vertex}.

Once $\mathbf{I}_{\wt\nu}^{(1)}$ and $\mathbf{I}_{\wt\nu}^{(2)}$ are obtained, we get the full expression of the seed integral $\mathcal{Y}(r_1,r_2)$, which can be further simplified and separated into the  signal and the background, namely, $\mathcal{Y}(r_1,r_2)=\mathcal{Y}_\text{S}(r_1,r_2)+\mathcal{Y}_\text{BG}(r_1,r_2)$.
The simplification is spelled out in App. \ref{appd_vertex}. Here, we only show the final result for the background $\mathcal{Y}_\text{BG}(r_1,r_2)$, since the signal $\mathcal{Y}_\text{S}(r_1,r_2)$ has been given in (\ref{eq_YSresult}):
\begin{align}\label{4pt_BG_signaldispersion_-2-2}
  \mathcal{Y}_\text{BG}(r_1,r_2)
  =&~\mathcal{Y}_{\text{BG},>}(r_1,r_2)\theta(r_1-r_2)+\mathcal{Y}_{\text{BG},>}(r_2,r_1)\theta(r_2-r_1);\n\\
  \mathcal{Y}_{\rm{BG},>}(r_1,r_2)
  =&~\FR{\ii 2^{\ii\wt\nu}\coth(\pi\wt\nu)}{\sqrt{2\pi}}  \sum_{n=0}^{\infty}\FR{(1+n-\ii\wt\nu)_{n-1/2}}{n!}   {\,}_2\mathcal{F}_1\biggl[\bgm1,\fr{1}{2}-2n+\ii\nn\\\fr{3}{2}-2n+\ii\nn\edm\bigg|-\FR{r_1}{r_2}\biggr]\n\\
  &\times{}_2\mathcal{F}_1\biggl[\bgm\fr{1}{4}+\fr{\ii\nn}{2},\fr{3}{4}+\fr{\ii\nn}{2}\\1+\ii\nn\edm\bigg|r_2^2\biggr]r_1 \Big(\FR{r_2}{2}\Big)^{2n}  +(\nn\to-\nn).
\end{align}
This expression appears different from the known results in the literature \cite{Qin:2022fbv}, but a direct numerical check shows that they agree with each other. Therefore, we have successfully bootstrapped the 4-point correlation functions with single massive exchange by dispersion integrals.

As we can see, for this particular 4-point example, performing the dispersion integral is by no means simpler than performing the nest time integral directly \cite{Qin:2022fbv}. Rather, our calculation here serves as a proof of principle, and shows that the dispersion relations really work for correlators with more complicated kinematics than 3-point single-exchange diagram. On the other hand, we can anticipate that the use of dispersion relation can bring significant simplification to the 4-point correlators at 1-loop level. We will pursue this 1-loop calculation in a future work.

\section{Analytical Structure on the Complex Line-Energy Plane}
\label{sec_line}

In the previous two sections, we considered the analytical properties and dispersion relations of inflation correlators in the complex plane of a vertex energy. Starting from this section, we are going to study the analytical properties of inflation correlators from a different perspective, by treating a line energy as a complex variable. In general, inflation correlators with massive exchanges also develop branch cuts on the complex plane of line energies. Therefore, it is possible to develop a different type of dispersion relations on the line energy plane, which we call \emph{line dispersion relations}. As we shall show, branch cuts on the complex plane of a line energy can all be connected to the \emph{nonlocal signal} of the inflation correlator. Therefore, a line dispersion relation allows us to compute the entire inflation correlator from its nonlocal signal alone. 

At the first sight, it may appear trivial that the branch cuts on a line energy plane can be entirely attributed to the nonlocal signal. Indeed, recall that the nonlocal signal with respect to a line energy $K_i$ refers to the part of the correlator which develops complex powers in $K_i$ in the soft $K_i$ limit: 
\bge
\label{eq_GsoftKlimit}
  \lim_{K_i\to 0}\wt{\mathcal{G}}\Big(\{E_\ell\},\{K_j\}\Big)\sim f\Big(\{E_\ell\},\{K_j\}\Big)K_i^{\pm\ii\wt\nu}+g\Big(\{E_\ell\},\{K_j\}\Big),
\ede
where both $f$ and $g$ are analytic at $K_i=0$, i.e., they have ordinary Taylor expansions at $K_i=0$. Therefore, the nonlocal signal, by definition, is associated with the branch point at $K=0$ generated by the complex-power term $f(\{E_\ell\},\{K_j\})K_i^{\pm\ii\wt\nu}$. However, things are less trivial than they appear: The functions $f$ and $g$ are analytic in $K_i$ only within a finite domain around $K_i=0$ where their Taylor expansions converge. Outside the convergence domain, these two functions could well develop new nonanalytic behaviors, including branch cuts, on the $K_i$ plane. These new nonanalyticities, in particular the ones in $g$, are not obviously related to the nonlocal signal. Therefore, it is quite remarkable that all branch cuts on the $K_i$ plane, including those not generated by nonlocal signals, can actually be connected to the nonlocal signal alone. In this section, we will spell out the details of reducing the entire correlator to its nonlocal signal. In this sense, we may say that the nonlocal signal by itself knows all about the whole correlator. 

Recall from the previous two sections that a vertex dispersion integral relates an inflation correlator with its \emph{signal}, both local and nonlocal. On the other hand, the line dispersion enables the recovery of full correlator from the nonlocal signal alone. Therefore, we see that the line dispersion is more ``economic'' than vertex dispersion in that it can generate the full correlator from a smaller set of data. This may have a practical advantage for bootstrapping inflation correlators, since the nonlocal signal appears easier to identify and to compute than the local signal, especially at the loop level \cite{Qin:2022lva,Qin:2023bjk,Qin:2023nhv}. Therefore, we may expect that the line dispersion relation may be a useful tool to bootstrap some complicated loop correlators whose full analytical results remain out of reach with currently known methods. 
 
\paragraph{Defining the nonlocal signal}
Clearly, the nonlocal signal plays a central role in the line dispersion relation. By definition, the nonlocal signal is a term in the correlator that develops complex powers $K^{\pm\ii\wt\nu}$ in the soft line energy limit $K\to 0$, namely the $f(E_\ell,K_j)K_i^{\pm\ii\wt\nu}$ term in (\ref{eq_GsoftKlimit}). Now let us identify this piece in the four-point seed integral $\mathcal{I}^{p_1p_2}(k_{12},k_{34},k_s)$ in (\ref{seed_tree_4pt_full}) without really computing it. 

When we fix the two vertex energies $k_{12}$ and $k_{34}$ in their physical domain and let $k_s\to 0$, the seed integral $\mathcal{I}^{p_1p_2}(k_{12},k_{34},k_s)$ is well convergent in the early time limit. Thus, its analytical behavior at $k_s=0$ is fully determined by the analytical property of the integrand in $k_s$, which in turn is determined by the bulk propagator $D_{\aa\bb}(k_s;\tau_1,\tau_2)$. Clearly, all four bulk propagators listed in (\ref{eq_Dmp})-(\ref{eq_Dpmpm}) are constructed from a pair of Hankel functions $\rm{H}_{+\ii\wt\nu}^{(1)}$ and $\rm{H}_{-\ii\wt\nu}^{(2)}$. Thus, we can regroup these Hankel functions to separate all bulk propagators into a piece analytic at $k_s=0$ and a piece that contains complex powers in $k_s$. In practice, this can be neatly done by rewriting each Hankel function as a linear combination of Bessel function of the first kind $\rm{J}_{\pm\ii\wt\nu}$; See (\ref{eq_HinJ}).  Then, the Hankel product in the propagator can be rewritten as:
\begin{align}
\label{eq_HHtoJJ}
    \rm{H}&^{(2)}_{-\ii\nn}(-k_s\tau_1)\rm{H}^{(1)}_{\ii\nn}(-k_s\tau_2)\n\\
    =\ &\rm{csch}(\pi\nn)^2\rm{J}_{\ii\nn}(-k_s\tau_1)\rm{J}_{-\ii\nn}(-k_s\tau_2)+\bigl(1+\coth(\pi\nn)\bigr)^2\rm{J}_{-\ii\nn}(-k_s\tau_1)\rm{J}_{\ii\nn}(-k_s\tau_2)\n\\
    &-\rm{csch}(\pi\nn)\bigl(1+\coth(\pi\nn)\bigr)\Big[\rm{J}_{\ii\nn}(-k_s\tau_1)\rm{J}_{\ii\nn}(-k_s\tau_2)+\rm{J}_{-\ii\nn}(-k_s\tau_1)\rm{J}_{-\ii\nn}(-k_s\tau_2)\Big].
\end{align}
In this expression, we have two types of terms: One involves a product of two $\rm{J}_{\pm\ii\wt\nu}$ with opposite orders, namely $\rm{J}_{\pm\ii\wt\nu}\rm{J}_{\mp\ii\wt\nu}$, as listed in the first line on the right hand side of (\ref{eq_HHtoJJ}); The other type involves a product of two $\rm{J}_{\pm\ii\wt\nu}$ with the same order, namely $\rm{J}_{\pm\ii\wt\nu}\rm{J}_{\pm\ii\wt\nu}$, listed in the last line of (\ref{eq_HHtoJJ}). By expanding these Bessel J functions in the $k_s\to 0$ limit, it is straightforward to see that the opposite-order terms $\rm{J}_{\pm\ii\wt\nu}\rm{J}_{\mp\ii\wt\nu}$ are analytic as $k_s\to 0$, while the same-order terms $\rm{J}_{\pm\ii\wt\nu}\rm{J}_{\pm\ii\wt\nu}$ behaves like $k_s^{\pm2\ii\wt\nu}$ as $k_s\to 0$. Thus, the same-order terms in the propagators precisely give rise to the nonlocal-signal part of the seed integral, while the opposite-order terms contain no nonlocal signal. Either by more careful inspection of the integral or by direct calculation, one can confirm that the opposite-order terms correspond to the local signal and the background, but we will not need this detailed separation between the local signal and the background in this section. Incidentally, from the boundary viewpoint, the same-order part can be viewed as the two-point correlator of a given conformal block with dimension $\Delta=\fr32\pm\ii\wt\nu$, while the opposite-order part the correlator between a conformal block and its shadow. 

Based on the above observation, we now separate all four bulk propagators $D_{\aa\bb}(k;\tau_1,\tau_2)$ according to their analytic property at $k\to 0$ in the following way:
\begin{align}
  D_{\aa\bb} (k;\tau_1,\tau_2)
  =\Sigma(k;\tau_1,\tau_2)+\Omega_{\aa\bb}(k;\tau_1,\tau_2).
\end{align}
Here the \emph{same-order propagators} $\Sigma(k;\tau_1,\tau_2)$ involve terms with same-order Bessel-J products, and thus are nonanalytic at $k=0$:
\begin{align}
\label{eq_SigmaProp}
  \Sigma(k;\tau_1,\tau_2)
  \equiv&-\FR{\pi(\tau_1\tau_2)^{3/2}}{4\sinh(\pi\wt\nu)}\Big[\rm{J}_{\ii\nn}(-k\tau_1)\rm{J}_{\ii\nn}(-k\tau_2)+\rm{J}_{-\ii\nn}(-k\tau_1)\rm{J}_{-\ii\nn}(-k\tau_2)\Big],
\end{align}
while the four \emph{opposite-order propagators} $\Omega_{\aa\bb}(k;\tau_1,\tau_2)$ involve terms with opposite-order Bessel-J products, and thus are analytic at $k=0$:
\begin{align}\label{eq_OmegaProp}
  \Omega_{\pm\mp}(k;\tau_1,\tau_2)
  \equiv&~\FR{\pi(\tau_1\tau_2)^{3/2}}{4\sinh(\pi\wt\nu)}\Big[\big(\coth(\pi\wt\nu)-1\big)\rm{J}_{\pm\ii\nn}(-k\tau_1)\rm{J}_{\mp\ii\nn}(-k\tau_2)-(\wt\nu\to-\wt\nu)\Big],\n\\
  \Omega_{\pm\pm}(k;\tau_1,\tau_2)
  \equiv&~\Omega_{\mp\pm}(k;\tau_1,\tau_2)\theta(\tau_1-\tau_2)+\Omega_{\mp\pm}(k;\tau_1,\tau_2)\theta(\tau_2-\tau_1).
\end{align}
We have deliberately removed the SK indices in the same-order propagators $\Sigma(k;\tau_1,\tau_2)$, to highlight the fact that this propagator is actually independent of the SK contours: All four choices of the SK labels $\aa,\bb=\pm$ yield the same expression $\Sigma(k;\tau_1,\tau_2)$. This is closely tied to the fact that the nonanalytic part of the propagator is real and symmetric in the two time variables $\tau_1$ and $\tau_2$. In particular, the symmetry under $\tau_1\leftrightarrow \tau_2$ renders the time-ordering $\theta$ functions ineffective in the same-sign propagators. 

However, let us immediately clarify that the same-order propagator $\Sigma(k;\tau_1,\tau_2)$ is \emph{not} the symmetrization of the original bulk propagator $D_{\aa\bb}(k;\tau_1,\tau_2)$ with respect to $\tau_1\leftrightarrow r_2$. As one can directly check, the opposite-order propagator $\Omega_{\pm\pm}(k;\tau_1,\tau_2)$ also contains a piece that is symmetric with respect to $\tau_1\leftrightarrow r_2$ but is nevertheless analytic at $k_s=0$. In fact, this additional piece corresponds to a part of the local signal that is symmetric in $k_{12}\leftrightarrow k_{34}$.

Now, we can put the above separated bulk propagator back into the seed integral, and separate the seed integral accordingly:
\begin{align}
\label{eq_ItoPQ}
  \mathcal{I}_{\aa\bb}^{p_1p_2}(k_{12},k_{34},k_s)=&~\mathcal{P}_{\aa\bb}^{p_1p_2}(k_{12},k_{34},k_s)+\mathcal{Q}_{\aa\bb}^{p_1p_2}(k_{12},k_{34},k_s),
\end{align}
where $\mathcal{P}_{\aa\bb}^{p_1p_2}(k_{12},k_{34},k_s)$ and $\mathcal{Q}_{\aa\bb}^{p_1p_2}(k_{12},k_{34},k_s)$ are respectively nonanalytic and analytic at $k_s\to 0$ when $k_{12}$ and $k_{34}$ staying in the interior of their physical domain, whose definitions are:
\begin{align}
\label{eq_Pab}
  \mathcal{P}_{\aa\bb}^{p_1p_2}(k_{12},k_{34},k_s)
  \equiv&-\mathsf{ab}\ k_s^{5+p_{12}}\int_{-\infty}^{0}\di\tau_1\di\tau_2\,(-\tau_1)^{p_1}(-\tau_2)^{p_2}e^{\ii\mathsf{a}k_{12}\tau_1+\ii\mathsf{b}k_{34}\tau_2}\Sigma(k_s;\tau_1,\tau_2), \\
\label{eq_Qab}
  \mathcal{Q}_{\aa\bb}^{p_1p_2}(k_{12},k_{34},k_s)
  \equiv&-\mathsf{ab}\ k_s^{5+p_{12}}\int_{-\infty}^{0}\di\tau_1\di\tau_2\,(-\tau_1)^{p_1}(-\tau_2)^{p_2}e^{\ii\mathsf{a}k_{12}\tau_1+\ii\mathsf{b}k_{34}\tau_2}\Omega_{\aa\bb}(k_s;\tau_1,\tau_2).
\end{align}
We note that, although the same-order propagator itself $\Sigma(k;\tau_1,\tau_2)$ is independent of SK indices, the nonanalytic integrals $\mathcal{P}_{\aa\bb}^{p_1p_2}(k_{12},k_{34},k_s)$ still have nontrivial dependences on $\aa,\bb=\pm$ through the exponential factors $e^{\ii\aa k_{12}\tau_1+\ii\bb k_{34}\tau_2}$. 

To complete our list of new definitions, we can also define the integrals with SK indices summed:
\begin{align}
\label{eq_Pint}
  &\mathcal{P}^{p_1p_2}(k_{12},k_{34},k_s)\equiv\sum_{\aa,\bb=\pm}\mathcal{P}_{\aa\bb}^{p_1p_2}(k_{12},k_{34},k_s); \\
\label{eq_Qint}
  &\mathcal{Q}^{p_1p_2}(k_{12},k_{34},k_s)\equiv\sum_{\aa,\bb=\pm}\mathcal{Q}_{\aa\bb}^{p_1p_2}(k_{12},k_{34},k_s).
\end{align}
From the above discussion, we see that $\mathcal{P}^{p_1p_2}$ is nothing but the nonlocal signal, while $\mathcal{Q}^{p_1p_2}$ is the sum of the local signal and the background:
\begin{align}
\label{eq_PisINS}
  \mathcal{P}^{p_1p_2}(k_{12},k_{34},k_s)=&~\mathcal{I}_\text{NS}^{p_1p_2}(k_{12},k_{34},k_s),\\
  \mathcal{Q}^{p_1p_2}(k_{12},k_{34},k_s)=&~\mathcal{I}_\text{LS}^{p_1p_2}(k_{12},k_{34},k_s)+\mathcal{I}_\text{BG}^{p_1p_2}(k_{12},k_{34},k_s).
\end{align}

\paragraph{Same-order integral} Now let us briefly look at the two integrals defined in (\ref{eq_Pint}) and (\ref{eq_Qint}). First, consider the same-order integral $\mathcal{P}^{p_1p_2}(k_{12},k_{34},k_s)$. Combining (\ref{eq_SigmaProp}), (\ref{eq_Pab}), and (\ref{eq_Pint}), we see that the nonlocal signal can be directly expressed as a sum of factorized time integrals: 
\begin{align}
\label{eq_INS}
  \mathcal{I}_\text{NS}^{p_1p_2}(k_{12},k_{34},k_s)
  =\FR{\pi}{4\sinh(\pi\wt\nu)}\sum_{\aa,\bb,\cc=\pm}\aa\bb\mathcal{V}_\cc^{p_1}(\aa k_{12},k_s)\mathcal{V}_\cc^{p_2}(\bb k_{34},k_s),
\end{align}
where we have introduced two single-layer integrals $\mathcal{V}_{\pm}^p(E,K)$, defined by:
\begin{align}\label{mathcalV_definition}
  \mathcal{V}_{\pm}^p(E,K)
  \equiv&~ K^{5/2+p}\int_{-\infty}^0\di\tau\,(-\tau)^{3/2+p}e^{+\ii E\tau}\rm{J}_{\pm\ii\wt\nu}(-K\tau).
\end{align}
This integral can be directly done and the result is expressed in terms of the (dressed) Gauss's hypergeometric function: (See App. \ref{appd_function} for our definition of the dressed hypergeometric functions.)
\begin{align}\label{mathcalV_result}
  \mathcal{V}_{\pm}^p(E,K)=&~\FR{2^{3/2+p}}{\sqrt\pi}\Big(\FR{K}{\ii E}\Big)^{5/2+p\pm\ii\wt\nu}{}_2\mathcal{F}_1\left[\bgm \fr12(\fr52+p\pm\ii\wt\nu),\fr12(\fr72+p\pm\ii\wt\nu)\\ 1\pm\ii\wt\nu\edm\middle|\FR{K^2}{E^2}\right].
\end{align}
Thus, the computation of the nonlocal signal involves only single-layer integrals, which is a direct consequence of the nonlocal-signal cutting rule studied in the literature \cite{Tong:2021wai,Qin:2023bjk,Qin:2023nhv}.

\paragraph{Parity of the opposite-order integral} Next, let us turn to the opposite-order integrals $\mathcal{Q}_{\aa\bb}^{p_1p_2}$. Unlike the nonlocal signal, these integrals involve genuine time orderings that cannot be removed, resulting in final expressions of higher ``transcendental weight'' \cite{Fan:2024iek}, and thus are more difficult to compute. We are going to compute them using dispersion relations below. Here, without computing them directly, we point out that the integral $\mathcal{Q}_{\aa\bb}^{p_1p_2}$ has a very useful property: It possesses a fixed parity under the parity transformation of the line energy: $k_s\to -k_s$. 

To see this point, we make use of a property of Bessel J function, given in App.\ \ref{appd_function}, which shows that $\rm{J}_{\pm\ii\wt\nu}(e^{\ii\pi}z)\rm{J}_{\mp\ii\wt\nu}(e^{\ii\pi}w)=\rm{J}_{\pm\ii\wt\nu}(z)\rm{J}_{\mp\ii\wt\nu}(w)$. As a result, the opposite-order propagator $\Omega_{\aa\bb}(k;\tau_1,\tau_2)$ is invariant under the sign flip of its energy:
\bge
  \Omega_{\aa\bb}(-k;\tau_1,\tau_2)=\Omega_{\aa\bb}(k;\tau_1,\tau_2). 
\ede
With this property and the definition of the opposite-order integral in (\ref{eq_Qab}), it is straightforward to see that $\mathcal{Q}_{\aa\bb}^{p_1p_2}$ has a fixed parity $(-1)^{1+p_{12}}$ under the $k_s$-parity transformation $k_s\to -k_s$:
\bge
\label{eq_Qparity}
  \mathcal{Q}_{\aa\bb}^{p_1p_2}(k_{12},k_{34},-k_s)=(-1)^{1+p_{12}}\mathcal{Q}_{\aa\bb}^{p_1p_2}(k_{12},k_{34},k_s),
\ede
where $(-1)^{1+p_{12}}$ comes entirely from the prefactor $k_s^{5+p_{12}}$ in our definition of $\mathcal{Q}_{\aa\bb}^{p_1p_2}$. This property will be very useful for our following derivation of the line dispersion relation.

\paragraph{Analyticity along the positive real axis}

After a brief analysis of same-order and opposite-order integrals, now let us come back to the main goal of this section, namely, to diagnose the nonanalyticity of the seed integral $\mathcal{I}^{p_1p_2}(k_{12},k_{34},k_s)$ on the complex $k_s$ plane. 

The strategy is similar to what we adopted in Sec.\ \ref{sec_vertex_disp}, namely, to use the contour-deformation method. With this method, we will show that the seed integral $\mathcal{I}^{p_1p_2}(k_{12},k_{34},k_s)$ is analytic everywhere on the complex $k_s$ plane, expect for a possible branch cut lying on the whole negative real axis. In the next part, we shall relate the discontinuity of this branch cut to the one in the nonlocal signal $\mathcal{I}_\text{NS}^{p_1p_2}$. 

Similar to the behavior in the vertex energy plane, the seed integral is obviously analytic in $k_s$ for $\text{Im}\,k_s\neq 0$, a direct consequence of contour deformation argument. More nontrivial is the following fact:
\begin{align}
\label{eq_discI_positiveks}
  \theta(k_s)\mathop{\text{Disc}}_{k_s}\mathcal{I}^{p_1p_2}_{\aa\bb}(k_{12},k_{34},k_s)=0.
\end{align} 
That is, the seed integral is analytic in $k_s$ for all $k_s>0$. This is quite remarkable because the region $k_s>0$ is not entirely physical: Physically allowed $k_s$ satisfies $0\leq k_s\leq\text{min}\{k_{12},k_{34}\}$. Thus, the statement (\ref{eq_discI_positiveks}) in particular implies that seed integrals $\mathcal{I}^{p_1p_2}_{\aa\bb}$ are regular on the boundaries of the physical region when $k_s=k_{12}$ and $k_s=k_{34}$. As is well known, the absence of singularities at these \emph{folded} configurations is a consequence of choosing the Bunch-Davies initial state for all fluctuating modes in the bulk. Now let us prove (\ref{eq_discI_positiveks}) rigorously using our contour-deformation method. 

Once again, the analytical behavior of seed integrals on the complex $k_s$ plane is governed by the UV behavior of the integrands, namely the convergence of the integral as $
\tau_{1,2}\to-\infty$. So let us look at these UV regions for the opposite-sign integrals $\mathcal{I}_{\pm\mp}^{p_1p_2}$ and same-sign integrals $\mathcal{I}_{\pm\pm}^{p_1p_2}$, respectively. 

First, we consider the opposite-sign integral: 
\begin{align}
  \mathcal{I}_{+-}^{p_1p_2}(k_{12},k_{34},k_s)
  =&~\FR{\pi e^{-\pi\wt\nu}}{4} k_s^{5+p_{12}} \int_{-\infty}^0\di\tau_1\di\tau_2\,(-\tau_1)^{3/2+p_1}(-\tau_2)^{3/2+p_2}\n\\
  &~\times e^{\ii k_{12}\tau_1-\ii k_{34}\tau_2}\rm{H}_{-\ii\wt\nu}^{(2)}(-k_s\tau_1)\rm{H}_{\ii\wt\nu}^{(1)}(-k_s\tau_2).
\end{align}
Clearly, the integrand is well defined in the entire integration region for all $k_s>0$. Furthermore, we only consider IR finite processes so that the integral is convergent as $\tau_{1,2}\to 0$. Thus, any potential singularity of the integral on the complex $k_s$ plane must come from the UV divergences when $\tau_{1,2}\to -\infty$. However, it is easy to see this never happens for $k_s>0$. In fact, using the asymptotic behavior of the Hankel functions (\ref{eq_Hankel_asymptotic}), we see that the integrand behaves like:
\bge
  e^{+\ii(k_{12}+k_s)\tau_1}e^{-\ii(k_{34}+k_s)\tau_2},
\ede  
up to irrelevant power factors of $\tau_1$ and $\tau_2$. We see that, for physical values of $k_{12}$ and $k_{34}$, and for any $k_s>0$, the phases of these factors never change sign or hit zero. Thus, we conclude that, with the original choices of the integration contour (with proper $\ii \ep$ prescriptions), the opposite-sign integral $\mathcal{I}_{+-}^{p_1p_2}(k_{12},k_{34},k_s)$ is regular for any $k_s>0$. With completely the same argument, we can also show that the other opposite-sign integral $\mathcal{I}_{-+}^{p_1p_2}(k_{12},k_{34},k_s)$ is regular for any $k_s>0$ as well. 

The two same-sign seed integrals can be analyzed similarly. Let us consider the all-plus integral:
\begin{align}
\label{eq_Ipp}
  &\mathcal{I}_{++}^{p_1p_2}(k_{12},k_{34},k_s)
  =-\FR{\pi e^{-\pi\wt\nu}}{4} k_s^{5+p_{12}} \int_{-\infty}^0\di\tau_1\di\tau_2\,(-\tau_1)^{3/2+p_1}(-\tau_2)^{3/2+p_2}e^{\ii k_{12}\tau_1+\ii k_{34}\tau_2}\n\\
  &~\times \Big[\rm{H}_{\ii\wt\nu}^{(1)}(-k_s\tau_1)\rm{H}_{-\ii\wt\nu}^{(2)}(-k_s\tau_2)\theta(\tau_1-\tau_2)+\rm{H}_{-\ii\wt\nu}^{(2)}(-k_s\tau_1)\rm{H}_{\ii\wt\nu}^{(1)}(-k_s\tau_2)\theta(\tau_2-\tau_1)\Big].
\end{align}
Parallel to the previous argument, the integral can develop singular behaviors on the complex $k_s$ plane only through UV divergences. They occur when either the earlier time variable or both the time variables go to $-\infty$. Taking the $\theta(\tau_1-\tau_2)$ part of (\ref{eq_Ipp}) as an example:
\begin{align}
\label{eq_Ipp_theta12}
 -\FR{\pi e^{-\pi\wt\nu}}{4} k_s^{5+p_{12}} \int_{-\infty}^0\di\tau_2\int_{\tau_2}^0\di\tau_1\,(-\tau_1)^{3/2+p_1}(-\tau_2)^{3/2+p_2}e^{\ii k_{12}\tau_1+\ii k_{34}\tau_2} 
    \rm{H}_{\ii\wt\nu}^{(1)}(-k_s\tau_1)\rm{H}_{-\ii\wt\nu}^{(2)}(-k_s\tau_2).
\end{align}
In the UV limit $\tau_{1},\tau_2\to-\infty$, the integrand behaves like $e^{+\ii(k_{12}-k_s)\tau_1}e^{+\ii(k_{34}+k_s)\tau_2}$ up to unimportant power factors. Then, after finishing the $\tau_1$ integral, we get two terms, each behaves in the $\tau_2\to-\infty$ limit as $e^{+\ii(k_{12}+k_{34})\tau_2}$ and $e^{+\ii(k_{34}+k_s)\tau_2}$, respectively. Clearly, both phases stay positive for all $k_s>0$, so that the integral is well convergent with its original contour. One can similarly analyze the $\theta(\tau_2-\tau_1)$ term in (\ref{eq_Ipp}) and gets the same result. The analysis for $\mathcal{I}_{--}^{p_1p_2}$ is also the same. Thus we conclude that the same-sign seed integrals are also regular for all $k_s>0$. This completes the proof of (\ref{eq_discI_positiveks}).

\paragraph{Discontinuity in the line energy}

\begin{figure}
\centering 
\includegraphics[width=0.5\textwidth]{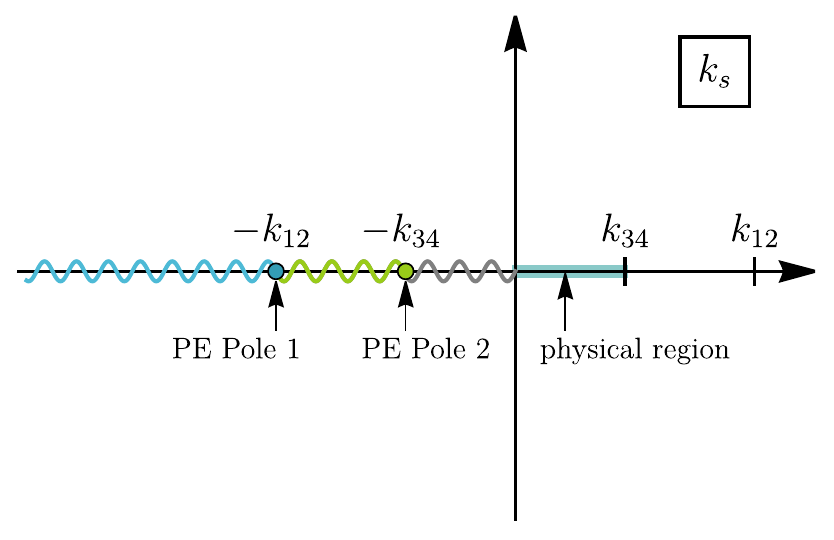} 
\caption{The analytical structure of the seed integral $\mathcal{I}^{p_1p_2}(k_{12},k_{34},k_s)$ in (\ref{seed_tree_4pt_full}) on the complex $k_{s}$ plane, with $k_{12}$ and $k_{34}$ staying in their physical regions $k_{12},k_{34}>0$. The physical region of $k_{s}$ ($0\leq k_s\leq \text{min}\{k_{12},k_{34}\}$) is marked in green. The seed integral is analytic in $k_{s}$ everywhere except when $k_{s}\in (-\infty,0]$. In this interval, we have three branch points: two partial-energy (PE) poles at $k_s=-k_{12}$ and $k_{s}=-k_{34}$, and the signal branch point at $k_{12}=0$. The total-energy pole $k_{1234}=0$ is never met for finite $k_{s}$ when $k_{12}\neq-k_{34}$, and can be thought of as sitting at $k_{s}=-\infty$ due to the scale invariance of the seed integral. The branch cut can be chosen to be lying in the entire interval $k_{12}\in(-\infty,-k_s]$, with the discontinuity itself being discontinuous at the two partial-energy poles.  }
  \label{fig_ks}
\end{figure}

From (\ref{eq_discI_positiveks}) we see that any possible branch cuts of the seed integrals $\mathcal{I}_{\aa\bb}^{p_1p_2}$ must lie in the negative real axis. In other words, we have:
\bge
\label{eq_DiscItoDiscPQ}
  \mathop{\text{Disc}}_{k_s}\mathcal{I}^{p_1p_2}(k_{12},k_{34},k_s)
  = \theta(-k_s)\mathop{\text{Disc}}_{k_s}\Big[\mathcal{I}_\text{NS}^{p_1p_2}(k_{12},k_{34},k_s)+\mathcal{Q}^{p_1p_2}(k_{12},k_{34},k_s)\Big].
\ede
Here we have used (\ref{eq_ItoPQ}) with all SK indices summed, as well as  (\ref{eq_PisINS}), namely, the same-order integral $\mathcal{P}^{p_1p_2}$ with all SK indices summed is nothing but the nonlocal signal $\mathcal{I}^{p_1p_2}_\text{NS}$.

 Now we are ready to relates this discontinuity with that of the nonlocal signal. To this end, we make use of the above result (\ref{eq_discI_positiveks}), written in terms of $\mathcal{I}^{p_1p_2}_\text{NS}$ and $\mathcal{Q}^{p_1p_2}$:
\bge
  \theta(k_s)\mathop{\text{Disc}}_{k_s}\Big[\mathcal{I}_\text{NS}^{p_1p_2}(k_{12},k_{34},k_s)+\mathcal{Q}^{p_1p_2}(k_{12},k_{34},k_s)\Big]=0.
\ede
Then, using the parity of the opposite-order integral $\mathcal{Q}$ in (\ref{eq_Qparity}), we get:
\bge
  \theta(k_s)\mathop{\text{Disc}}_{k_s}\Big[\mathcal{I}_\text{NS}^{p_1p_2}(k_{12},k_{34},k_s)-(-1)^{1+p_{12}}\mathcal{Q}^{p_1p_2}(k_{12},k_{34},-k_s)\Big]=0.
\ede
This can be equivalently written as:
\bge
  \theta(-k_s)\mathop{\text{Disc}}_{k_s}\mathcal{Q}^{p_1p_2}(k_{12},k_{34},k_s) = (-1)^{1+p_{12}}\theta(-k_s)\mathop{\text{Disc}}_{k_s}\mathcal{I}_\text{NS}^{p_1p_2}(k_{12},k_{34},-k_s) .
\ede
Substituting this relation back to (\ref{eq_DiscItoDiscPQ}), we finally get:
\begin{keyeqn}
\begin{align}
\label{eq_discI_ks}
  \mathop{\text{Disc}}_{k_s}\mathcal{I}^{p_1p_2}(k_{12},k_{34},k_s)
  =&~\mathop{\text{Disc}}_{k_s}\Big[\mathcal{I}_\text{NS}^{p_1p_2}(k_{12},k_{34},k_s)-(-1)^{p_{12}}\mathcal{I}_\text{NS}^{p_1p_2}(k_{12},k_{34},-k_s)\Big]\theta(-k_s).
\end{align}
\end{keyeqn}
That is, the discontinuity of the full seed integral can be completely related to that of the nonlocal signal alone. This is the central result of the current section, and forms the basis for the line energy dispersion relation, to be discussed below.

\section{Bootstrapping Correlators with Line Dispersion Relation}
\label{sec_linedisp}

\begin{figure}
\centering 
\includegraphics[width=0.5\textwidth]{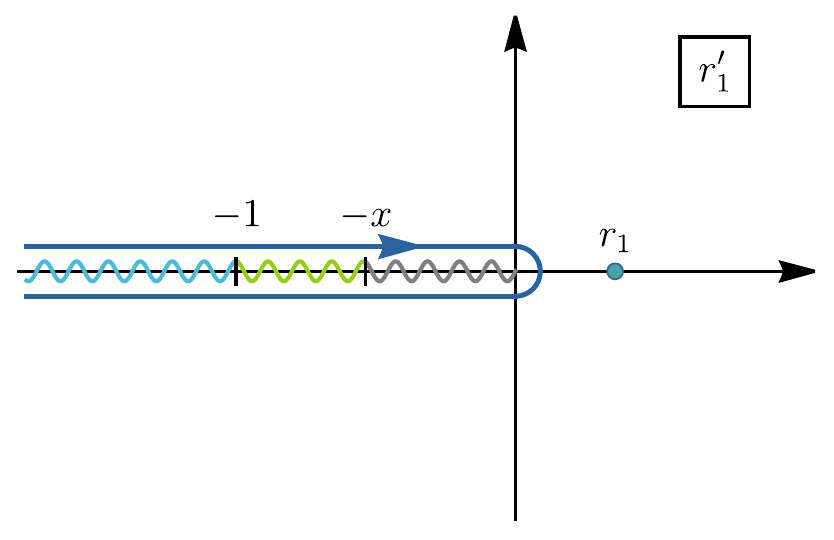} 
\caption{The analytical structure of the 4-point seed integral $\mathcal{Z}(r_1',x)$ on the complex $r_1'$ plane, with $x\in(0,\infty)$ fixed in the interior of its physical region. There are three branch points at $r_1'=0$ (signal branch point), $r_1'=-1$ (partial-energy pole $k_{s}=-k_{12}$), and $r_1'=-x$ (partial-energy pole $k_s=-k_{34}$), with a branch cut connecting them on the whole negative real axis. In this plot, we also show the insertion of a pole at $r_1'=r_1$ at which we compute the seed integral via the line dispersion integral (\ref{dispersion_int_line}), whose contour is shown by the blue curve surrounding the branch cut. }
  \label{fig_line_disp}
\end{figure}

From the analytical structure of the seed integral on the complex $k_s$ plane, it is straightforward to construct dispersion integrals, which relate the whole seed integral with its nonlocal signal alone. For clarity, let us still specialize to the case of $p_{1}=p_2=-2$. Once again, we use the fact that the dimensionless seed integral depends only on two independent momentum ratios, and we have freedom to choose them. A convenient choice is $r_1=k_{s}/k_{12}$ and $x\equiv k_{34}/k_{12}$, so that the analytical structure of the seed integral in the line energy $k_s$ is manifest on the complex $r_1$ plane. Again, to avoid potential confusions, we introduce a new variable for the seed integral with this particular choice of arguments:
\bge\label{eq_ZtoI}
  \mathcal{Z}(r_1=k_s/k_{12},x=k_{34}/k_{12})\equiv \mathcal{I}^{-2,-2}(k_{12},k_{34},k_s).
\ede
The integral for the nonlocal signal $\mathcal{Z}_\text{NS}(r_1,x)$ is likewise defined. Then, with this new notation, the discontinuity of the seed integral (\ref{eq_discI_ks}) can be rewritten as:
\begin{align}
\label{eq_diskZr1}
  \mathop{\text{Disc}}_{r_1}\mathcal{Z}(r_1,x)
  = \mathop{\text{Disc}}_{r_1}\Big[ \mathcal{Z}_\text{NS}(r_1,x)-\mathcal{Z}_\text{NS}(-r_1,x)\Big] \theta(-r_1),
\end{align}
where $x$ remains in the physical region $x>0$. We show this result in Fig.\ \ref{fig_line_disp}.

From this result we learn a lesson: The singularity structure of the seed integral as a function of one momentum ratio, say $r_1$, is crucially dependent on how we choose and fix other ratios. This is made clear by comparing Fig.\;\ref{fig_4pt_r1plane}, where we fix $r_2$,  and Fig.\;\ref{fig_line_disp}, where we fix $x=r_1/r_2$. There is nothing mysterious here: In the most general situation, the seed integral is to be treated as a function of multiple complex variables, whose singularity structure on a multidimensional complex space can be quite complicated. The dispersion relations considered in this work, on the other hand, are always formulated on a fixed complex dimension-1 submanifold, where we only see the projections of higher dimensional singularities. By fixing different ratios, we are working on different complex dimension-1 submanifolds, and it is not surprising that the projections of singularities on these submanifolds are different. 

With the discontinuity given in (\ref{eq_diskZr1}), we can directly write down a dispersion integral for the seed integral $\mathcal{Z}(r_1,x)$:
\begin{align}\label{dispersion_int_line}
   \mathcal{Z}(r_1,x)=\FR{r_1}{2\pi\ii}\int_{-\infty}^0\FR{\di r}{r(r-r_1)}\mathop{\text{Disc}}_{r_1}\Big[ \mathcal{Z}_\text{NS}(r_1,x)-\mathcal{Z}_\text{NS}(-r_1,x)\Big].
\end{align}
Here we have introduced a first-order subtraction. This choice follows from the asymptotic behavior of the seed integral $\mathcal{Z}(r_1,x)$ in the limit $|r_1|\to\infty$. Note that $r_1=\infty$ is the total energy limit where $\mathcal{Z}(r_1,x)$ diverges at most logarithmically by power counting of time in the time integral. So, a first-order subtraction is sufficient to make the dispersion integral well defined. 

The nonlocal signal of the seed integral has been presented in (\ref{eq_INS}), and here we rewrite it as a function of $r_1$ and $x$:
\begin{align} 
  \mathcal{Z}_\text{NS}(r_1=k_s/k_{12},x=k_{34}/k_{12})
  =\FR{\pi}{4\sinh(\pi\wt\nu)}\sum_{\aa,\bb,\cc=\pm}\aa\bb\mathcal{V}_\cc^{-2}(\aa k_{12},k_s)\mathcal{V}_\cc^{-2}(\bb k_{34},k_{s}),
\end{align}
Using the result for $\mathcal{V}^{-2}_{\mathsf{c}}$ in (\ref{mathcalV_result}), we get an explicit expression for $\mathcal{Z}_{\rm{NS}}$ as:
\begin{align}\label{I_NL_result_r1x}
  \mathcal{Z}_{\rm{NS}}(r_1,x)=\FR{1-\ii\sinh(\pi\nn)}{2\pi}x^{-1/2+\ii\nn}r_1^{1-2\ii\nn}\mathbf{F}_{\nn}^{-2}(r_1/x)\mathbf{F}_{\nn}^{-2}(r_1)+(\nn\to-\nn),
\end{align}
where $\mathbf{F}_{\nn}^p$ is defined in (\ref{eq_mbFp}).

\begin{figure}
\centering 
\includegraphics[width=0.58\textwidth]{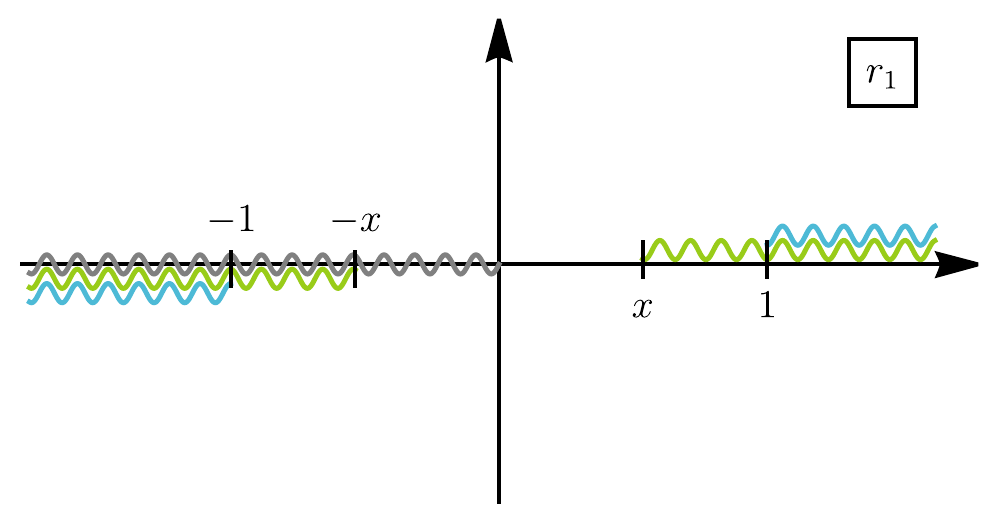} 
\caption{The analytical structure of the nonlocal signal $\mathcal{Z}_\text{NS}(r_1,x)$ on the complex $r_1$ plane, with $x\in(0,\infty)$ fixed in the interior of its physical region. The nonlocal signal is discontinuous for all real $r_1$ except in the physical interval $0<x<\text{min}\{x,1\}$ in which $\mathcal{Z}_\text{NS}(r_1,x)$ is real. The gray, green, and cyan wiggly lines show the branch cut arising from $r^{\pm2\ii\wt\nu}$, $\mb{F}_{\pm\wt\nu}^{-2}(\pm r_1/x)$, and $\mb{F}_{\pm\wt\nu}^{-2}(\pm r_1)$, respectively. All branch cuts are on the real axis. However, we offset the green and cyan cuts to highlight our description for computing the discontinuity of a product of functions, and also for clearer illustration. }
  \label{fig_ZNS}
\end{figure}

\paragraph{Discontinuity of the nonlocal signal}
To evaluate the dispersion integral (\ref{dispersion_int_line}), we need the discontinuity of the nonlocal signal $\mathcal{Z}_{\rm{NS}}$. It is possible to get this discontinuity by analyzing the integral expression for $\mathcal{V}_\pm^{p}$ without really evaluating it, like we did for $\mathcal{U}_\pm^p$ before. However, here we choose to present the discontinuity directly by known analytical properties of power functions and Gauss's hypergeometric functions. Notice that the power function $r_1^{-2\ii\wt\nu}$ has a branch cut in the negative real axis, and that the Gauss's hypergeometric function in $\mathbf{F}_{\nn}^{-2}$ factors has a branch cut when its argument $z\in(1,\infty)$. Thus, all three $r_1$-dependent factors in (\ref{I_NL_result_r1x}) make contributions to the discontinuity of the nonlocal signal. More explicitly:
\begin{itemize}
  \item $r_1^{\pm2\ii\nn}$ contributes a branch cut for $r_1\in(-\infty,0)$.
  \item $\mathbf{F}_{\pm\nn}^{-2}(r_1/x)$ contributes a branch cut for $r_1\in(-\infty,-x)\cup (x,\infty)$. 
  \item $\mathbf{F}_{\pm\nn}^{-2}(r_1)$ contributes a branch cut for $r_1\in(-\infty,-1)\cup (1,\infty)$.
\end{itemize}
Then, it is straightforward to see that the nonlocal signal $\mathcal{Z}_\text{NS}(r_1,x)$ in (\ref{I_NL_result_r1x}) has a branch cut for any real value of $r_1$ except when $0<r_1<\text{min}\{x,1\}$, in which $\mathcal{Z}_\text{NS}(r_1,x)$ is real. In addition, the discontinuity across the branch cut is itself discontinuous at $r_1=\pm x$ and $r_1=\pm 1$. We show these branch cuts in Fig.\;\ref{fig_ZNS}, where we make manifest the contributions from different factors in (\ref{I_NL_result_r1x}).

Incidentally, when we compute a quantity like $\mathop{\text{Disc}}\limits_z [f(z)g(z)]$ where both $f$ and $g$ have discontinuities, there are multiple equivalent ways to express it in terms of the discontinuity of individual factor. For instance, when $z>0$, we have
\begin{align}
  \mathop{\text{Disc}}_z \Big[f(z)g(z)\Big]
  = f(z^+)\mathop{\text{Disc}}_z\Big[g(z)\Big]+\mathop{\text{Disc}}_z\Big[f(z)\Big]g(z^-)\n\\
  =f(z^-)\mathop{\text{Disc}}_z\Big[g(z)\Big]+\mathop{\text{Disc}}_z\Big[f(z)\Big]g(z^+).
\end{align}
Here and below, we introduce the shorthand notation $z^\pm\equiv z e^{\pm\ii\ep}$ with $\ep$ an infinitesimal positive real. Thus, when computing the discontinuity of products of functions, one can make various choices. To fix our choice, we infinitesimally displace some branch cuts into complex plane, as shown in Fig.\ \ref{fig_ZNS}. According to this prescription, for instance, when we compute the discontinuity across the cyan branch cut (from $\mb{F}_{\wt\nu}^{-2}(r_1)$ factor) in the negative real axis, we should evaluate the other two factors on the lower edges of the gray and green branch cut, namely, we take $(r_1^+)^{1-2\ii\wt\nu}$ and $\mb{F}_{\wt\nu}^{-2}(r_1^+/x)$.\footnote{Note that, for negative $r_1$, $r_1^+$ corresponds to the lower edge of the branch cut and $r^-$ to the upper edge.}

The structure of the branch cut suggests that we should break the line dispersion integral (\ref{dispersion_int_line}) into three pieces, each corresponding to the branch cut from a given factor, and also to a wiggly line of a given color in Fig.\;\ref{fig_ZNS}. Below we work out the discontinuity across each of these branch cuts. 

First, the gray branch cut in Fig.\;\ref{fig_ZNS} is contributed by the power factor $r_1^{1-2\ii\wt\nu}$ in (\ref{dispersion_int_line}). We define the discontinuity from this branch cut into the following function:
\begin{align}\label{mathcalF_1}
  \mb{D}_{\nn}^{(1)}(r_1,x)
  \equiv&~ \mathop{\text{Disc}}_{r_1}\Big[r_1^{1-2\ii\wt\nu}\Big]\mathbf{F}^{-2}_{\nn}(r_1^-/x)\mathbf{F}^{-2}_{\nn}(r_1^-)\n\\
  =&-2\sinh(2\pi\wt\nu)(-r_1)^{1-2\ii\wt\nu}\mathbf{F}^{-2}_{\nn}(r_1^-/x)\mathbf{F}^{-2}_{\nn}(r_1^-).
\end{align}
Here the two hypergeometric factors are taking values from the upper edges of their branch cuts on $r_1$ plane, consistent with our displacement of the branch cuts in Fig.\ \ref{fig_ZNS}.

Second, the green branch cut in Fig.\;\ref{fig_ZNS} is contributed by the hypergeometric factor $\mb{F}_{\pm\wt\nu}^{-2}(\pm r_1/x)$ in (\ref{dispersion_int_line}). For negative $r_1$, the discontinuity across  this branch cut is given by the following function:
\begin{align} \label{mathcalF_2}
  \mb{D}_{\nn}^{(2)}(r_1,x)
  \equiv&~ (r_1^+)^{1-2\ii\nn}\mathop{\text{Disc}}_{r_1}\Big[\mathbf{F}^{-2}_{\nn}(r_1/x)\Big]\mathbf{F}^{-2}_{\nn}(r_1^-)\n\\
  =&-e^{-2\pi\wt\nu}(-r_1)^{1-2\ii\wt\nu}\mb{G}_{\wt\nu}(r_1/x)\mb{F}_{\wt\nu}^{-2}(r_1^-),
\end{align}
where we have defined the discontinuity of $\mb{F}_{\wt\nu}^{-2}(z)$ along its branch cut to be $\mb{G}_{\wt\nu}$. Using the known property of the Gauss's hypergeometric function in (\ref{disc_2f1}), we can find an explicit expression for $\mb{G}_{\wt\nu}$:
\begin{align}\label{eq_mbG}
    \mathbf{G}_{\nn}(z)\equiv\mathop{\text{Disc}}_{z}\mathbf{F}_{\nn}^{-2}(z)=-\sqrt{2\pi^3}\,\rm{csch}\,(\pi\nn)\,\2F1\biggl[\bgm\frac{1}{4}-\fr{\ii\nn}{2},\fr{3}{4}-\fr{\ii\nn}{2}\\1\edm\bigg|1-z^2\biggr].~~~~(z<-1)
\end{align}

Third, the cyan branch cut in Fig.\;\ref{fig_ZNS} is contributed by the hypergeometric factor $\mb{F}_{\pm\wt\nu}^{-2}(\pm r_1)$ in (\ref{dispersion_int_line}). For negative $r_1$, we define the discontinuity across this branch cut into the following function:
\begin{align}\label{mathcalF_3}
  \mb{D}_{\nn}^{(3)}(r_1,x)
  \equiv&~ (r_1^+)^{1-2\ii\nn}\mathbf{F}^{-2}_{\nn}(r_1^+/x)\mathop{\text{Disc}}_{r_1}\Big[\mathbf{F}^{-2}_{\nn}(r_1)\Big]\n\\
  =&-e^{-2\pi\wt\nu}(-r_1)^{1-2\ii\wt\nu}\Big[\mathbf{F}^{-2}_{\nn}(r_1^-/x)-\mb{G}_{\wt\nu}(r_1/x)\Big]\mb{G}_{\wt\nu}(r_1) , 
\end{align}
where we have used the relation $\mathbf{F}^{-2}_{\nn}(r_1^+/x)=\mathbf{F}^{-2}_{\nn}(r_1^-/x)+\mathbf{G}_{\nn}(r_1/x)$ to rewrite a hypergeometric factor in terms of its value across the branch cut.

In summary, for negative values of $r_1$, the discontinuity across the branch cuts of the nonlocal signal $\mathcal{Z}_\text{NS}(r_1,x)$ is given by:
\begin{align}
\label{eq_DiscZNSn}
  \mathop{\text{Disc}}_{r_1}\mathcal{Z}_\text{NS}(r_1<0,x)=\mb{D}_{\nn}^{(1)}(r_1,x)\theta(-r_1)+\mb{D}_{\nn}^{(2)}(r_1,x)\theta(-r_1-x)+\mb{D}_{\nn}^{(3)}(r_1,x)\theta(-r_1-1). 
\end{align}

On the other hand, as shown in Fig.\;\ref{fig_ZNS}, the green and cyan branch cuts also extend to positive real values of $r_1$. However, the discontinuities for these ``positive'' branch cuts are not independent, since the two ``positive'' branch cuts can be related to the corresponding two ``negative'' branch cuts by a $180^\circ$ rotation around the origin $r_1=0$ via the \emph{lower} plane (in order not to cross the gray branch cut). Now, using the facts that the function $\mathbf{F}^{-2}_{\nn}(r_1)$ is even in $r_1$, and that $(e^{-\ii\pi}r_1)^{1-2\ii\wt\nu}=-e^{-2\pi\wt\nu}r_1^{1-2\ii\wt\nu}$ for $r_1>0$, we have:
\begin{align}
\label{eq_DiscZNSp}
  \mathop{\text{Disc}}_{r_1}\mathcal{Z}_\text{NS}(r_1>0,x)=e^{2\pi\wt\nu}\mb{D}_{\nn}^{(2)}(r_1,x)\theta(r_1-x)+e^{2\pi\wt\nu}\mb{D}_{\nn}^{(3)}(r_1,x)\theta(r_1-1).
\end{align}
Thus we have found the explicit expressions for all five branch cuts shown in Fig.\;\ref{fig_ZNS}.
 
\paragraph{Line dispersion integral and the result} 
Based on the previous analysis of the branch cut of the nonlocal signal, we are now ready to find the explicit expression for the line dispersion integral. Combining (\ref{eq_DiscZNSn}) and (\ref{eq_DiscZNSp}), we see that the dispersion integral (\ref{dispersion_int_line}) now boils down to three integrals $\mb J_{i}^{\wt\nu}$ $(i=1,2,3)$: 
\begin{align}\label{mathcalZ=sum_of_boldsymbolI_L}
    {\mathcal{Z}}(r_1,x)=-\FR{\ii+\sinh(\pi\nn)}{4\pi^2}x^{-1/2+\ii\nn}{r_1}\Bigl[\mathbf{J}_{\wt\nu}^{(1)}(r_1,x)+\mathbf{J}_{\wt\nu}^{(2)}(r_1,x)+\mathbf{J}_{\wt\nu}^{(3)}(r_1,x)\Bigr]+(\nn\to-\nn),
\end{align}
where the three terms $(\mb{J}_{\wt\nu}^{(1)},\mb{J}_{\wt\nu}^{(2)},\mb{J}_{\wt\nu}^{(3)})$ correspond to integrals around the gray, green, and cyan branch cuts in Fig.\ \ref{fig_ZNS}, respectively:
\begin{align}
\label{boldsymbolI_L_1}
    \mathbf{J}_{\wt\nu}^{(1)}(r_1,x)\equiv&\int_{-\infty}^{0}\di r\FR{\mb{D}_{\nn}^{(1)}(r,x)}{r(r-r_1)},\\
\label{boldsymbolI_L_2}
    \mathbf{J}_{\wt\nu}^{(2)}(r_1,x)\equiv&\int_{-\infty}^{-x}\di r\FR{(1-e^{2\pi\nn})\mb{D}_{\nn}^{(2)}(r,x)}{r(r-r_1)},\\
\label{boldsymbolI_L_3}
    \mathbf{J}_{\wt\nu}^{(3)}(r_1,x)\equiv&\int_{-\infty}^{-1}\di r\FR{(1-e^{2\pi\nn})\mb{D}_{\nn}^{(3)}(r,x)}{r(r-r_1)}.
\end{align}
These three integrals can be computed analytically via PMB representation, although the details are quite lengthy. We collect them in App.\ \ref{appd_line}, and present the results below. 

As mentioned many times before, the 4-point seed integral $\mathcal{Z}(r_1,x)$ can be written as a sum of nonlocal signal (NS), the local signal (LS), and the background (BG): 
\bge
  \mathcal{Z}(r_1,x)=\mathcal{Z}_\text{NS}(r_1,x)+\mathcal{Z}_\text{LS}(r_1,x)+\mathcal{Z}_\text{BG}(r_1,x).
\ede
It turns out that the nonlocal signal $\mathcal{Z}_\text{NS}$ is contributed only by the integral around the gray branch cut, namely $\mathbf{J}_{\wt\nu}^{(1)}$. The result is simply identical to our input (\ref{I_NL_result_r1x}), which we collect here for completeness:
\begin{align} 
  \mathcal{Z}_{\rm{NS}}(r_1,x)=\FR{1-\ii\sinh(\pi\nn)}{2\pi}x^{-1/2+\ii\nn}r_1^{1-2\ii\nn}\mathbf{F}_{\nn}^{-2}(r_1/x)\mathbf{F}_{\nn}^{-2}(r_1)+(\nn\to-\nn),
\end{align}
The local signal $\mathcal{Z}_\text{LS}$ receives contributions from the integrals around the gray and the green branch cuts, namely $\mathbf{J}_{\wt\nu}^{(1)}$ and $\mathbf{J}_{\wt\nu}^{(2)}$, respectively. The result is:
\begin{align}
  \mathcal{Z}_{\rm{LS}}(r_1,x)=&~\mathcal{Z}_{\rm{LS},>}(r_1,x)\theta\big(1-|x|\big)+\mathcal{Z}_{\rm{LS},>}(r_1/x,1/x)\theta\big(|x|-1\big),\\
  \mathcal{Z}_{\rm{LS},>}(r_1,x)=&~\FR{1-\ii\sinh(\pi\nn)}{2\pi}x^{-1/2-\ii\nn}r_1\mathbf{F}^{-2}_{-\nn}(r_1/x)\mathbf{F}_{\nn}^{-2}(r_1)+(\nn\to-\nn).
\end{align}
Finally, the background $\mathcal{Z}_\text{BG}$  receives contributions from the integrals around all three branch cuts, namely $\mathbf{J}_{\wt\nu}^{(i)}$ $(i=1,2,3)$, whose result can be simplified into the following form: 
\begin{align}
  \mathcal{Z}_{\rm{BG}}(r_1,x)=&~\mathcal{Z}_{\rm{BG},>}(r_1,x)\theta\big(1-|x|\big)+\mathcal{Z}_{\rm{BG},>}(r_1/x,1/x)\theta\big(|x|-1\big),\\
  \mathcal{Z}_{\rm{BG},>}(r_1,x)=&\sum_{n=0}^{\infty}\frac{8(-x)^n r_1}{(1+2n)^2+4\nn^2}\,\3F2\biggl[\bgm1,\frac{1}{2}+\frac{n}{2},1+\frac{n}{2}\\\frac{5}{4}+\frac{n}{2}-\frac{\ii\nn}{2},\frac{5}{4}+\frac{n}{2}+\frac{\ii\nn}{2}\edm\bigg|r_1^2\biggr].
\end{align}
This expression for $\mathcal{Z}_{\rm{BG}}$ has a different look from known results in \cite{{Qin:2022fbv}}, but is identical to the latter. In fact, the background part is a two-variable hypergeometric function known as Kampé de Fériet function and allows for many different series representations \cite{Arkani-Hamed:2018kmz,Qin:2022fbv,Xianyu:2022jwk}.

\section{Conclusions and Outlooks}\label{sec_conclusion}

As the dS counterparts of flat-space scattering amplitudes, inflation correlators possess distinct analytical structure. For general massive-exchange processes, branch cuts usually appear in the complex plane of appropriate kinematic variables, which connect physics in the UV and IR regions. In the IR regions, such branch cuts are closely related to logarithmic oscillations of the correlators in the physical regions, known as CC signals in the context of Cosmological Collider physics. The CC signals have received many studies in recent years. In particular, it has been shown that, although the computation of general inflation correlators is difficult, we can apply the cutting rule and the factorization theorem to extract CC signals in the squeezed limit \cite{Tong:2021wai,Qin:2023bjk,Qin:2023nhv}. In comparison, the corresponding branch cuts on the complex domain beyond the physical region are less understood and deserve more studies. 

In this work, we explore the analytical properties of massive inflation correlators as functions of two types of kinematic variables: the vertex energies and the line energies. For both types of energies of a tree-level correlator, we identified the total-energy and partial-energy poles, the signal branch point, together with the branch cuts connecting them. Based on this structure, we developed two distinct dispersion relations: a vertex dispersion relation which relates the correlator to its full signal, and a line dispersion relation which relates the correlator to the nonlocal signal alone. With these dispersion relations, we have successfully bootstrapped a few tree-level and 1-loop massive inflation correlators. At 1-loop level, our method is manifestly UV finite and free from any regularization procedure. This allows us to neatly single out the renormalization-independent part of the correlator, which is unambiguously determined by analyticity. 

Although there have been scattered studies on analytical properties of inflation correlators (and the related wavefunction coefficients), to our best knowledge, the dispersion relations have not been used to bootstrap the full massive inflation correlators. Our work filled this gap by providing a few proof-of-principle calculations. While the computation itself can often become lengthy compared to other existing methods for simple examples, it nevertheless shows the potential power of the dispersion techniques in bootstrapping more complicated diagrams. Thus, we consider this work a first step in carrying out a more extensive program of dispersive bootstrap. Naturally, many directions are open to further explorations, and we conclude this work by mentioning some of them.

A natural first task is to chart all nonanalyticities of a given tree diagram, beyond the 4-point single exchange. This includes not only the locations of poles and branch cuts, but also the discontinuities across all branch cuts. With these data, we can imagine to recursively bootstrap more complicated diagrams from simple sub-diagrams, either analytically or numerically. 

Next, it would be very interesting to explore the potential of dispersive bootstrap for  loop diagrams. We have seen that dispersion relations could be advantageous in bootstrapping one-loop diagrams, including the absence of UV divergences, the simplified expressions, and the separation of renormalization dependent and independent parts. They encourage us to consider more complicated loop processes. As a concrete first step, we may try to combine the dispersive and spectral methods and bootstrap 1-loop bubble processes with spinning exchanges and derivative couplings, and this will be explored in a follow-up work. Beyond the bubble topology, it is not immediately clear that techniques like spectral decomposition are still available. Nevertheless, it looks promising to us to numerically implement the dispersive techniques for loop processes. We plan to investigate this route in a future work. 

As we pointed out many times, the dispersive bootstrap at its core is an idea to reconstruct the whole diagram from a knowledge of sub-diagrams. In this regard, what we have considered in this work is the most straightforward realization, namely, exploiting the complex energy planes. It is also interesting to search for ``dispersion relations'' with not only complex energies, but also other complex parameters. In flat space, it has been very fruitful to consider scattering amplitudes on complex planes of mass, angular momentum, and even spacetime dimensions. We can imagine that the analytical structures in these complex parameters could also bring us new insights and new methods for inflation correlators. Also, it has been shown recently that the parity-odd part of a cosmological correlator (or a wavefunction coefficient) automatically factorized under rather general conditions. \cite{Liu:2019fag,Cabass:2022rhr,Stefanyszyn:2023qov,Stefanyszyn:2024msm} Thus, it would be very interesting to develop dispersion techniques for parity-violating theories. 

Last but not least, the dispersion relations in flat spacetime or in CFT are usually tied to nonperturbative properties of amplitudes, and are used to make nonperturbative statements about the unitarity and positivity of the theory. On the other hand, in this work, we only apply the dispersion techniques at the diagrammatic level. How are the two approaches related? From pure diagrammatic analysis, is it possible to gain insights applicable to all orders in perturbation theory? Similar to flat-space situations, we believe that, at least for simple kinematics with full dS isometries, it is possible to make progress along these directions. We leave all these interesting topics for future studies.

\paragraph{Acknowledgments} We thank Xingang Chen, Enrico Pajer, Carlos Duaso Pueyo, Sébastien Renaux-Petel, Xi Tong, Lian-Tao Wang, Yi Wang, Denis Werth, Jiayi Wu, Hongyu Zhang, and Yuhang Zhu for useful discussions. This work is supported by NSFC under Grants No.\ 12275146 and No.\ 12247103, the National Key R\&D Program of China (2021YFC2203100), and the Dushi Program of Tsinghua University.

\newpage
\begin{appendix}

\section{Notations}\label{appd_notation}

In this appendix, for readers' convenience, we collect some frequently used variables in Table \ref{tab_notations}, together with the numbers of equations where they are defined or first appear.
\begin{table}[tbph] 
  \centering
   \caption{List of selected symbolic notations }
   \vspace{2mm}
  \begin{tabular}{lll}
   \toprule[1.5pt]
    Notation 
    &\multicolumn{1}{c}{Description} & Equation \\ \hline 
    $E_i$ & External energy & Above (\ref{mathcalG_of_energy})\\
    $K_i$ & Internal energy & Above (\ref{mathcalG_of_energy})\\
    $u$ & Momentum ratio $u\equiv2k_3/k_{123}$ & Above (\ref{eq_XtoI})\\
    $r_1,r_2$ & Momentum ratios $r_1\equiv k_s/k_{12}$ and $r_2\equiv k_s/k_{34}$ & (\ref{def_r})\\
    $x$ & Momentum ratio $x\equiv k_{34}/k_{12}$ & Above (\ref{eq_ZtoI}) \\
    $C_{\mathsf{a}}(k;\tau)$ & External propagator of conformal scalar $\phi_c$ & (\ref{eq_CSProp})\\
    $D_{\mathsf{ab}}(k;\tau_1,\tau_2)$& Internal propagator of massive scalar $\sigma$ & (\ref{eq_Dmp}), (\ref{eq_Dpm}), and (\ref{eq_Dpmpm})\\
    $\Sigma(k;\tau_1,\tau_2)$ & Same-order part of bulk propagator & (\ref{eq_SigmaProp})\\
    $\Omega_{\mathsf{ab}}(k;\tau_1,\tau_2)$ & Opposite-order part of bulk propagator & (\ref{eq_OmegaProp})\\
    $\mathbf{Q}_{\mathsf{ab}}(k;\tau_1,\tau_2)$ & 1-loop bubble function of $\si$ & (\ref{loop_propagator})\\
    $\mathsf{a},\mathsf{b},\cdots$ & SK or non-SK indices taking value from $\pm1$ & (\ref{eq_GraphInt})\\
    $\mathcal{U}^p_{\pm}(E_i,K_i)$ & Subgraph integral with one Hankel H & (\ref{eq_Uplus})\\
    $\mathcal{V}^p_{\pm}(E_i,K_i)$ & Subgraph integral with one Bessel J & (\ref{mathcalV_definition})\\
    $\mathcal{G}$ & A general tree graph for $\phi_c$ correlators & (\ref{eq_GraphInt})\\
    $\mathcal{I}_{\mathsf{ab}}(k_{12},k_{34},k_s)$ & 4-point single-exchange seed integral & (\ref{seed_tree_4pt})\\
    $\mathcal{P}_{\mathsf{ab}}$ & Same-order part of $\mathcal{I}_{\aa\bb}$ & (\ref{eq_Pab})\\
    $\mathcal{Q}_{\mathsf{ab}}$ & Opposite-order part of $\mathcal{I}_{\aa\bb}$ & (\ref{eq_Qab})\\
    $\mathcal{X}(u)$ & 3-point tree integral in $u$ & (\ref{eq_XtoI})\\
    $\mathcal{J}(u)$ & 3-point 1-loop bubble integral in $u$ & (\ref{seed_loop})\\
    $\mathcal{Y}(r_1,r_2)$ & 4-point tree integral in $r_1$ and $r_2$ & (\ref{eq_YtoI})\\
    $\mathcal{Z}(r_1,x)$ & 4-point tree integral in $r_1$ and $x$ & (\ref{eq_ZtoI})\\
    $\mb{F}_{\wt\nu}^{p}(z)$ & A rescaled Gauss's hypergeometric funtion  & (\ref{eq_mbFp})\\ 
    $\mb{G}_{\wt\nu} (z)$ &  Discontinuity of $\mb{F}^{-2}_{\nn}(z)$ when $z<-1$  & (\ref{eq_mbG}) \\
   \bottomrule[1.5pt] 
  \end{tabular}
  \label{tab_notations}
\end{table}

\section{Useful Functions and Properties}\label{appd_function}

In this appendix, we collect a few special functions and their properties used in the main text. These are standard material, and we quote them from \cite{nist:dlmf,wolfram}.

\paragraph{Euler $\mb{Gamma}$ products and fractions} In this work we use the following shorthand notation for the productions and fractions of Euler $\Gamma$ functions:
\bge
  \Gamma\big[a_1,\cdots,a_n\big]\equiv \Gamma(a_1)\cdots \Gamma(a_n);
\ede
\begin{align}
    \Gamma\biggl[\bgm a_1,a_2,\cdots,a_m\\b_1,b_2,\cdots,b_n\edm\biggr]\equiv\FR{\Gamma(a_1)\Gamma(a_2)\cdots\Gamma(a_m)}{\Gamma(b_1)\Gamma(b_2)\cdots\Gamma(b_n)}.
\end{align}
With this notation, the Pochhammer symbol $(a)_n$ is defined as
\begin{align}
    (a)_n\equiv\Gamma\biggl[\bgm a+n\\a\edm\biggr].
\end{align}

\paragraph{Hypergeometric functions}
The (generalized) hypergeometric function is used in this work, whose standard form is defined by the following series when convergent, and by analytical continuation otherwise:
\begin{align}
    _{p}\rm{F}_{q}\biggl[\bgm a_1,a_2,\cdots,a_p\\b_1,b_2,\cdots,b_q\edm\bigg|z\biggr]\equiv\sum_{n=0}^{\infty}\FR{(a_1)_n(a_2)_n\cdots(a_p)_n}{(b_1)_n(b_2)_n\cdots(b_q)_n}\FR{z^n}{n!}.
\end{align} 
In particular, ${}_{2}\text{F}_1$ is known as the Gauss's or ordinary hypergeometric function. 

There are a few useful variations whose definitions are different from the standard form only in prefactors. First, the \emph{regularized} hypergeometric function ${}_{p}\wt{\rm{F}}_{q}$ is defined by:
\begin{align}
    {}_{p}\wt{\rm{F}}_{q}\biggl[\bgm a_1,a_2,\cdots,a_p\\b_1,b_2,\cdots,b_q\edm\bigg|z\biggr]\equiv\FR{1}{\Gamma[b_1,b_2,\cdots,b_q]}{}_{p}\rm{F}_{q}\biggl[\bgm a_1,a_2,\cdots,a_p\\b_1,b_2,\cdots,b_q\edm\bigg|z\biggr].
\end{align}
It is called regularized, because, when the argument $z$ is not at the singular points, the regularized hypergeometric function is an entire function of all the parameters $(a_1,\cdots,a_p,b_1,\cdots,b_q)$.

Second, we frequently use the ``dressed" hypergeometric function ${}_{p}{\mathcal{F}}_{q}$ in the main text, because it simplifies a lot of expressions:
\begin{align}
    {}_{p}{\mathcal{F}}_{q}\biggl[\bgm a_1,a_2,\cdots,a_p\\b_1,b_2,\cdots,b_q\edm\bigg|z\biggr]\equiv\FR{\Gamma[a_1,a_2,\cdots,a_p]}{\Gamma[b_1,b_2,\cdots,b_q]}{}_{p}\rm{F}_{q}\biggl[\bgm a_1,a_2,\cdots,a_p\\b_1,b_2,\cdots,b_q\edm\bigg|z\biggr].
\end{align}

It is useful to note that the Gauss's hypergeometric function ${}_2\text{F}_1[\cdots|z]$ in general has two branch points at $z=1$ and $z=\infty$. It is our convention to choose the branch cut connecting these two points to lie in the interval $z\in(1,\infty)$ on the real axis. We define the value of ${}_2\text{F}_1[\cdots|z]$ when $z>1$ by its value on the \emph{lower} edge of the branch cut. Then, the value on the upper edge is determined by the discontinuity across the branch cut. More explicitly:
\begin{equation}\label{disc_2f1}
    \left\{
    \begin{aligned}
       &{}_2\rm{F}_1\biggl[\bgm a,b\\c\edm\bigg|z^+\biggr] \\
        =&\ 2\pi\ii e^{\pi\ii(a+b-c)}\Gamma\biggl[\bgm c\\a+b-c+1,c-a,c-b\edm\biggr]{}_2\rm{F}_1\biggl[\bgm a,b\\a+b-c+1\edm\bigg|1-z\biggr]\\
        &+e^{2\pi\ii(a+b-c)}{}_2\rm{F}_1\biggl[\bgm a,b\\c\edm\bigg|z\biggr],\\
      &{}_2\rm{F}_1\biggl[\bgm a,b\\c\edm\bigg|z^-\biggr]={}_2\rm{F}_1\biggl[\bgm a,b\\c\edm\bigg|z\biggr].
    \end{aligned}
    \right.
\end{equation}

For power functions with non-integer powers, we can get a branch cut along the negative real axis by restricting the argument of variable in $(-\pi,\pi]$.
\begin{align}\label{disc of power}
    \begin{aligned}
        &(ze^{\ii\pi})^{p}=e^{\ii\pi p}z^{p},\\
        &(ze^{-\ii\pi})^{p}=e^{-\ii\pi p}z^{p}.~~~~(z>0)
    \end{aligned}
\end{align}

\paragraph{Bessel functions} In this work, we used the standard Bessel J function, especially its analytical property. Generically, a Bessel J function $J_{\nu}(z)$ has a branch cut on the negative real axis, connecting $z=0$ and $z=-\infty$. The discontinuity across this branch cut is conveniently captured by the following identity:
\begin{align}\label{disc_besselJ}
  \rm{J}_{\pm\ii\wt\nu}(e^{m\pi\ii}z)=e^{\mp m\wt\nu\pi}\rm{J}_{\pm\ii\wt\nu}(z).~~~~(z>0)
\end{align}

More frequently appeared in the main text are Hankel functions $\rm{H}_{\nu}^{(1)}$ and $\rm{H}_{\nu}^{(2)}$, which can be expressed in terms of Bessel J function as:
\begin{align}\label{eq_HinJ}
  \begin{aligned}
  &\rm{H}^{(1)}_{\ii\nn}(z)=\bigl(1+\coth(\pi\nn)\bigr)\rm{J}_{\ii\nn}(z)-\rm{csch}(\pi\nn)\rm{J}_{-\ii\nn}(z),\\
  &\rm{H}^{(2)}_{-\ii\nn}(z)=-\rm{csch}(\pi\nn)\rm{J}_{\ii\nn}(z)+\bigl(1+\coth(\pi\nn)\bigr)\rm{J}_{-\ii\nn}(z).
  \end{aligned}
\end{align}
Consequently, the Hankel functions $\rm{H}_{\nu}^{(j)}(z)$ $(j=1,2)$ possess branch cuts on the negative real axis of $z$, whose discontinuity can be found from the following identities: 
\begin{equation}\label{disc of hankel}
    \begin{aligned}
        &\rm{H}^{(1)}_{\ii\nn}(ze^{\ii\pi})=-e^{\pi\nn}\rm{H}^{(2)}_{\ii\nn}(z),\\
        &\rm{H}^{(1)}_{\ii\nn}(ze^{-\ii\pi})= 2\cosh(\pi\nn)\rm{H}^{(1)}_{\ii\nn}(z)+e^{\pi\nn}\rm{H}^{(2)}_{\ii\nn}(z),\\
        &\rm{H}^{(2)}_{-\ii\nn}(ze^{\ii\pi})=e^{\pi\nn}\rm{H}^{(1)}_{-\ii\nn}(z)+ 2\cosh(\pi\nn) \rm{H}^{(2)}_{-\ii\nn}(z),\\
        &\rm{H}^{(2)}_{-\ii\nn}(ze^{-\ii\pi})=-e^{\pi\nn}\rm{H}^{(1)}_{-\ii\nn}(z).~~~~(z>0)
\end{aligned}
\end{equation}

\section{Vertex Dispersion Integral with PMB Representation}\label{appd_vertex}

In this section we collect some details of  computing the 4-point single-exchange correlator from the vertex dispersion integral (\ref{dispersion_int_vertex_4pt}). As shown in Sec. \ref{subsec_4ptTree_VB}, the vertex dispersion integral for the 4-point tree seed integral can be reduced to  (\ref{eq_Yr1r2dispInt}), which in turns amount to the computation of two integrals $\mb{I}^{(j)}_{\wt\nu}$ ($j=1,2$), which we collect here again:
\begin{align}
  &\mathbf{I}_{\nn}^{(1)}(r_1,r_2)\equiv\int_{-r_2}^{0}\di r\FR{(-r)^{1/2-\ii\nn}\mathbf{F}_{\nn}^{-2}(-r)}{r(r-r_1)},\\
  \label{eq_I2app}
  &\mathbf{I}_{\nn}^{(2)}(r_1,r_2)\equiv\int_{-1}^{-r_2}\di r\FR{(-r)^{1/2-\ii\nn}\mathbf{F}_{\nn}^{-2}(-r)}{r(r-r_1)}.
\end{align}

\paragraph{Computing $\mathbf{I}_{\nn}^{(1)}$ and $\mathbf{I}_{\nn}^{(2)}$} Now we compute the two integrals above with PMB representation. For $\mathbf{I}_{\nn}^{(1)}$, we take the MB representation of $\mathbf{F}_{\nn}^{-2}(-r)$, which is given by:
\begin{align}
    \mathbf{F}_{\nn}^{-2}(-r)=\int_{-\ii\infty}^{\ii\infty}\FR{\di s}{2\pi\ii}(-r)^{-s}\mathbb{F}_{\nn}(s),
\end{align}
where 
\begin{align}\label{mathbbF}
    \mathbb{F}_{\nn}(s)\equiv -\FR{\ii\pi2^{-1+s+\ii\nn}e^{-\ii\pi s/2}}{\sinh(\pi\nn)}\Gamma\biggl[\bgm\fr{s}{2},\fr{1}{2}-s-\ii\nn\\1-\fr{s}{2}-\ii\nn\edm\biggr].
\end{align}
then the original integral $ \mb{I}_{\nn}^{(1)}$ becomes:
\begin{align}
    \mb{I}_{\nn}^{(1)}=\int_{-r_2}^{0}\di r\int_{-\ii\infty}^{\ii\infty}\FR{\di s}{2\pi\ii}\FR{(-r)^{1/2-s-\ii\nn}}{r(r-r_1)}\mathbb{F}_{\nn}(s).
\end{align}
The integral over $r$ can then be finished directly, which gives:
\begin{align}
    \int_{-r_2}^{0}\di r\FR{(-r)^{1/2-s-\ii\nn}}{r(r-r_1)}=\FR{r_2^{1/2-s-\ii\nn}}{r_1}{}_2\mathcal{F}_1\biggl[\bgm\fr{1}{2}-s-\ii\nn,1\\\fr{3}{2}-s-\ii\nn\edm\bigg|-\FR{r_2}{r_1}\biggr].
\end{align}
Now, the original integral $ \mb{I}_{\nn}^{(1)}$ has been recasted into an integral over a Mellin variable $s$:
\begin{align}
    \mb{I}_{\nn}^{(1)}=\int_{-\ii\infty}^{\ii\infty}\FR{\di s}{2\pi\ii}\FR{r_2^{1/2-s-\ii\nn}}{r_1}{}_2\mathcal{F}_1\biggl[\bgm\fr{1}{2}-s-\ii\nn,1\\\fr{3}{2}-s-\ii\nn\edm\bigg|-\FR{r_2}{r_1}\biggr]\mathbb{F}_{\nn}(s).
\end{align}
Again, we use residue theorem to compute the integral over $s$. For $r_2\in(0,1)$, we need to close the contour from the left side, and get a set of poles coming from $\Gamma[s/2]$ in $\mathbb{F}_{\nn}(s)$:
\begin{align}
    s=-2n.~~~~(n=0,1,2,\cdots)
\end{align}
Summing up all residues we get:\footnote{Here we introduce the notation $\rm{Res}(\mathbf{I},s)$ to represent the residue of integral $\mathbf{I}$ of the pole at $s$, multiplying an extra factor $2\pi\ii$ for simplicity. For example, if for integral $\mathbf{I}$ there is only one pole $s$ inside the contour, then the final result is simply $\mathbf{I}=\pm\rm{Res}(\mathbf{I},s)$, where the plus/minus sign depends on the direction of the contour.}
\begin{align}
    \mb{I}_{\nn}^{(1)}=\sum_{n=0}^{\infty}\rm{Res}(\mb{I}_{\nn}^{(1)},-2n),
\end{align}
where 
\begin{align}\label{res_I_V_1_-2n}
    \rm{Res}(\mb{I}_{\nn}^{(1)},-2n)=-\FR{\ii\pi2^{-2n+\ii\nn}r_2^{{1}/{2}+2n-\ii\nn}}{\sinh(\pi\nn)r_1}\Gamma\biggl[\bgm\fr{1}{2}+2n-\ii\nn\\1+n,1+n-\ii\nn\edm\biggr]{}_2\mathcal{F}_1\biggl[\bgm1,\fr{1}{2}+2n-\ii\nn\\\fr{3}{2}+2n-\ii\nn\edm\bigg|-\FR{r_2}{r_1}\biggr].
\end{align}
This completes the computation of $\mb{I}_{\nn}^{(1)}$.

Next we consider $\mb{I}_{\nn}^{(2)}$ in (\ref{eq_I2app}). Again we use the MB representation of $\mathbf{F}^{-2}_{\nn}$ (\ref{mathbbF}) and get:
\begin{align}
    \mb{I}_{\nn}^{(2)}=\int_{-1}^{-r_2}\di r\int_{-\ii\infty}^{\ii\infty}\FR{\di s}{2\pi\ii}\FR{(-r)^{1/2-s-\ii\nn}}{r(r-r_1)}\mathbb{F}_{\nn}(s).
\end{align}
So the integral over $r$ can be done:
\begin{align}\label{I_V_2_over_r}
   & \int_{-1}^{-r_2}\di r\ \FR{(-r)^{\frac{1}{2}-s-\ii\nn}}{r(r-r_1)}= \Gamma\Bigl[\fr{1}{2}+s+\ii\nn\Bigr]\n\\
    &\times\biggl({r_2^{-\frac{1}{2}-s-\ii\nn}}{}_2\wt{\rm{F}}_1\biggl[\bgm1,\frac{1}{2}+s+\ii\nn\\\frac{3}{2}+s+\ii\nn\edm\bigg|-\FR{r_1}{r_2}\biggr]-{}_2\wt{\rm F}_1\biggl[\bgm1,\frac{1}{2}+s+\ii\nn\\\frac{3}{2}+s+\ii\nn\edm\bigg|-r_1\biggr]\biggr),
\end{align}
where ${}_2\wt{\rm F}_1[\cdots]$ is the regularized Gauss's hypergeometric function whose definition is collected in App. \ref{appd_function}.
Again the integral over $s$ can be finished via residue theorem. Closing the contour from the left side, there are two sets of poles: one from $\Gamma[s/2]$ in $\mathbb{F}_{\nn}(s)$, another from $\Gamma[1/2+s+\ii\nn]$ contributed by the integral over $r$ (\ref{I_V_2_over_r}):
\begin{align}
    \left\{\begin{aligned}
    &s=-2n,\\
    &s=-\fr{1}{2}-n-\ii\nn.~~~~(n=0,1,2,\cdots)
    \end{aligned}\right.
\end{align}
Then we get
\begin{align}
    \mb{I}_{\nn}^{(2)}=\sum_{n=0}^{\infty}\rm{Res}(\mb{I}_{\nn}^{(2)},-2n)+\sum_{n=0}^{\infty}\rm{Res}(\mb{I}_{\nn}^{(2)},-\fr{1}{2}-n-\ii\nn),
\end{align}
where
\begin{align}\label{res_I_V_2_-2n}
   &\rm{Res}(\mb{I}_{\nn}^{(2)},-2n)= -\FR{\ii\pi2^{-2n+\ii\nn}}{\sinh(\pi\nn)}\Gamma\biggl[\bgm\fr{1}{2}+2n-\ii\nn\\1+n,1+n-\ii\nn\edm\biggr]\n\\
    &\times\biggl(r_2^{-1/2+2n-\ii\nn}{}_2\mathcal{F}_1\biggl[\bgm1,\fr{1}{2}-2n+\ii\nn\\\fr{3}{2}-2n+\ii\nn\edm\bigg|-\FR{r_1}{r_2}\biggr]-{}_2\mathcal{F}_1\biggl[\bgm1,\fr{1}{2}-2n+\ii\nn\\\fr{3}{2}-2n+\ii\nn\edm\bigg|-r_1\biggr]\biggr),
\end{align}
and
\begin{align}
    \rm{Res}(\mb{I}_{\nn}^{(2)},-\fr{1}{2}-n-\ii\nn)=&\ \FR{\pi2^{-3/2-n}e^{\ii\pi(-1-2n+2\ii\nn)/4}}{\sinh(\pi\nn)}\Gamma\biggl[\bgm-\fr{1}{4}-\fr{n}{2}-\fr{\ii\nn}{2}\\\fr{5}{4}+\fr{n}{2}-\fr{\ii\nn}{2}\edm\biggr]\n\\
    &\times\biggl(r_2^{n}{}_2\wt{\rm{F}}_1\biggl[\bgm1,-n\\1-n\edm\bigg|-\FR{r_1}{r_2}\biggr]-{}_2\wt{\rm{F}}_1\biggl[\bgm1,-n\\1-n\edm\bigg|-r_1\biggr]\biggr).
\end{align}
Note that 
\begin{align}
    {}_2\wt{\rm{F}}_1\biggl[\bgm1,-n\\1-n\edm\bigg|x\biggr]=\Gamma[n](-x)^n,~~~~(n=0,1,2,\cdots)
\end{align}
so residues from the second set of poles actually vanish:
\begin{align}
  \rm{Res}(\mb{I}_{\nn}^{(2)},-\fr{1}{2}-n-\ii\nn)=0.
\end{align}
This completes the computation of  $\mb{I}_{\nn}^{(2)}$. Let us collect the explicit results for both integrals $\mb{I}_{\nn}^{(j)}$ $(j=1,2)$ here for future reference:
\begin{align}
\label{eq_In1result}
  \mb{I}_{\nn}^{(1)}(r_1,r_2)=&\sum_{n=0}^{\infty}-\FR{\ii\pi2^{-2n+\ii\nn}r_2^{{1}/{2}+2n-\ii\nn}}{\sinh(\pi\nn)r_1}\Gamma\biggl[\bgm\fr{1}{2}+2n-\ii\nn\\1+n,1+n-\ii\nn\edm\biggr]{}_2\mathcal{F}_1\biggl[\bgm1,\fr{1}{2}+2n-\ii\nn\\\fr{3}{2}+2n-\ii\nn\edm\bigg|-\FR{r_2}{r_1}\biggr],\\
\label{eq_In2result}
  \mb{I}_{\nn}^{(2)}(r_1,r_2)=&\ \sum_{n=0}^{\infty}-\FR{\ii\pi2^{-2n+\ii\nn}}{\sinh(\pi\nn)}\Gamma\biggl[\bgm\fr{1}{2}+2n-\ii\nn\\1+n,1+n-\ii\nn\edm\biggr]\n\\
  &\times\biggl(r_2^{-1/2+2n-\ii\nn}{}_2\mathcal{F}_1\biggl[\bgm1,\fr{1}{2}-2n+\ii\nn\\\fr{3}{2}-2n+\ii\nn\edm\bigg|-\FR{r_1}{r_2}\biggr]-{}_2\mathcal{F}_1\biggl[\bgm1,\fr{1}{2}-2n+\ii\nn\\\fr{3}{2}-2n+\ii\nn\edm\bigg|-r_1\biggr]\biggr).
\end{align}

\paragraph{Simplifying the result} In principle, we can just substitute the above results for $\mb{I}_{\nn}^{(j)}$ $(j=1,2)$ into (\ref{dispersion_int_vertex_4pt}) to get an analytical expression for the tree seed integral, but this expression is obviously to be simplified. Now we describe how to massage this expression, using various functional identities, to get a reasonably simplified result. 

For definiteness, we consider the case $0<r_1<r_2<1$ without loss of generality. 
Given this relation, the result can be separated into the signal and the background without introducing $\theta$ factors. 

Then, for $\mb{I}_{\nn}^{(1)}$ shown in (\ref{eq_In1result}), we use the following relation:
\begin{align}
  \2F1\biggl[\bgm a,b\\b+1\edm\bigg|-x\biggr]=x^{-b}\Gamma\biggl[\bgm a-b,1+b\\a\edm\biggr]-\FR{b\times x^{-a}}{a-b}\2F1\biggl[\bgm a,a-b\\1+a-b\edm\bigg|-\FR{1}{x}\biggr],
\end{align}
through which we get
\begin{align}
  {}_2\mathcal{F}_1\biggl[\bgm1,\fr{1}{2}+2n-\ii\nn\\\fr{3}{2}+2n-\ii\nn\edm\bigg|-\FR{r_2}{r_1}\biggr]=\pi~\rm{sech}(\pi\nn)\Bigl(\FR{r_1}{r_2}\Bigr)^{1/2+2n-\ii\nn}-\FR{r_1}{r_2}{}_2\mathcal{F}_1\biggl[\bgm1,\fr{1}{2}-2n+\ii\nn\\\fr{3}{2}-2n+\ii\nn\edm\bigg|-\FR{r_1}{r_2}\biggr].
\end{align}
Then
\begin{align}
\label{boldsymbolI_V_1=sum_mathbfI}
  \mb{I}_{\nn}^{(1)}(r_1,r_2)=&-\sum_{n=0}^{\infty}\FR{\ii\pi2^{-2n+\ii\nn}}{\sinh(\pi\nn)}\Gamma\biggl[\bgm\fr{1}{2}+2n-\ii\nn\\1+n,1+n-\ii\nn\edm\biggr]\n\\
  &\times\biggl(\pi~\rm{sech}(\pi\nn){r_1}^{-1/2+2n-\ii\nn}-{r_2}^{-1/2+2n-\ii\nn}{}_2\mathcal{F}_1\biggl[\bgm1,\fr{1}{2}-2n+\ii\nn\\\fr{3}{2}-2n+\ii\nn\edm\bigg|-\FR{r_1}{r_2}\biggr]\biggr)\n\\
  =&~\mathbf{\Lambda}_{1}^{\nn}(r_1,r_2)-\mathbf{\Lambda}_{2}^{\nn}(r_1,r_2),
\end{align} 
where
\begin{align}\label{mathbfI_V_I}
  \mathbf{\Lambda}_{1}^{\nn}(r_1,r_2)\equiv&\sum_{n=0}^{\infty}-\FR{\ii\pi^22^{-2n+\ii\nn}{r_1}^{-1/2+2n-\ii\nn}}{\sinh(\pi\nn)\cosh(\pi\nn)}\Gamma\biggl[\bgm\fr{1}{2}+2n-\ii\nn\\1+n,1+n-\ii\nn\edm\biggr],\\\label{mathbfI_V_II}
  \mathbf{\Lambda}_{2}^{\nn}(r_1,r_2)\equiv&\sum_{n=0}^{\infty}-\FR{\ii\pi2^{-2n+\ii\nn}{r_2}^{-1/2+2n-\ii\nn}}{\sinh(\pi\nn)}\Gamma\biggl[\bgm\fr{1}{2}+2n-\ii\nn\\1+n,1+n-\ii\nn\edm\biggr]{}_2\mathcal{F}_1\biggl[\bgm1,\fr{1}{2}-2n+\ii\nn\\\fr{3}{2}-2n+\ii\nn\edm\bigg|-\FR{r_1}{r_2}\biggr].
\end{align}

Also, if we define
\begin{align}\label{mathbfI_V_III}
  \mathbf{\Lambda}_{3}^{\nn}(r_1,r_2)\equiv\sum_{n=0}^{\infty}-\FR{\ii\pi2^{-2n+\ii\nn}}{\sinh(\pi\nn)}\Gamma\biggl[\bgm\fr{1}{2}+2n-\ii\nn\\1+n,1+n-\ii\nn\edm\biggr]{}_2\mathcal{F}_1\biggl[\bgm1,\fr{1}{2}-2n+\ii\nn\\\fr{3}{2}-2n+\ii\nn\edm\bigg|-r_1\biggr],
\end{align}
then the second integral $\mb{I}_{\nn}^{(2)}$ can be expressed as:
\begin{align}\label{boldsymbolI_V_2=sum_mathbfI}
  \mb{I}_{\nn}^{(2)}(r_1,r_2)=\mathbf{\Lambda}_{2}^{\nn}(r_1,r_2)-\mathbf{\Lambda}_{3}^{\nn}(r_1,r_2).
\end{align}
Now, substituting (\ref{boldsymbolI_V_1=sum_mathbfI}) and (\ref{boldsymbolI_V_2=sum_mathbfI}) into (\ref{eq_Yr1r2dispInt}), we get
\begin{align}\label{mathcalY=sum_mathbfI_V_appd}
  \mathcal{Y}(r_1,r_2)=&\ \biggl[\Bigl(\FR{1-\ii\sinh(\pi\nn)}{2\pi^2}\cosh(\pi\nn){r_1}\Bigl[\mathbf{\Lambda}_{1}^{\nn}(r_1,r_2)-\mathbf{\Lambda}_{2}^{\nn}(r_1,r_2)\Bigr]+(\nn\to-\nn)\Bigr)r_2^{1/2-\ii\nn}\mathbf{F}_{\nn}^{-2}(r_2)\n\\
  &+\Bigl({r_1}\Bigl[\mathbf{\Lambda}_{2}^{\nn}(r_1,r_2)-\mathbf{\Lambda}_{3}^{\nn}(r_1,r_2)\Bigr]+(\nn\to-\nn)\Bigr)\FR{1-\ii\sinh(\pi\nn)}{2\pi^2}\cosh(\pi\nn)r_2^{1/2-\ii\nn}\mathbf{F}_{\nn}^{-2}(r_2)\biggr]\n\\
  &+(\nn\to-\nn).
\end{align}
\paragraph{Signal} One can show that the terms associated with $\mathbf{\Lambda}_{1}^{\pm\nn}$ give the signal part. In fact, the summation in (\ref{mathbfI_V_I}) can be done:
\begin{align}
  \mathbf{\Lambda}_{1}^{\nn}(r_1,r_2)&=\sum_{n=0}^{\infty}-\FR{\ii\pi^22^{-2n+\ii\nn}{r_1}^{-1/2+2n-\ii\nn}}{\sinh(\pi\nn)\cosh(\pi\nn)}\Gamma\biggl[\bgm\fr{1}{2}+2n-\ii\nn\\1+n,1+n-\ii\nn\edm\biggr]\n\\
  &=\pi~\rm{sech}(\pi\nn)r_1^{-1/2-\ii\nn}\mathbf{F}_{\nn}^{-2}(r_1),
\end{align}
As a result, the terms involving (\ref{mathbfI_V_I}) in (\ref{mathcalY=sum_mathbfI_V_appd}) gives:
\begin{align}
  &\mathcal{Y}_{\rm{S},>}(r_1,r_2)=\Bigl(\FR{1-\ii\sinh(\pi\nn)}{2\pi}r_1^{1/2-\ii\nn}\mathbf{F}_{\nn}^{-2}(r_1)+(\nn\to-\nn)\Bigr)\Bigl(r_2^{1/2-\ii\nn}\mathbf{F}_{\nn}^{-2}(r_2)+(\nn\to-\nn)\Bigr).
\end{align}
This is exactly the signal part of the 4-point tree seed integral. 

\paragraph{Background} It follows that all other terms besides the $\mathbf{\Lambda}_{1}^{\nn}$ terms give rise to the background:
\begin{align}\label{mathcalY_BG=sum_mathbfI_V_appd}
  \mathcal{Y}&_{\rm{BG},>}(r_1,r_2)\n\\
  =&\ \biggl[\Bigl(\FR{1-\ii\sinh(\pi\nn)}{2\pi^2}\cosh(\pi\nn){r_1}\Bigl[-\mathbf{\Lambda}_{2}^{\nn}(r_1,r_2)\Bigr]+(\nn\to-\nn)\Bigr)r_2^{1/2-\ii\nn}\mathbf{F}_{\nn}^{-2}(r_2)\n\\
  &+\Bigl({r_1}\Bigl[\mathbf{\Lambda}_{2}^{\nn}(r_1,r_2)-\mathbf{\Lambda}_{3}^{\nn}(r_1,r_2)\Bigr]+(\nn\to-\nn)\Bigr)\FR{1-\ii\sinh(\pi\nn)}{2\pi^2}\cosh(\pi\nn)r_2^{1/2-\ii\nn}\mathbf{F}_{\nn}^{-2}(r_2)\biggr]\n\\
  &+(\nn\to-\nn).
\end{align}
This expression can be further simplified thanks to several cancellations. First, it is easy to see that all terms including $\mathbf{\Lambda}_{2}^{\pm\nn}(r_1,r_2)\mathbf{F}^{-2}_{\pm\nn}(r_2)$ cancel out. Second, it can be shown that all terms involving $\mathbf{\Lambda}_{3}^{\pm\nn}(r_1,r_2)$ cancel out. As a result, the background of 4-point seed integral can be simplified into:
\begin{align}
  \mathcal{Y}_{\rm{BG},>}(r_1,r_2)=&\ \FR{\ii}{\pi^2}\sinh(\pi\nn)\cosh(\pi\nn)\times{r_1}\times r_2^{1/2+\ii\nn}\mathbf{\Lambda}_{2}^{\nn}(r_1,r_2)\mathbf{F}^{-2}_{-\nn}(r_2)+(\nn\to-\nn)\n\\
  =&\ \sum_{n=0}^{\infty}\biggl(\FR{\ii\coth(\pi\nn)}{\pi^{1/2}2^{1/2+2n-\ii\nn}}\times r_1\times r_2^{2n}\times\Gamma\biggl[\bgm\fr{1}{2}+2n-\ii\nn\\1+n,1+n-\ii\nn\edm\biggr]\n\\
  &\times{}_2\mathcal{F}_1\biggl[\bgm1,\fr{1}{2}-2n+\ii\nn\\\fr{3}{2}-2n+\ii\nn\edm\bigg|-\FR{r_1}{r_2}\biggr]{}_2\mathcal{F}_1\biggl[\bgm\fr{1}{4}+\fr{\ii\nn}{2},\fr{3}{4}+\fr{\ii\nn}{2}\\1+\ii\nn\edm\bigg|r_2^2\biggr]\biggr)+(\nn\to-\nn).
\end{align}
This is still not the expression found in previous works, but we have checked numerically that it agrees with known results, as mentioned in the main text. There are a large number of functional identities and resummation tricks with which one may prove the agreement analytically, but we shall not pursue this pure mathematical exercise in this work. 

Instead, in the rest of this appendix, we prove the cancellation of $\mathbf{\Lambda}_{3}^{\nn}$ terms. More precisely, we shall prove $\mathbf{\Lambda}_{3}^{\nn}(r_1,r_2)+\mathbf{\Lambda}_{3}^{-\nn}(r_1,r_2)=0$.  

To this end, we use the standard series representation for the dressed hypergeometric function in $\mathbf{\Lambda}_{3}^{\nn}$:
\begin{align}
  {}_2\mathcal{F}_1\biggl[\bgm a,b\\c\edm\bigg|x\biggr]=\sum_{m=0}^{\infty}\Gamma\biggl[\bgm a+m,b+m\\c+m,1+m\edm\biggr]x^m.
\end{align}
Then, the expression (\ref{mathbfI_V_III}) for $\mathbf{\Lambda}_{3}^{\nn}$ can be rewritten as: 
\begin{align}
  \mathbf{\Lambda}_{3}^{\nn}(r_1,r_2)=\sum_{n=0}^{\infty}\sum_{m=0}^{\infty}-\FR{\ii\pi2^{-2n+\ii\nn}}{\sinh(\pi\nn)}\Gamma\biggl[\bgm\fr{1}{2}+2n-\ii\nn\\1+n,1+n-\ii\nn\edm\biggr]\FR{(-r_1)^m}{\fr{1}{2}-2n+m+\ii\nn}.
\end{align}
In this expression, the sum over $n$ can be directly finished:
\begin{align}
  \mathbf{\Lambda}_{3}^{\nn}(r_1,r_2)=\sum_{m=0}^{\infty}-\FR{\ii\pi2^{1+\ii\nn}(-r_1)^m}{(1+2m+2\ii\nn)\sinh(\pi\nn)}\3F2\biggl[\bgm\fr{1}{4}-\fr{\ii\nn}{2},\fr{3}{4}-\fr{\ii\nn}{2},-\fr{1}{4}-\fr{m}{2}-\frac{\ii\nn}{2}\\\frac{3}{4}-\fr{m}{2}+\fr{\ii\nn}{2},1-\ii\nn\edm\bigg|1\biggr].
\end{align}
Then, we use the following identity of $\3F2$ (Eq.\;(4.3.4) of \cite{Slater:1966}):
\begin{align}\label{identity_3F2}
  \3F2\biggl[\bgm a,b,c\\d,e\edm\bigg|1\biggr]=&\ \Gamma\biggl[\bgm1-a,d,e,c-b\\e-b,d-b,1+b-a,c\edm\biggr]\3F2\biggl[\bgm b,1+b-d,1+b-e\\1+b-a,1+b-c\edm\bigg|1\biggr]\n\\
  &+\Gamma\biggl[\bgm1-a,d,e,b-c\\e-c,d-c,1+c-a,b\edm\biggr]\3F2\biggl[\bgm c,1+c-d,1+c-e\\1+c-a,1+c-b\edm\bigg|1\biggr].
\end{align}
Then $\mathbf{\Lambda}_{3}^{\nn}$ can be rewritten as:
\begin{align}
  \mathbf{\Lambda}_{3}^{\nn}(r_1,r_2)=&\ \sum_{m=0}^{\infty}\FR{\ii(-r_1)^m}{\sinh(\pi\nn)}\bigg\{2^{-7/2-m}\Bigl((-1)^{m+1}-1\Bigr)\Gamma\Bigl[1+m,-\fr{1}{4}-\fr{m}{2}+\fr{\ii\nn}{2},-\fr{1}{4}-\fr{m}{2}-\fr{\ii\nn}{2}\Bigr]\n\\
  &+\FR{2^{3/2}\pi^2}{(2+m)\cosh(\pi\nn)\Gamma\bigl[\fr{1}{4}+\fr{\ii\nn}{2},\fr{1}{4}-\fr{\ii\nn}{2}\bigr]}\3F2\biggl[\bgm1+\fr{m}{2},\fr{3}{4}+\fr{\ii\nn}{2},\fr{3}{4}-\fr{\ii\nn}{2}\\\fr{3}{2},2+\fr{m}{2}\edm\bigg|1\biggr]\bigg\}.
\end{align}
From this expression, it is easy to see $\mathbf{\Lambda}_{3}^{-\nn}(r_1,r_2)=-\mathbf{\Lambda}_{3}^{\nn}(r_1,r_2)$. 

\section{Line Dispersion Integral with PMB Representation}\label{appd_line}

In this appendix, we spell out the details of computing the three integrals (\ref{boldsymbolI_L_1})-(\ref{boldsymbolI_L_3}) arising from the line dispersion relation for the 4-point seed integral (\ref{dispersion_int_line}). The strategy is again the PMB representation. 

\paragraph{Computing $\mathbf{J}_{\nn}^{(1)}$} For the first integral $\mathbf{J}_{\nn}^{(1)}$, we take the MB representation for $\mb{D}_{\wt\nu}^{(1)}$ appeared in the integrand, whose expression is given in (\ref{mathcalF_1}).\footnote{A fine point is that the arguments of the two hypergeometric factors in (\ref{mathcalF_1}) are taken values from the lower edge of the branch cut, where our MB representation for $\mb{F}_{\wt\nu}^{-2}$ in (\ref{mathbbF}) is valid. On the contrary, if we want to evaluate the $\mb{F}_{\wt\nu}^{-2}$ on the upper edge of the branch cut, such as $\mb{F}_{\wt\nu}^{-2}(z^+)$ when $z>1$, we need to begin with $\mb{F}_{\wt\nu}^{-2}(z^-)$, and add back the discontinuity across the branch cut using (\ref{disc_2f1}).} Then we get:
\begin{align}
    \mb{J}_{\nn}^{(1)}(r_1,x)={-2\sinh(2\pi\nn)}\int_{-\infty}^{0}\di r\int_{-\ii\infty}^{\ii\infty}\FR{\di s_1}{2\pi\ii} \FR{\di s_2}{2\pi\ii}\FR{(-r)^{1-s_1-s_2-2\ii\nn}x^{s_2}}{r(r-r_1)}\mathbb{F}_{\nn}(s_1)\mathbb{F}_{\nn}(s_2),
\end{align}
where $\mathbb{F}_{\wt\nu}(s)$ is given in (\ref{mathbbF}). Then, the $r$ integral is again directly done, giving:
\begin{align}\label{boldsymbolI_L_1_PMB}
    \mb{J}_{\nn}^{(1)}(r_1,x)=-2\pi\sinh(2\pi\nn)\int_{-\ii\infty}^{\ii\infty}\FR{\di s_1}{2\pi\ii} \FR{\di s_2}{2\pi\ii}\FR{r_1^{-s_1-2\ii\nn}(r_1/x)^{-s_2}}{\sin\bigl(\pi(s_1+s_2+2\ii\nn)\bigr)}\mathbb{F}_{\nn}(s_1)\mathbb{F}_{\nn}(s_2).
\end{align}
When applying the residue theorem to compute this integral, we meet a subtlety here due to the ``mixed poles'' such as those from $1/\sin[\pi(s_1+s_2+2\ii\nn)]$, and we need to deal with these poles carefully. Below we spell out some details.

First, consider the $s_1$-integral. Since we want to obtain a result for physical $r_1\in (0,1)$, the factor $r_1^{-s_1}$ says that we should close the contour from the left side on the $s_1$ plane. There are two sets of left poles in $s_1$, respectively from $\Gamma(s_1/2)$ and $1/\sin[\pi(s_1+s_2+2\ii\wt\nu)]$:
\bge
\label{eq_s1poles1}
  \begin{cases}
    s_1=-2n_1,\\
    s_1=-s_2-n_1-2\ii\wt\nu.~~~~(n_1=0,1,2,\cdots)
  \end{cases}
\ede
After evaluating the Mellin integrand in (\ref{boldsymbolI_L_1_PMB}) on these two sets of poles respectively, we are left with an $s_2$ integral. The analysis of the $s_2$ integral depends on which sets of $s_1$-poles we take. 

Suppose we take the first set of poles $s_1=-2n_1$. Then, we are left with a factor of $(r_1/x)^{-s_2}=r_2^{-s_2}$. For physical $r_2\in(0,1)$, we should pick up left $s_2$-poles of the integrand. Examining the integrand in (\ref{boldsymbolI_L_1_PMB}), we see that there are two sets of left poles:\footnote{One may naively think that the second left poles $s_2=2n_1-n_2+2\ii\wt\nu$ with fixed $n_1$ and $n_2\in\mathbb{N}$ are not all ``left,'' in the sense that some of these poles have positive real part when $2n_1-n_2>0$. However, we emphasize that the criterion for a pole being left or right is not the sign of its real part, but rather the sign in front of the natural number $n_2$ that parameterize the set of poles. Therefore, in this case, all poles with a $-n_2$ term should be counted as left poles.}
\bge
\label{eq_s2pole1}
  \begin{cases}
  s_2=-2n_2,\\
  s_2=2n_1-n_2-2\ii\wt\nu.~~~~(n_2=0,1,2,\cdots)
  \end{cases}
\ede
However, some of ``poles'' from the second set (e.g., the second line in (\ref{eq_s2pole1})) coincide with zeros from the factor $1/\Gamma(1-s_2/2-\ii\wt\nu)$ in $\mathbb{F}_{\wt\nu}(s_2)$, which locate at  $s_2=2n+2-2\ii\wt\nu$ with $n=0,1,2,\cdots$. Thus, the poles in the second line of (\ref{eq_s2pole1}) clash with these zeros if $2n_1-n_2$ happens to be a positive even integer. As a result, among all poles in  (\ref{eq_s2pole1}), only the following ones make nonzero contributions to the final results:
\bge
  \begin{cases}
  s_2=-2n_2,~~~~&(n_2=0,1,2,\cdots)\\
  s_2=2n_1-(2n_2+1)-2\ii\wt\nu, &(n_2=0,1,2,\cdots)\\
  s_2=2n_1-2n_2-2\ii\wt\nu. &(n_2=n_1,n_1+1,n_1+2,\cdots)
  \end{cases}
\ede

So much for the poles involving $s_1=-2n_1$. Now let us return to (\ref{eq_s1poles1}) and consider the second set of $s_1$-poles, namely $s_1=-s_2-n_1-2\ii\wt\nu$. After evaluating the Mellin integrand (\ref{boldsymbolI_L_1_PMB}) at these poles, we get a factor $x^{s_2}$. Since we are considering the region $k_{12}>k_{34}$, namely $x=k_{34}/k_{12}<1$, the factor $x^{s_2}$ suggests that we should take the right $s_2$-poles. There is only one set of right poles from the factor $\Gamma[1/2-s_2-\ii\nn]$ in $\mathbb{F}_{\nn}(s_2)$: 
\begin{align}
    s_2=\fr{1}{2}+n_2-\ii\nn.~~~~(n_{1,2}=0,1,2,\cdots)
\end{align} 
Naively, one may expect that there is another set of right poles coming from the factor $\Gamma(s_1/2)=\Gamma[(-s_2-n_1-2\ii\nn)/2]$ in $\mathbb{F}_{\nn}(s_1)$, since we are now evaluating $s_1$ at $s_1=-s_2-n_1-2\ii\wt\nu$. However, this is overcounting since the pole from the $\Gamma(s_1/2)$ factor has already been included in (\ref{eq_s1poles1}). So, we should not include them again here.

In summary, to compute $\mb{J}_{\nn}^{(1)}(r_1,x)$ we need to pick up the following four sets of poles from the Mellin integrand in (\ref{boldsymbolI_L_1_PMB}):
\begin{align}\label{pole_nonlocal_1}
    \left\{\begin{aligned}
    &s_1=-2n_1,s_2=-2n_2,~~~~&&(n_{1},n_2\in\mathbb{N})\\
    &s_1=-2n_1,s_2=2n_1-(2n_2+1)-2\ii\nn,&&(n_{1},n_2\in\mathbb{N})\\
    &s_1=-2n_1,s_2=2n_1-2n_2-2\ii\nn,&&(n_{1}\in\mathbb{N};n_2-n_1\in \mathbb{N})\\
    &s_1=-s_2-n_1-2\ii\nn,s_2=\fr{1}{2}+n_2-\ii\nn. &&(n_{1},n_2\in\mathbb{N})
    \end{aligned}\right.
\end{align}
The contributions from these poles in order are:
{\allowdisplaybreaks
\begin{align}\label{mathbfUp_1}
  \mathbf{\Upsilon}_{1}^{\nn}(r_1,x)=&\ \sum_{n_1=0}^{\infty}\sum_{n_2=0}^{\infty}-\ii\times2^{1-2n_1-2n_2+2\ii\wt{\nu}}\pi^3 \rm{csch}(\pi\wt{\nu})^2r_1^{2n_1+2n_2-2\ii\wt{\nu}}x^{-2n_2}\n\\
  &\times\Gamma\biggl[\bgm\fr{1}{2}+2n_1-\ii\wt{\nu},\fr{1}{2}+2n_2-\ii\wt{\nu}\\1+n_1,1+n_2,1+n_1-\ii\wt{\nu},1+n_2-\ii\wt{\nu}\edm\biggr],\\
\label{mathbfUp_2}
  \mathbf{\Upsilon}_{2}^{\nn}(r_1,x)=&\ \sum_{n_1=0}^{\infty}\sum_{n_2=0}^{\infty}(-1)^{3/2+n_1+n_2+\ii\wt{\nu}}2^{-2n_2}\pi^2\coth(\pi\wt{\nu})r_1^{1+2n_2}x^{-1+2n_1-2n_2-2\ii\nn}\n\\
  &\times\Gamma\biggl[\bgm\fr{1}{2}+2n_1-\ii\wt{\nu},-\fr{1}{2}+n_1-n_2-\ii\wt{\nu},\fr{3}{2}-2n_1+2n_2+\ii\wt{\nu}\\1+n_1,\fr{3}{2}-n_1+n_2,1+n_1-\ii\wt{\nu}\edm\biggr],\\
\label{mathbfUp_3}
  \mathbf{\Upsilon}_{3}^{\nn}(r_1,x)=&\ \sum_{n_1=0}^{\infty}\sum_{n_2=n_1}^{\infty}(-1)^{-n_1+n_2}2^{1-2n_2}\pi^2e^{-\pi\wt{\nu}}\coth(\pi\nn)r_1^{2n_2}x^{2n_1-2n_2-2\ii\nn}\n\\
  &\times\Gamma\biggl[\bgm\fr{1}{2}+2n_1-\ii\wt{\nu},n_1-n_2-\ii\wt{\nu},\fr{1}{2}-2n_1+2n_2+\ii\wt{\nu}\\1+n_1,1-n_1+n_2,1+n_1-\ii\wt{\nu}\edm\biggr],\\
\label{mathbfUp_4}
  \mathbf{\Upsilon}_{4}^{\nn}(r_1,x)=&\ \sum_{n_1=0}^{\infty}\sum_{n_2=0}^{\infty}\ii^{2n_2-n_1}2^{-n_1}\pi^2e^{-\pi\wt{\nu}}\coth(\pi\wt{\nu})r_1^{n_1}x^{1/2+n_2-\ii\nn}\n\\
  &\times\Gamma\biggl[\bgm 1+n_1+n_2,-\fr{1}{4}-\fr{n_1}{2}-\fr{n_2}{2}-\fr{\ii\wt{\nu}}{2},\fr{1}{4}+\fr{n_2}{2}-\fr{\ii\wt{\nu}}{2}\\1+n_2,\fr{3}{4}-\fr{n_2}{2}-\fr{\ii\wt{\nu}}{2},\fr{5}{4}+\fr{n_1}{2}+\fr{n_2}{2}-\fr{\ii\wt{\nu}}{2}\edm\biggr].
\end{align}
}
Then we get the full result for the integral $\mb{J}_{\nn}^{(1)}$ as:
\begin{align}
  \mb{J}_{\nn}^{(1)}(r_1,x)=\mathbf{\Upsilon}_{1}^{\nn}(r_1,x)+\mathbf{\Upsilon}_{2}^{\nn}(r_1,x)+\mathbf{\Upsilon}_{3}^{\nn}(r_1,x)+\mathbf{\Upsilon}_{4}^{\nn}(r_1,x).
\end{align}

\paragraph{Computing $\mathbf{J}_{\nn}^{(2)}$}  For the second integral $\mb{J}_{\nn}^{(2)}$ in (\ref{boldsymbolI_L_2}), we again take the MB representation of its numerator $\mb{D}_{\wt\nu}^{(2)}$ in (\ref{mathcalF_2}). Then, the integral $\mb{J}_{\nn}^{(2)}$ becomes:
\begin{align}
    \mb{J}_{\nn}^{(2)}(r_1,x)
    =&\int_{-\infty}^{-x}\di r\FR{(1-e^{-2\pi\nn})(-r)^{1-2\ii\nn}}{r(r-r_1)}\mathbf{G}_{\nn}(r/x)\mathbf{F}_{\nn}^{-2}(r)\n\\
   =&~(1-e^{-2\pi\nn})\int_{-\infty}^{-x}\di r\int_{-\ii\infty}^{\ii\infty}\FR{\di s_1}{2\pi\ii} \FR{\di s_2}{2\pi\ii}\FR{(-r)^{1-s_1-s_2-2\ii\nn}x^{s_1}}{r(r-r_1)}\mathbb{G}_{\nn}(s_1)\mathbb{F}_{\nn}(s_2),
\end{align}
where $\mathbb{G}_{\nn}(s_1)$ is the MB representation of $\mathbf{G}_{\nn}(r/x)$:
\begin{align}
    \mathbb{G}_{\nn}(s_1)=-2^{2s_1+\ii\nn}\coth(\pi\nn)\Gamma\Bigl[s_1,s_1+\ii\nn,\fr{1}{2}-2s_1-\ii\nn\Bigr].
\end{align}
After finishing the integral over $r$, we get:
\begin{align}\label{boldsymbolI_2_PMB}
    \mb{J}_{\nn}^{(2)}(r_1,x)=&\ (1-e^{-2\pi\nn})\int_{-\ii\infty}^{\ii\infty}\FR{\di s_1}{2\pi\ii} \FR{\di s_2}{2\pi\ii}x^{-s_2-2\ii\nn}\Gamma[s_1+s_2+2\ii\nn]\n\\
    &\times{}_2\wt{\rm{F}}_1\biggl[\bgm1,s_1+s_2+2\ii\nn\\1+s_1+s_2+2\ii\nn\edm\bigg|-\FR{r_1}{x}\biggr]\mathbb{G}_{\nn}(s_1)\mathbb{F}_{\nn}(s_2).
\end{align}
The analysis of poles are very similar to the previous case and here we only list the result. That is, there are two sets of poles contributing to the integral when all momentum ratios taking values from their physical region, together with the condition $k_{12}>k_{34}$:
\begin{align}\label{pole_nonlocal_2}
    \left\{\begin{aligned}
    &s_1=\fr{1}{2}+n_1-\ii\nn,s_2=-2n_2,\\
    &s_1=\fr{1}{2}+n_1-\ii\nn,s_2=-\fr{1}{2}-n_1-n_2-\ii\nn.~~~~(n_{1,2}=0,1,2,\cdots)
    \end{aligned}\right.
\end{align}
The contribution from the first set of poles is:
\begin{align}\label{mathbfUp_5}
  \mathbf{\Upsilon}_{5}^{\nn}(r_1,x)=&\ \sum_{n_1=0}^{\infty}\sum_{n_2=0}^{\infty}(-1)^{1/2+n_1}2^{{1}/{2}+n_1-2n_2+\ii\wt{\nu}}{\pi}e^{-\pi\nn}\coth(\pi\nn)\n\\
  &\times\Gamma\biggl[\bgm\fr{1}{2}+2n_2-\ii\nn,\fr{1}{4}+\fr{n_1}{2}-\fr{\ii\nn}{2},\fr{1}{4}+\fr{n_1}{2}+\fr{\ii\nn}{2}\\1+n_1,1+n_2,1+n_2-\ii\nn\edm\biggr]\n\\
  &\times{}_2\mathcal{F}_1\biggl[\bgm1,\frac{1}{2}+n_1-2n_2+\ii\nn\\\fr{3}{2}+n_1-2n_2+\ii\nn\edm\bigg|-\FR{r_1}{x}\biggr]x^{2n_2-2\ii\nn},
\end{align}
and the contribution from the second set of poles is:
\begin{align}\label{mathbfUp_6}
  \mathbf{\Upsilon}_{6}^{\nn}(r_1,x)=&\ \sum_{n_1=0}^{\infty}\sum_{n_2=0}^{\infty}{\ii^{{3}/{2}+3n_2+3n_1}}2^{-1-n_1}{\pi}e^{-\fr{3}{2}\pi\nn}\coth(\pi\nn)r_1^{n_1}x^{1/2+n_2-\ii\nn}\n\\
  &\times\Gamma\biggl[\bgm 1+n_1+n_2,\fr{1}{4}+\fr{n_2}{2}-\fr{\ii\nn}{2},\fr{1}{4}+\fr{n_2}{2}+\fr{\ii\nn}{2},-\fr{1}{4}-\fr{n_1}{2}-\fr{n_2}{2}-\fr{\ii\nn}{2}\\1+n_2,\fr{5}{4}+\fr{n_1}{2}+\fr{n_2}{2}-\fr{\ii\nn}{2}\edm\biggr].
\end{align}
Then, the second integral $\mb{J}_{\nn}^{(2)}$ can be expressed as:
\begin{align}
  \mb{J}_{\nn}^{(2)}(r_1,x)=\mathbf{\Upsilon}_{5}^{\nn}(r_1,x)+\mathbf{\Upsilon}_{6}^{\nn}(r_1,x).
\end{align}

\paragraph{Computing $\mathbf{J}_{\nn}^{(3)}$} Finally, we consider the third integral  $\mb{J}_{\nn}^{(3)}$ in (\ref{boldsymbolI_L_3}). We again take the MB representation of the numerator $\mb{D}_{\wt\nu}^{(3)}$ in (\ref{mathcalF_3}) and finish the $r$ integral, which gives:
\begin{align}\label{boldsymbolI_L_3_PMB}
    \mb{J}_{\nn}^{(3)}(r_1,x)=\ &(1-e^{-2\pi\nn})\int_{-\ii\infty}^{\ii\infty}\FR{\di s_1}{2\pi\ii}\int_{-\ii\infty}^{\ii\infty}\FR{\di s_2}{2\pi\ii}x^{s_1}\Gamma[s_1+s_2+2\ii\nn]\n\\
    &\times{}_2\wt{\rm{F}}_1\biggl[\bgm1,s_1+s_2+2\ii\nn\\1+s_1+s_2+2\ii\nn\edm\bigg|-r_1\biggr]\mathbb{G}_{\nn}(s_2)\Bigl(\mathbb{F}_{\nn}(s_1)-\mathbb{G}_{\nn}(s_1)\Bigr).
\end{align}
After closing the contour from the right plane of $s_1$ then from the left plane of $s_2$, we get three sets of poles contributing residues:
\begin{align}\label{pole_nonlocal_3}
    \left\{\begin{aligned}
    &s_1=\fr{1}{2}+n_1-\ii\nn,s_2=-2n_2,\\
    &s_1=\fr{1}{2}+n_1-\ii\nn,s_2=-2n_2-2\ii\nn,\\
    &s_1=\fr{1}{2}+n_1-\ii\nn,s_2=-\fr{1}{2}-n_1-n_2-\ii\nn.~~~~(n_{1,2}=0,1,2,\cdots)
    \end{aligned}\right.
\end{align}
The contributions from these poles are given respectively by:
{\allowdisplaybreaks
\begin{align}\label{mathbfUp_7}
  \mathbf{\Upsilon}_{7}^{\nn}(r_1,x)=&\ \sum_{n_1=0}^{\infty}\sum_{n_2=0}^{\infty}(-1)^{n_1}2^{-1/2-2n_1+n_2+\ii\nn}\coth(\pi\nn)\bigl(\ii e^{-\pi\nn}+(-1)^{1+n_2}\bigr)\n\\
  &\ \times\Gamma\biggl[\bgm\fr{1}{2}+2n_1-\ii\nn,-n_1+\ii\nn,\fr{1}{4}+\fr{n_2}{2}-\fr{\ii\nn}{2},\fr{1}{4}+\fr{n_2}{2}+\fr{\ii\nn}{2}\\1+n_1,1+n_2\edm\biggr]\n\\    
  &\ \times{}_2\mathcal{{F}}_1\biggl[\bgm1,\fr{1}{2}-2n_1+n_2+\ii\nn\\\fr{3}{2}-2n_1+n_2+\ii\nn\edm\bigg|-r_1\biggr]x^{1/2+n_2-\ii\nn},\\
  \label{mathbfUp_8}
  \mathbf{\Upsilon}_{8}^{\nn}(r_1,x)=&\ \sum_{n_1=0}^{\infty}\sum_{n_2=0}^{\infty}(-1)^{n_1}2^{-1/2-2n_1+n_2-\ii\nn}\coth(\pi\nn)\bigl(\ii e^{-\pi\nn}+(-1)^{1+n_2}\bigr)\n\\
  &\ \times\Gamma\biggl[\bgm\frac{1}{2}+2n_1+\ii\nn,-n_1-\ii\nn,\fr{1}{4}+\fr{n_2}{2}-\fr{\ii\nn}{2},\fr{1}{4}+\fr{n_2}{2}+\fr{\ii\nn}{2}\\1+n_1,1+n_2\edm\biggr]\n\\
  &\ \times{}_2\mathcal{{F}}_1\biggl[\bgm1,\frac{1}{2}-2n_1+n_2-\ii\nn\\\frac{3}{2}-2n_1+n_2-\ii\nn\edm\bigg|-r_1\biggr]x^{1/2+n_2-\ii\nn},\\
  \label{mathbfUp_9}
  \mathbf{\Upsilon}_{9}^{\nn}(r_1,x)=&\ \sum_{n_1=0}^{\infty}\sum_{n_2=0}^{\infty}2^{-2-n_1}\coth(\pi\nn)\bigl(\ii e^{-\pi\nn}+(-1)^{1+n_2}\bigr)(-r_1)^{n_1}x^{1/2+n_2-\ii\nn}\n\\
  &\ \times\Gamma\biggl[\bgm 1+n_1+n_2,-\fr{1}{4}-\fr{n_1}{2}-\fr{n_2}{2}-\fr{\ii\nn}{2},-\fr{1}{4}-\fr{n_1}{2}-\fr{n_2}{2}+\fr{\ii\nn}{2}\\1+n_2\edm\biggr]\n\\
  &\ \times\Gamma\Bigl[\fr{1}{4}+\fr{n_2}{2}-\fr{\ii\nn}{2},\fr{1}{4}+\fr{n_2}{2}+\fr{\ii\nn}{2}\Bigr].
\end{align}
}
Then, the result for the integral $\mb{J}_{\nn}^{(3)}$ is:
\begin{align}
  \mb{J}_{\nn}^{(3)}(r_1,x)=\mathbf{\Upsilon}_{7}^{\nn}(r_1,x)+\mathbf{\Upsilon}_{8}^{\nn}(r_1,x)+\mathbf{\Upsilon}_{9}^{\nn}(r_1,x).
\end{align}

\paragraph{Simplifying the result} 
Now we have finished the computation of the three integrals $\mb{J}_{\wt\nu}^{(j)}$ $(j=1,2,3)$ in (\ref{boldsymbolI_L_1})-(\ref{boldsymbolI_L_3}). According to (\ref{mathcalZ=sum_of_boldsymbolI_L}), the final result of the seed integral is the sum of nine series $\mb\Upsilon_{\ell}^{\wt\nu}$ with $\ell=1,\cdots,9$. By looking at the dependence on various momentum ratios, it is straightforward to observe the following patterns: 
\begin{align}
  &\{\mb\Upsilon_1^{\wt\nu}\}\subset\text{nonlocal signal};\\
  &\{\mb\Upsilon_\ell^{\wt\nu};\ell=2,3,5\}\subset\text{local signal};\\
  &\{\mb\Upsilon_\ell^{\wt\nu};\ell=4,6,7,8,9\}\subset\text{background}.
\end{align}
Below, we will simply these 9 series according the above grouping. 

\paragraph{Nonlocal signal} The nonlocal signal, which only comes from $\mathbf{\Upsilon}^{\pm\nn}_{1}$, can be directly obtained by finishing the double sum in (\ref{mathbfUp_1}), and the result is:
\begin{align}
  \mathbf{\Upsilon}_{1}^{\nn}(r_1,x)=2\ii\pi\ r_1^{-2\ii\nn}\ \mathbf{F}_{\nn}^{-2}(r_1)\mathbf{F}_{\nn}^{-2}(r_1/x),
\end{align}
This gives exactly the nonlocal signal which is the starting point of the line dispersion integral:
\begin{align}
  \mathcal{Z}_{\rm{NS}}(r_1,x)=\FR{1-\ii\sinh(\pi\nn)}{2\pi}r_1^{1-2\ii\nn}x^{-1/2+\ii\nn}\mathbf{F}_{\nn}^{-2}(r_1)\mathbf{F}_{\nn}^{-2}(r_1/x)+(\nn\to-\nn).
\end{align}

\paragraph{Local signal} Local signal comes from $\mathbf\Upsilon_{2}^{\pm\nn}$, $\mathbf{\Upsilon}_{3}^{\pm\nn}$, and $\mathbf\Upsilon_{5}^{\pm\nn}$, whose explicit results are respectively (\ref{mathbfUp_2}), (\ref{mathbfUp_3}), and (\ref{mathbfUp_5}). For $\mathbf{\Upsilon}_{3}^{\nn}$, its double sum can be directly cpmputed, and the result is
\begin{align}
  \mathbf{\Upsilon}_{3}^{\nn}(r_1,x)=\FR{2\ii\pi}{1+\tanh(\pi\nn)}x^{-2\ii\nn}\mathbf{F}_{\nn}^{-2}(r_1)\mathbf{F}^{-2}_{-\nn}(r_1/x).
\end{align}
The simplification of $\mathbf{\Upsilon}_{5}^{\nn}$ is more complicated. We first expand the ${}_2\mathcal{F}_1$ factor in (\ref{mathbfUp_5}) and get
\begin{align}
  \mathbf{\Upsilon}_{5}^{\nn}(r_1,x)=&\ \sum_{n_1=0}^{\infty}\sum_{n_2=0}^{\infty}\sum_{n_3=0}^{\infty}\FR{(-1)^{1/2+n_1+n_3}2^{{1}/{2}+n_1-2n_2+\ii\wt{\nu}}{\pi}}{\fr{1}{2}+n_1-2n_2+n_3+\ii\nn}e^{-\pi\nn}\coth(\pi\nn)r_1^{n_3}x^{2n_2-n_3-2\ii\nn}\n\\
  &\times\Gamma\biggl[\bgm\fr{1}{2}+2n_2-\ii\nn,\fr{1}{4}+\fr{n_1}{2}-\fr{\ii\nn}{2},\fr{1}{4}+\fr{n_1}{2}+\fr{\ii\nn}{2}\\1+n_1,1+n_2,1+n_2-\ii\nn\edm\biggr].
\end{align}
Then the sum over $n_1$ can be finished:
\begin{align}
  \mathbf{\Upsilon}_{5}^{\nn}(r_1,x)=&\ \ii2^{-1/2-2n_2+\ii\nn}\pi e^{-\pi\nn}\coth(\pi\nn)(-r_1)^{n_3}x^{2n_2-n_3-2\ii\nn}\Gamma\biggl[\bgm\fr{1}{2}+2n_2-\ii\nn\\1+n_2,1+n_2-\ii\nn\edm\biggr]\n\\
  &\times\bigg\{{}_3\mathcal{F}_2\biggl[\bgm\fr{1}{4}-\fr{\ii\nn}{2},\fr{1}{4}+\fr{\ii\nn}{2},\fr{1}{4}-n_2+\fr{n_3}{2}+\fr{\ii\nn}{2}\\\fr{1}{2},\fr{5}{4}-n_2+\fr{n_3}{2}+\fr{\ii\nn}{2}\edm\bigg|1\biggr]\n\\
  &-2\times {}_3\mathcal{F}_2\biggl[\bgm\fr{3}{4}-\fr{\ii\nn}{2},\fr{3}{4}+\fr{\ii\nn}{2}\fr{3}{4}-n_2+\fr{n_3}{2}+\fr{\ii\nn}{2}\\\fr{3}{2},\fr{7}{4}-n_2+\fr{n_3}{2}+\fr{\ii\nn}{2}\edm\bigg|1\biggr]\bigg\}.
\end{align}
Thereafter, we use the formula (\ref{identity_3F2}) again and get
\begin{align}
  \mathbf{\Upsilon}_{5}^{\nn}(r_1,x)=&\ \sum_{n_2=0}^{\infty}\sum_{n_3=0}^{\infty}\ii\pi^32^{1-n_3}e^{-\pi\nn}\rm{csch}(\pi\nn)(-r_1)^{n_3}x^{2n_2-n_3-2\ii\nn}\n\\
  &\times\Gamma\biggl[\bgm\fr{1}{2}+2n_2-\ii\nn,\fr{1}{2}-2n_2+n_3+\ii\nn\\1+n_2,1+n_2-\ii\nn,1-n_2+\fr{n_3}{2},1-n_2+\fr{n_3}{2}+\ii\nn\edm\biggr].
\end{align}
The next key step is to devide this result into two parts, based on the parity of the summation index $n_3$. When $n_3$ is odd, we replace $n_3$ by $2n_3+1$ and get
\begin{align}\label{mathbfUp_5odd}
  \mathbf{\Upsilon}_{5,\rm{odd}}^{\nn}(r_1,x)=&\ \sum_{n_2=0}^{\infty}\sum_{n_3=0}^{\infty}-\ii\pi^3 4^{-n_3}e^{-\pi\nn}\rm{csch}(\pi\nn)r_1^{1+2n_3}x^{-1+2n_2-2n_3-2\ii\nn}\n\\
  &\times\Gamma\biggl[\bgm\fr{1}{2}+2n_2-\ii\nn,\fr{3}{2}-2n_2+2n_3+\ii\nn\\1+n_2,1+n_2-\ii\nn,\fr{3}{2}-n_2+n_3,\fr{3}{2}-n_2+n_3+\ii\nn\edm\biggr].
\end{align}
When $n_3$ is even, we replace $n_3$ by $2n_3$ and get
\begin{align}\label{mathbfUp_5even}
  \mathbf{\Upsilon}_{5,\rm{even}}^{\nn}(r_1,x)=&\ \sum_{n_2=0}^{\infty}\sum_{n_3=n_2}^{\infty}\ii\pi^3 2^{1-2n_3}e^{-\pi\nn}\rm{csch}(\pi\nn)r_1^{2n_3}x^{2n_2-2n_3-2\ii\nn}\n\\
  &\times\Gamma\biggl[\bgm\fr{1}{2}+n_2-\ii\nn,\fr{1}{2}-2n_2+2n_3+\ii\nn\\1+n_2,1+n_2-\ii\nn,1-n_2+n_3,1-n_2+n_3+\ii\nn\edm\biggr].
\end{align}
Note that the sum over $n_3$ is not from $0$ to $\infty$ as some terms vanish because of the factor $\Gamma[1-n_2+n_3]$ in the denominator. This explains our split of $\mathbf{\Upsilon}_{5}^{\nn}$ into two parts:
Comparing (\ref{mathbfUp_5odd}) with (\ref{mathbfUp_2}), we find
\begin{align}
  \mathbf{\Upsilon}_{2}^{\nn}(r_1,x)+\mathbf{\Upsilon}_{5,\rm{odd}}^{\nn}(r_1,x)=0.
\end{align}
Then we only need to compute $\mathbf{\Upsilon}_{5,\rm{even}}^{\nn}$, which can be directly done:
\begin{align}
  \mathbf{\Upsilon}_{5,\rm{even}}^{\nn}(r_1,x)=\FR{2\ii\pi}{1+\coth(\pi\nn)}x^{-2\ii\nn}\mathbf{F}_{\nn}^{-2}(r_1)\mathbf{F}^{-2}_{-\nn}(r_1/x).
\end{align}
Then we get the whole local signal:
\begin{align}
  \mathbf{\Upsilon}_{3}^{\nn}(r_1,x)+\mathbf{\Upsilon}_{5,\rm{even}}^{\nn}(r_1,x)=2\ii\pi\ x^{-2\ii\nn}\ \mathbf{F}_{\nn}^{-2}(r_1)\mathbf{F}^{-2}_{-\nn}(r_1/x).
\end{align}
Consequently, 
\begin{align}
  \mathcal{Z}_{\rm{LS}}(r_1,x)=\FR{1-\ii\sinh(\pi\nn)}{2\pi}r_1x^{-1/2-\ii\nn}\mathbf{F}_{\nn}^{-2}(r_1)\mathbf{F}^{-2}_{-\nn}(r_1/x)+(\nn\to-\nn).
\end{align}

\paragraph{Background} The background comes from 5 terms: $\mathbf{\Upsilon}_{4}^{\pm\nn}$, $\mathbf{\Upsilon}_{6}^{\pm\nn}$, $\mathbf{\Upsilon}_{7}^{\pm\nn}$, $\mathbf{\Upsilon}_{8}^{\pm\nn}$, and $\mathbf{\Upsilon}_{9}^{\pm\nn}$. First, for $\mathbf{\Upsilon}_{4}^{\pm\nn}$, one can directly finish the sum over $n_1$ in (\ref{mathbfUp_4}) and get:
\begin{align}
  \mathbf{\Upsilon}_{4}^{\nn}(r_1,x)=&\ \sum_{n=0}^{\infty}(-2)^n\pi^{5/2}e^{-\pi\nn}\coth(\pi\nn)x^{1/2+n-\ii\nn}\Gamma\biggl[\bgm\fr{1}{4}+\fr{n}{2}-\fr{\ii\nn}{2}\\1+n,\fr{3}{4}-\fr{n}{2}-\fr{\ii\nn}{2}\edm\biggr]\n\\
  &\times\bigg\{\ii r_1\sec\bigl[\fr{\pi}{4}(1+2n+2\ii\nn)\bigr]{}_3\mathcal{F}_2\biggl[\bgm1,1+\fr{n}{2},\fr{3}{2}+\fr{n}{2}\\\fr{7}{4}+\fr{n}{2}-\fr{\ii\nn}{2},\fr{7}{4}+\fr{n}{2}+\fr{\ii\nn}{2}\edm\bigg|r_1^2\biggr]\n\\
  &-\csc\bigl[\fr{\pi}{4}(1+2n+2\ii\nn)\bigr]{}_3\mathcal{F}_2\biggl[\bgm1,\fr{1}{2}+\fr{n}{2},1+\fr{n}{2}\\\fr{5}{4}+\fr{n}{2}-\fr{\ii\nn}{2},\fr{5}{4}+\fr{n}{2}+\fr{\ii\nn}{2}\edm\bigg|r_1^2\biggr]\bigg\}.
\end{align}
We can also deal with $\mathbf{\Upsilon}_{6}^{\nn}$ and $\mathbf{\Upsilon}_{9}^{\nn}$ in a similar way. Adding these three terms together, we get
\begin{align}\label{Up_4+6+9}
  &\mathbf{\Upsilon}_{4}^{\nn}(r_1,x)+\mathbf{\Upsilon}_{6}^{\nn}(r_1,x)+\mathbf{\Upsilon}_{9}^{\nn}(r_1,x)\n\\
  &=\FR{(-1)^{1+n}16\pi^2\rm{csch}(\pi\nn)}{(1+2n)^2+4\nn^2}x^{1/2+n-\ii\nn}\3F2\biggl[\bgm1,\fr{1}{2}+\fr{n}{2},1+\fr{n}{2}\\\fr{5}{4}+\fr{n}{2}-\fr{\ii\nn}{2},\fr{5}{4}+\fr{n}{2}+\fr{\ii\nn}{2}\edm\bigg|r_1^2\biggr].
\end{align}
For the rest two terms $\mathbf{\Upsilon}_{7}^{\nn}$ and $\mathbf{\Upsilon}_{8}^{\nn}$, we apply the same procedure used when simplifying $\mathbf{\Upsilon}_{5}^{\nn}$. Taking $\mathbf{\Upsilon}_{7}^{\nn}$ as an example, we expand the ${}_2\mathcal{F}_1$ factor in (\ref{mathbfUp_7}) and get
\begin{align}
  \mathbf{\Upsilon}_{7}^{\nn}(r_1,x)=&\ \sum_{n_1=0}^{\infty}\sum_{n_2=0}^{\infty}\sum_{n_3=0}^{\infty}\FR{(-1)^{1+n_1+n_3}2^{-1/2-2n_1+n_2+\ii\nn}}{\fr{1}{2}-2n_1+n_2+n_3+\ii\nn}\bigl((-1)^{n_2}-\ii e^{-\pi\nn}\bigr)\coth(\pi\nn)\n\\
  &\times r_1^{n_3}x^{1/2+n_2-\ii\nn}\Gamma\biggl[\bgm-n_1+\ii\nn,\fr{1}{2}+2n_1-\ii\nn\fr{1}{4}+\fr{n_2}{2}-\fr{\ii\nn}{2},\fr{1}{4}+\fr{n_2}{2}+\fr{\ii\nn}{2}\\1+n_1,1+n_2\edm\biggr].
\end{align}
Then the sum over $n_1$ can be finished:
\begin{align}
  &\mathbf{\Upsilon}_{7}^{\nn}(r_1,x)= \sum_{n_2=0}^{\infty}\sum_{n_3=0}^{\infty}2^{-2+n_2}\sqrt{\pi}\bigl(\ii(-1)^{1+n_2}-e^{-\pi\nn}\bigr)\coth(\pi\nn)\rm{csch}(\pi\nn)(-r_1)^{n_3}x^{1/2+n_2-\ii\nn}\n\\
  &\times\Gamma\biggl[\bgm\fr{1}{4}+\fr{n_2}{2}-\fr{\ii\nn}{2},\fr{1}{4}+\fr{n_2}{2}+\fr{\ii\nn}{2}\\1+n_2\edm\biggr]{}_3\mathcal{F}_2\biggl[\bgm\fr{1}{4}-\fr{\ii\nn}{2},\fr{3}{4}-\fr{\ii\nn}{2},-\fr{1}{4}-\fr{n_2}{2}-\fr{n_3}{2}-\fr{\ii\nn}{2}\\\fr{3}{4}-\fr{n_2}{2}-\fr{n_3}{2}-\fr{\ii\nn}{2},1-\ii\nn\edm\bigg|1\biggr].
\end{align}
One can simplify $\mathbf{\Upsilon}_{8}^{\nn}$ in the same way and get:
\begin{align}
  &\mathbf{\Upsilon}_{8}^{\nn}(r_1,x)= \sum_{n_2=0}^{\infty}\sum_{n_3=0}^{\infty}2^{-2+n_2}\sqrt{\pi}\bigl(e^{-\pi\nn}+\ii(-1)^{n_2}\bigr)\coth(\pi\nn)\rm{csch}(\pi\nn)(-r_1)^{n_3}x^{1/2+n_2-\ii\nn}\n\\
  &\times\Gamma\biggl[\bgm\fr{1}{4}+\fr{n_2}{2}-\fr{\ii\nn}{2},\fr{1}{4}+\fr{n_2}{2}+\fr{\ii\nn}{2}\\1+n_2\edm\biggr]{}_3\mathcal{F}_2\biggl[\bgm\fr{1}{4}+\fr{\ii\nn}{2},\fr{3}{4}+\fr{\ii\nn}{2},-\fr{1}{4}-\fr{n_2}{2}-\fr{n_3}{2}+\fr{\ii\nn}{2}\\\fr{3}{4}-\fr{n_2}{2}-\fr{n_3}{2}+\fr{\ii\nn}{2},1+\ii\nn\edm\bigg|1\biggr].
\end{align}
After that, we use the transformation of $\3F2 $ in (\ref{identity_3F2}) again, and find:
\begin{align}
  {}_3\mathcal{F}_2\biggl[\bgm\fr{1}{4}-\fr{\ii\nn}{2},\fr{3}{4}-\fr{\ii\nn}{2},-\fr{1}{4}-\fr{n_2}{2}-\fr{n_3}{2}-\fr{\ii\nn}{2}\\\fr{3}{4}-\fr{n_2}{2}-\fr{n_3}{2}-\fr{\ii\nn}{2},1-\ii\nn\edm\bigg|1\biggr]-(\nn\to-\nn)=0,
\end{align}
which leads to a nontrivial result:
\begin{align}
  \mathbf{\Upsilon}_{7}^{\nn}(r_1,x)+\mathbf{\Upsilon}_{8}^{\nn}(r_1,x)=0.
\end{align}
Therefore, the background is totally from the three series in (\ref{Up_4+6+9}), and therefore we get: 
\begin{align}
  \mathcal{Z}_{\rm{BG},>}(r_1,x)=\sum_{n=0}^{\infty}\frac{8(-x)^n r_1}{(1+2n)^2+4\nn^2}\3F2\biggl[\bgm1,\frac{1}{2}+\frac{n}{2},1+\frac{n}{2}\\\frac{5}{4}+\frac{n}{2}-\frac{\ii\nn}{2},\frac{5}{4}+\frac{n}{2}+\frac{\ii\nn}{2}\edm\bigg|r_1^2\biggr].
\end{align}
This completes our computation of line dispersion integral for the 4-point tree seed integral. 

\section{Dispersion Integral for a Minkowski One-Loop Correlator}
\label{app_Mink} 

One feature of the dispersive bootstrap is that the UV divergence in the ordinary computation of 1-loop correlators is totally absent. This may be unfamiliar to some readers, so we use a simple example to connect our dispersion method with the more familiar traditional calculation by dimensional regularization.

\begin{figure}[t]
    \centering
    \includegraphics[width=0.35\textwidth]{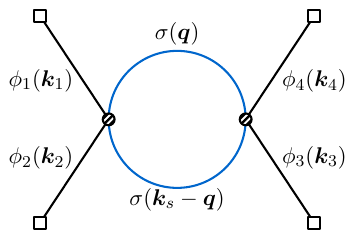}
    \caption{The 4-point correlator of four massive scalars $\phi_1,\cdots,\phi_4$ with 1-loop bubble exchange of a pair of massive scalar $\si$ in the $s$-channel.}
    \label{fd_4pt_1loop}
\end{figure}

Our example will be a 4-point 1-loop equal-time correlator of four scalar particles 
$\phi_i$ with masses $m_i$ ($i=1,2,3,4$), mediated by a pair of massive scalar $\si$ with mass $m$ running in a bubble loop, shown in Fig.\ \ref{fd_4pt_1loop}. We take the two vertices to be $\phi_1\phi_2\si^2$ and $\phi_3\phi_4\si^2$. Then, the diagram in Fig.\ \ref{fd_4pt_1loop} is computed by the following integral: 
\begin{align}
\label{eq_MinkBubble}
  \mathcal{G} 
  =&- \sum_{\aa,\bb=\pm}\aa\bb\int_{-\infty}^0\di t_1\di t_2\,D_{\aa}^{(m_1)}(k_1;\tau_1)D_{\aa}^{(m_2)}(k_2;\tau_1)D_{\bb}^{(m_3)}(k_3;\tau_2)D_{\bb}^{(m_4)}(k_4;\tau_2)\n\\
  &\times\int\FR{\di^d\bm q}{(2\pi)^d}D^{(m)}_{\aa\bb}(q;\tau_1,\tau_2)D^{(m)}_{\aa\bb}\Big(\big|\bm k_s-\bm q\big|;\tau_1,\tau_2\Big),
\end{align}
where $D_{\aa\bb}^{(m)}(k;\tau_1,\tau_2)$ is the bulk scalar propagator with mass $m$, and is given by: 
\begin{align}
  &D_{\pm\mp}^{(m)}(k;t_1,t_2)=\FR{e^{\pm\ii E(t_1-t_2)}}{2E}, \\
  &D_{\pm\pm}^{(m)}(k;t_1,t_2)=D_{\mp\pm}^{(m)}(k;t_1,t_2)\theta(\tau_1-\tau_2)+D_{\pm\mp}^{(m)}(k;t_1,t_2)\theta(\tau_2-\tau_1),
\end{align}
where $E\equiv\sqrt{k^2+m^2}$.

A computation of this diagram with dimensional regularization has been done in App.\;F of \cite{Xianyu:2022jwk}. Here we directly quote the result. By setting $d=3-\ep$ and let $\ep\to 0$, we have:
\begin{align}
\label{eq_GDR}
  &\mathcal{G}_\text{DR} 
  = \FR{1}{256\pi^2E_1E_2E_3E_4E_{1234}}\bigg[\FR{2}{\ep}-\ga_\text{E}+\log4\pi+2\n\\
  &+\FR{2}{E_{12}-E_{34}}\int_0^1\di\xi\,\bigg(E_{34}\log\FR{E_{12}+E_\text{min}}{\mu_R}-E_{12}\log\FR{E_{34}+E_\text{min}}{\mu_R}\bigg) \bigg]+\order{\ep},
\end{align}
where $\xi$ is a Feynman parameter, $E_\text{min}\equiv \sqrt{k_s^2+m^2/[\xi(1-\xi)]}$, $E_i=\sqrt{k_i^2+m_i^2}$, and $\mu_R$ is the renormalization scale. Note that the divergent term $\order{1/\ep}$ is proportional to $1/(E_1E_2E_3E_4E_{1234})$, and is what we would get by computing a contact diagram with $\phi_1\phi_2\phi_3\phi_4$ interaction. This is nothing but the local counterterm we should separate from the bare Lagrangian. The finite part of the counterterm is determined by a renormalization condition, and here we can choose the standard $\ob{\text{MS}}$ scheme, and remove the term proportional to $2/\ep-\gamma_\text{E}+\log4\pi$ in (\ref{eq_GDR}) altogether. Then, we get:
\begin{align}
   \mathcal{G}_{\ob{\text{MS}}} = &~\FR{1}{256\pi^2E_1E_2E_3E_4E_{1234}}\mathcal{I}(E_{12},E_{34},k_s),\n\\
   \mathcal{I}_{m}(E_{12},E_{34},k_s)\equiv&~ 
    2+\FR{2}{E_{12}-E_{34}}\int_0^1\di\xi\,\bigg(E_{34}\log\FR{E_{12}+E_\text{min}}{\mu_R}-E_{12}\log\FR{E_{34}+E_\text{min}}{\mu_R}\bigg).
\end{align}

To make things even simpler, we set the loop mass $m=0$, and so that the integral over $\xi$ can be done, leading to the following expression:
\begin{align}
  \mathcal{I}_{0}(E_{12},E_{34},k_s) 
  = 2+\FR{2}{E_{12}-E_{34}} \bigg(E_{34}\log\FR{E_{12}+k_s}{\mu_R}-E_{12}\log\FR{E_{34}+k_s}{\mu_R}\bigg) .
\end{align}

Now, let us consider the dispersion integral for $\mathcal{I}_{0}(E_{12},E_{34},k_s) 
$ on the complex-$E_{12}$ plane, with $E_{34}$ and $k_s$ fixed in the physical region. Clearly the only discontinuity comes from the first logarithmic factor:
\bge
\mathop{\text{Disc}}_{E_{12}}\mathcal{I}_{0}(E_{12})=\FR{4\pi \ii  E_{34}}{E_{12}-E_{34}},~~~~~E_{12}\in(-\infty,-k_s).
\ede
Then, we can use this discontinuity to form a dispersion integral. To determine the subtraction order, we note that the full correlator $\mathcal{I}_{0}$ approaches to a constant as $E_{12}\to\infty$:
\bge
  \lim_{E_{12}\to\infty}\mathcal{I}_{0}(E_{12})=2-2\log\FR{E_{34}+k_s}{\mu_R}.
\ede
Therefore our dispersion integral should have a first-order subtraction:
\begin{align}
  \mathcal{I}_0(E_{12})=\mathcal{I}_{0}(0)+\FR{E_{12}}{2\pi\ii}\int_{-\infty}^{-k_s}\di E\,\FR{\mathop{\text{Disc}}_{E}\mathcal{I}_{0}(E)}{E(E-E_{12})}.
\end{align}
This equality can be directly verified by finishing the integral. The lesson to be learned here is that the dispersion integral itself is independent of $\mu_R$ and is convergent. Similarly, had we started from the discontinuity of the regularized version (\ref{eq_GDR}) to do the dispersion integral, we will not get any term $\propto 1/\ep$. So, the dispersion method here is free of UV regularization procedure; On the other hand, the renormalization-scale dependence cannot be removed. In the dispersion calculation, this dependence is introduced by the subtraction point $\mathcal{I}_0(0)$. Clearly, our `` further modified'' $\ob{\text{MS}}$ subtraction corresponds to choosing $\mathcal{I}_0(0)=2 -2\log(k_s/\mu_R)$.

\end{appendix}

\newpage
\bibliography{CosmoCollider} 

\providecommand{\href}[2]{#2}\begingroup\raggedright\begin{thebibliography}{100}

\bibitem{Baumann:2022jpr}
D.~Baumann, D.~Green, A.~Joyce, E.~Pajer, G.~L. Pimentel, C.~Sleight, and
  M.~Taronna, ``{Snowmass White Paper: The Cosmological Bootstrap},'' in {\em
  {2022 Snowmass Summer Study}}.
\newblock 3, 2022.
\newblock \href{http://arxiv.org/abs/2203.08121}{{\ttfamily arXiv:2203.08121
  [hep-th]}}.

\bibitem{Maldacena:2011nz}
J.~M. Maldacena and G.~L. Pimentel, ``{On graviton non-Gaussianities during
  inflation},'' \href{http://dx.doi.org/10.1007/JHEP09(2011)045}{{\em JHEP}
  {\bfseries 09} (2011) 045}, \href{http://arxiv.org/abs/1104.2846}{{\ttfamily
  arXiv:1104.2846 [hep-th]}}.

\bibitem{Assassi:2012zq}
V.~Assassi, D.~Baumann, and D.~Green, ``{On Soft Limits of Inflationary
  Correlation Functions},''
  \href{http://dx.doi.org/10.1088/1475-7516/2012/11/047}{{\em JCAP} {\bfseries
  11} (2012) 047}, \href{http://arxiv.org/abs/1204.4207}{{\ttfamily
  arXiv:1204.4207 [hep-th]}}.

\bibitem{Arkani-Hamed:2017fdk}
N.~Arkani-Hamed, P.~Benincasa, and A.~Postnikov, ``{Cosmological Polytopes and
  the Wavefunction of the Universe},''
  \href{http://arxiv.org/abs/1709.02813}{{\ttfamily arXiv:1709.02813
  [hep-th]}}.

\bibitem{Baumann:2017jvh}
D.~Baumann, G.~Goon, H.~Lee, and G.~L. Pimentel, ``{Partially Massless Fields
  During Inflation},'' \href{http://dx.doi.org/10.1007/JHEP04(2018)140}{{\em
  JHEP} {\bfseries 04} (2018) 140},
\href{http://arxiv.org/abs/1712.06624}{{\ttfamily arXiv:1712.06624 [hep-th]}}.

\bibitem{Arkani-Hamed:2018bjr}
N.~Arkani-Hamed and P.~Benincasa, ``{On the Emergence of Lorentz Invariance and
  Unitarity from the Scattering Facet of Cosmological Polytopes},''
  \href{http://arxiv.org/abs/1811.01125}{{\ttfamily arXiv:1811.01125
  [hep-th]}}.

\bibitem{Arkani-Hamed:2018kmz}
N.~Arkani-Hamed, D.~Baumann, H.~Lee, and G.~L. Pimentel, ``{The Cosmological
  Bootstrap: Inflationary Correlators from Symmetries and Singularities},''
  \href{http://dx.doi.org/10.1007/JHEP04(2020)105}{{\em JHEP} {\bfseries 04}
  (2020) 105}, \href{http://arxiv.org/abs/1811.00024}{{\ttfamily
  arXiv:1811.00024 [hep-th]}}.

\bibitem{Baumann:2019oyu}
D.~Baumann, C.~Duaso~Pueyo, A.~Joyce, H.~Lee, and G.~L. Pimentel, ``{The
  cosmological bootstrap: weight-shifting operators and scalar seeds},''
  \href{http://dx.doi.org/10.1007/JHEP12(2020)204}{{\em JHEP} {\bfseries 12}
  (2020) 204}, \href{http://arxiv.org/abs/1910.14051}{{\ttfamily
  arXiv:1910.14051 [hep-th]}}.

\bibitem{Baumann:2020dch}
D.~Baumann, C.~Duaso~Pueyo, A.~Joyce, H.~Lee, and G.~L. Pimentel, ``{The
  Cosmological Bootstrap: Spinning Correlators from Symmetries and
  Factorization},'' \href{http://dx.doi.org/10.21468/SciPostPhys.11.3.071}{{\em
  SciPost Phys.} {\bfseries 11} (2021) 071},
  \href{http://arxiv.org/abs/2005.04234}{{\ttfamily arXiv:2005.04234
  [hep-th]}}.

\bibitem{Sleight:2019mgd}
C.~Sleight, ``{A Mellin Space Approach to Cosmological Correlators},''
  \href{http://dx.doi.org/10.1007/JHEP01(2020)090}{{\em JHEP} {\bfseries 01}
  (2020) 090}, \href{http://arxiv.org/abs/1906.12302}{{\ttfamily
  arXiv:1906.12302 [hep-th]}}.

\bibitem{Sleight:2019hfp}
C.~Sleight and M.~Taronna, ``{Bootstrapping Inflationary Correlators in Mellin
  Space},'' \href{http://dx.doi.org/10.1007/JHEP02(2020)098}{{\em JHEP}
  {\bfseries 02} (2020) 098}, \href{http://arxiv.org/abs/1907.01143}{{\ttfamily
  arXiv:1907.01143 [hep-th]}}.

\bibitem{Sleight:2020obc}
C.~Sleight and M.~Taronna, ``{From AdS to dS exchanges: Spectral
  representation, Mellin amplitudes, and crossing},''
  \href{http://dx.doi.org/10.1103/PhysRevD.104.L081902}{{\em Phys. Rev. D}
  {\bfseries 104} no.~8, (2021) L081902},
  \href{http://arxiv.org/abs/2007.09993}{{\ttfamily arXiv:2007.09993
  [hep-th]}}.

\bibitem{Sleight:2021iix}
C.~Sleight and M.~Taronna, ``{On the consistency of (partially-)massless matter
  couplings in de Sitter space},''
  \href{http://dx.doi.org/10.1007/JHEP10(2021)156}{{\em JHEP} {\bfseries 10}
  (2021) 156}, \href{http://arxiv.org/abs/2106.00366}{{\ttfamily
  arXiv:2106.00366 [hep-th]}}.

\bibitem{Sleight:2021plv}
C.~Sleight and M.~Taronna, ``{From dS to AdS and back},''
  \href{http://dx.doi.org/10.1007/JHEP12(2021)074}{{\em JHEP} {\bfseries 12}
  (2021) 074}, \href{http://arxiv.org/abs/2109.02725}{{\ttfamily
  arXiv:2109.02725 [hep-th]}}.

\bibitem{Hillman:2019wgh}
A.~Hillman, ``{Symbol Recursion for the dS Wave Function},''
  \href{http://arxiv.org/abs/1912.09450}{{\ttfamily arXiv:1912.09450
  [hep-th]}}.

\bibitem{Pajer:2020wnj}
E.~Pajer, D.~Stefanyszyn, and J.~Supe\l{}, ``{The Boostless Bootstrap:
  Amplitudes without Lorentz boosts},''
  \href{http://dx.doi.org/10.1007/JHEP12(2020)198}{{\em JHEP} {\bfseries 12}
  (2020) 198}, \href{http://arxiv.org/abs/2007.00027}{{\ttfamily
  arXiv:2007.00027 [hep-th]}}. [Erratum: JHEP 04, 023 (2022)].

\bibitem{Pajer:2020wxk}
E.~Pajer, ``{Building a Boostless Bootstrap for the Bispectrum},''
  \href{http://dx.doi.org/10.1088/1475-7516/2021/01/023}{{\em JCAP} {\bfseries
  01} (2021) 023}, \href{http://arxiv.org/abs/2010.12818}{{\ttfamily
  arXiv:2010.12818 [hep-th]}}.

\bibitem{Goodhew:2020hob}
H.~Goodhew, S.~Jazayeri, and E.~Pajer, ``{The Cosmological Optical Theorem},''
  \href{http://dx.doi.org/10.1088/1475-7516/2021/04/021}{{\em JCAP} {\bfseries
  04} (2021) 021}, \href{http://arxiv.org/abs/2009.02898}{{\ttfamily
  arXiv:2009.02898 [hep-th]}}.

\bibitem{Jazayeri:2021fvk}
S.~Jazayeri, E.~Pajer, and D.~Stefanyszyn, ``{From locality and unitarity to
  cosmological correlators},''
  \href{http://dx.doi.org/10.1007/JHEP10(2021)065}{{\em JHEP} {\bfseries 10}
  (2021) 065}, \href{http://arxiv.org/abs/2103.08649}{{\ttfamily
  arXiv:2103.08649 [hep-th]}}.

\bibitem{Melville:2021lst}
S.~Melville and E.~Pajer, ``{Cosmological Cutting Rules},''
  \href{http://dx.doi.org/10.1007/JHEP05(2021)249}{{\em JHEP} {\bfseries 05}
  (2021) 249}, \href{http://arxiv.org/abs/2103.09832}{{\ttfamily
  arXiv:2103.09832 [hep-th]}}.

\bibitem{Goodhew:2021oqg}
H.~Goodhew, S.~Jazayeri, M.~H. Gordon~Lee, and E.~Pajer, ``{Cutting
  cosmological correlators},''
  \href{http://dx.doi.org/10.1088/1475-7516/2021/08/003}{{\em JCAP} {\bfseries
  08} (2021) 003}, \href{http://arxiv.org/abs/2104.06587}{{\ttfamily
  arXiv:2104.06587 [hep-th]}}.

\bibitem{Baumann:2021fxj}
D.~Baumann, W.-M. Chen, C.~Duaso~Pueyo, A.~Joyce, H.~Lee, and G.~L. Pimentel,
  ``{Linking the singularities of cosmological correlators},''
  \href{http://dx.doi.org/10.1007/JHEP09(2022)010}{{\em JHEP} {\bfseries 09}
  (2022) 010}, \href{http://arxiv.org/abs/2106.05294}{{\ttfamily
  arXiv:2106.05294 [hep-th]}}.

\bibitem{Gomez:2021qfd}
H.~Gomez, R.~L. Jusinskas, and A.~Lipstein, ``{Cosmological Scattering
  Equations},'' \href{http://dx.doi.org/10.1103/PhysRevLett.127.251604}{{\em
  Phys. Rev. Lett.} {\bfseries 127} no.~25, (2021) 251604},
  \href{http://arxiv.org/abs/2106.11903}{{\ttfamily arXiv:2106.11903
  [hep-th]}}.

\bibitem{Gomez:2021ujt}
H.~Gomez, R.~Lipinski~Jusinskas, and A.~Lipstein, ``{Cosmological scattering
  equations at tree-level and one-loop},''
  \href{http://dx.doi.org/10.1007/JHEP07(2022)004}{{\em JHEP} {\bfseries 07}
  (2022) 004}, \href{http://arxiv.org/abs/2112.12695}{{\ttfamily
  arXiv:2112.12695 [hep-th]}}.

\bibitem{Bonifacio:2021azc}
J.~Bonifacio, E.~Pajer, and D.-G. Wang, ``{From amplitudes to contact
  cosmological correlators},''
  \href{http://dx.doi.org/10.1007/JHEP10(2021)001}{{\em JHEP} {\bfseries 10}
  (2021) 001}, \href{http://arxiv.org/abs/2106.15468}{{\ttfamily
  arXiv:2106.15468 [hep-th]}}.

\bibitem{Meltzer:2021zin}
D.~Meltzer, ``{The inflationary wavefunction from analyticity and
  factorization},'' \href{http://dx.doi.org/10.1088/1475-7516/2021/12/018}{{\em
  JCAP} {\bfseries 12} no.~12, (2021) 018},
  \href{http://arxiv.org/abs/2107.10266}{{\ttfamily arXiv:2107.10266
  [hep-th]}}.

\bibitem{Hogervorst:2021uvp}
M.~Hogervorst, J.~a. Penedones, and K.~S. Vaziri, ``{Towards the
  non-perturbative cosmological bootstrap},''
  \href{http://dx.doi.org/10.1007/JHEP02(2023)162}{{\em JHEP} {\bfseries 02}
  (2023) 162}, \href{http://arxiv.org/abs/2107.13871}{{\ttfamily
  arXiv:2107.13871 [hep-th]}}.

\bibitem{DiPietro:2021sjt}
L.~Di~Pietro, V.~Gorbenko, and S.~Komatsu, ``{Analyticity and unitarity for
  cosmological correlators},''
  \href{http://dx.doi.org/10.1007/JHEP03(2022)023}{{\em JHEP} {\bfseries 03}
  (2022) 023}, \href{http://arxiv.org/abs/2108.01695}{{\ttfamily
  arXiv:2108.01695 [hep-th]}}.

\bibitem{Cabass:2021fnw}
G.~Cabass, E.~Pajer, D.~Stefanyszyn, and J.~Supe\l{}, ``{Bootstrapping large
  graviton non-Gaussianities},''
  \href{http://dx.doi.org/10.1007/JHEP05(2022)077}{{\em JHEP} {\bfseries 05}
  (2022) 077}, \href{http://arxiv.org/abs/2109.10189}{{\ttfamily
  arXiv:2109.10189 [hep-th]}}.

\bibitem{Wang:2021qez}
L.-T. Wang, Z.-Z. Xianyu, and Y.-M. Zhong, ``{Precision calculation of
  inflation correlators at one loop},''
  \href{http://dx.doi.org/10.1007/JHEP02(2022)085}{{\em JHEP} {\bfseries 02}
  (2022) 085}, \href{http://arxiv.org/abs/2109.14635}{{\ttfamily
  arXiv:2109.14635 [hep-ph]}}.

\bibitem{Premkumar:2021mlz}
A.~Premkumar, ``{Regulating loops in de Sitter spacetime},''
  \href{http://dx.doi.org/10.1103/PhysRevD.109.045003}{{\em Phys. Rev. D}
  {\bfseries 109} no.~4, (2024) 045003},
  \href{http://arxiv.org/abs/2110.12504}{{\ttfamily arXiv:2110.12504
  [hep-th]}}.

\bibitem{Hillman:2021bnk}
A.~Hillman and E.~Pajer, ``{A differential representation of cosmological
  wavefunctions},'' \href{http://dx.doi.org/10.1007/JHEP04(2022)012}{{\em JHEP}
  {\bfseries 04} (2022) 012}, \href{http://arxiv.org/abs/2112.01619}{{\ttfamily
  arXiv:2112.01619 [hep-th]}}.

\bibitem{Tong:2021wai}
X.~Tong, Y.~Wang, and Y.~Zhu, ``{Cutting rule for cosmological collider
  signals: a bulk evolution perspective},''
  \href{http://dx.doi.org/10.1007/JHEP03(2022)181}{{\em JHEP} {\bfseries 03}
  (2022) 181}, \href{http://arxiv.org/abs/2112.03448}{{\ttfamily
  arXiv:2112.03448 [hep-th]}}.

\bibitem{Heckelbacher:2022hbq}
T.~Heckelbacher, I.~Sachs, E.~Skvortsov, and P.~Vanhove, ``{Analytical
  evaluation of cosmological correlation functions},''
  \href{http://dx.doi.org/10.1007/JHEP08(2022)139}{{\em JHEP} {\bfseries 08}
  (2022) 139}, \href{http://arxiv.org/abs/2204.07217}{{\ttfamily
  arXiv:2204.07217 [hep-th]}}.

\bibitem{Pimentel:2022fsc}
G.~L. Pimentel and D.-G. Wang, ``{Boostless cosmological collider bootstrap},''
  \href{http://dx.doi.org/10.1007/JHEP10(2022)177}{{\em JHEP} {\bfseries 10}
  (2022) 177}, \href{http://arxiv.org/abs/2205.00013}{{\ttfamily
  arXiv:2205.00013 [hep-th]}}.

\bibitem{Wang:2022eop}
D.-G. Wang, G.~L. Pimentel, and A.~Ach\'ucarro, ``{Bootstrapping multi-field
  inflation: non-Gaussianities from light scalars revisited},''
  \href{http://dx.doi.org/10.1088/1475-7516/2023/05/043}{{\em JCAP} {\bfseries
  05} (2023) 043}, \href{http://arxiv.org/abs/2212.14035}{{\ttfamily
  arXiv:2212.14035 [astro-ph.CO]}}.

\bibitem{Qin:2022lva}
Z.~Qin and Z.-Z. Xianyu, ``{Phase information in cosmological collider
  signals},'' \href{http://dx.doi.org/10.1007/JHEP10(2022)192}{{\em JHEP}
  {\bfseries 10} (2022) 192}, \href{http://arxiv.org/abs/2205.01692}{{\ttfamily
  arXiv:2205.01692 [hep-th]}}.

\bibitem{Jazayeri:2022kjy}
S.~Jazayeri and S.~Renaux-Petel, ``{Cosmological bootstrap in slow motion},''
  \href{http://dx.doi.org/10.1007/JHEP12(2022)137}{{\em JHEP} {\bfseries 12}
  (2022) 137}, \href{http://arxiv.org/abs/2205.10340}{{\ttfamily
  arXiv:2205.10340 [hep-th]}}.

\bibitem{Qin:2022fbv}
Z.~Qin and Z.-Z. Xianyu, ``{Helical inflation correlators: partial
  Mellin-Barnes and bootstrap equations},''
  \href{http://dx.doi.org/10.1007/JHEP04(2023)059}{{\em JHEP} {\bfseries 04}
  (2023) 059}, \href{http://arxiv.org/abs/2208.13790}{{\ttfamily
  arXiv:2208.13790 [hep-th]}}.

\bibitem{Cabass:2022rhr}
G.~Cabass, S.~Jazayeri, E.~Pajer, and D.~Stefanyszyn, ``{Parity violation in
  the scalar trispectrum: no-go theorems and yes-go examples},''
  \href{http://dx.doi.org/10.1007/JHEP02(2023)021}{{\em JHEP} {\bfseries 02}
  (2023) 021}, \href{http://arxiv.org/abs/2210.02907}{{\ttfamily
  arXiv:2210.02907 [hep-th]}}.

\bibitem{Cabass:2022oap}
G.~Cabass, M.~M. Ivanov, and O.~H.~E. Philcox, ``{Colliders and ghosts:
  Constraining inflation with the parity-odd galaxy four-point function},''
  \href{http://dx.doi.org/10.1103/PhysRevD.107.023523}{{\em Phys. Rev. D}
  {\bfseries 107} no.~2, (2023) 023523},
  \href{http://arxiv.org/abs/2210.16320}{{\ttfamily arXiv:2210.16320
  [astro-ph.CO]}}.

\bibitem{Xianyu:2022jwk}
Z.-Z. Xianyu and H.~Zhang, ``{Bootstrapping one-loop inflation correlators with
  the spectral decomposition},''
  \href{http://dx.doi.org/10.1007/JHEP04(2023)103}{{\em JHEP} {\bfseries 04}
  (2023) 103}, \href{http://arxiv.org/abs/2211.03810}{{\ttfamily
  arXiv:2211.03810 [hep-th]}}.

\bibitem{Bonifacio:2022vwa}
J.~Bonifacio, H.~Goodhew, A.~Joyce, E.~Pajer, and D.~Stefanyszyn, ``{The
  graviton four-point function in de Sitter space},''
  \href{http://dx.doi.org/10.1007/JHEP06(2023)212}{{\em JHEP} {\bfseries 06}
  (2023) 212}, \href{http://arxiv.org/abs/2212.07370}{{\ttfamily
  arXiv:2212.07370 [hep-th]}}.

\bibitem{Salcedo:2022aal}
S.~A. Salcedo, M.~H.~G. Lee, S.~Melville, and E.~Pajer, ``{The Analytic
  Wavefunction},'' \href{http://dx.doi.org/10.1007/JHEP06(2023)020}{{\em JHEP}
  {\bfseries 06} (2023) 020}, \href{http://arxiv.org/abs/2212.08009}{{\ttfamily
  arXiv:2212.08009 [hep-th]}}.

\bibitem{Lee:2022fgr}
H.~Lee and X.~Wang, ``{Cosmological double-copy relations},''
  \href{http://dx.doi.org/10.1103/PhysRevD.108.L061702}{{\em Phys. Rev. D}
  {\bfseries 108} no.~6, (2023) L061702},
  \href{http://arxiv.org/abs/2212.11282}{{\ttfamily arXiv:2212.11282
  [hep-th]}}.

\bibitem{Qin:2023ejc}
Z.~Qin and Z.-Z. Xianyu, ``{Closed-form formulae for inflation correlators},''
  \href{http://dx.doi.org/10.1007/JHEP07(2023)001}{{\em JHEP} {\bfseries 07}
  (2023) 001}, \href{http://arxiv.org/abs/2301.07047}{{\ttfamily
  arXiv:2301.07047 [hep-th]}}.

\bibitem{Werth:2023pfl}
D.~Werth, L.~Pinol, and S.~Renaux-Petel, ``{Cosmological Flow of Primordial
  Correlators},'' \href{http://arxiv.org/abs/2302.00655}{{\ttfamily
  arXiv:2302.00655 [hep-th]}}.

\bibitem{Pinol:2023oux}
L.~Pinol, S.~Renaux-Petel, and D.~Werth, ``{The Cosmological Flow: A Systematic
  Approach to Primordial Correlators},''
  \href{http://arxiv.org/abs/2312.06559}{{\ttfamily arXiv:2312.06559
  [astro-ph.CO]}}.

\bibitem{Qin:2023bjk}
Z.~Qin and Z.-Z. Xianyu, ``{Inflation correlators at the one-loop order:
  nonanalyticity, factorization, cutting rule, and OPE},''
  \href{http://dx.doi.org/10.1007/JHEP09(2023)116}{{\em JHEP} {\bfseries 09}
  (2023) 116}, \href{http://arxiv.org/abs/2304.13295}{{\ttfamily
  arXiv:2304.13295 [hep-th]}}.

\bibitem{Qin:2023nhv}
Z.~Qin and Z.-Z. Xianyu, ``{Nonanalyticity and on-shell factorization of
  inflation correlators at all loop orders},''
  \href{http://dx.doi.org/10.1007/JHEP01(2024)168}{{\em JHEP} {\bfseries 01}
  (2024) 168}, \href{http://arxiv.org/abs/2308.14802}{{\ttfamily
  arXiv:2308.14802 [hep-th]}}.

\bibitem{Lee:2023jby}
M.~H.~G. Lee, C.~McCulloch, and E.~Pajer, ``{Leading loops in cosmological
  correlators},'' \href{http://dx.doi.org/10.1007/JHEP11(2023)038}{{\em JHEP}
  {\bfseries 11} (2023) 038}, \href{http://arxiv.org/abs/2305.11228}{{\ttfamily
  arXiv:2305.11228 [hep-th]}}.

\bibitem{Loparco:2023rug}
M.~Loparco, J.~Penedones, K.~Salehi~Vaziri, and Z.~Sun, ``{The
  K\"all\'en-Lehmann representation in de Sitter spacetime},''
  \href{http://dx.doi.org/10.1007/JHEP12(2023)159}{{\em JHEP} {\bfseries 12}
  (2023) 159}, \href{http://arxiv.org/abs/2306.00090}{{\ttfamily
  arXiv:2306.00090 [hep-th]}}.

\bibitem{AguiSalcedo:2023nds}
S.~Agui~Salcedo and S.~Melville, ``{The cosmological tree theorem},''
  \href{http://dx.doi.org/10.1007/JHEP12(2023)076}{{\em JHEP} {\bfseries 12}
  (2023) 076}, \href{http://arxiv.org/abs/2308.00680}{{\ttfamily
  arXiv:2308.00680 [hep-th]}}.

\bibitem{De:2023xue}
S.~De and A.~Pokraka, ``{Cosmology meets cohomology},''
  \href{http://dx.doi.org/10.1007/JHEP03(2024)156}{{\em JHEP} {\bfseries 03}
  (2024) 156}, \href{http://arxiv.org/abs/2308.03753}{{\ttfamily
  arXiv:2308.03753 [hep-th]}}.

\bibitem{Stefanyszyn:2023qov}
D.~Stefanyszyn, X.~Tong, and Y.~Zhu, ``{Cosmological Correlators Through the
  Looking Glass: Reality, Parity, and Factorisation},''
  \href{http://arxiv.org/abs/2309.07769}{{\ttfamily arXiv:2309.07769
  [hep-th]}}.

\bibitem{Xianyu:2023ytd}
Z.-Z. Xianyu and J.~Zang, ``{Inflation correlators with multiple massive
  exchanges},'' \href{http://dx.doi.org/10.1007/JHEP03(2024)070}{{\em JHEP}
  {\bfseries 03} (2024) 070}, \href{http://arxiv.org/abs/2309.10849}{{\ttfamily
  arXiv:2309.10849 [hep-th]}}.

\bibitem{Green:2023ids}
D.~Green, Y.~Huang, C.-H. Shen, and D.~Baumann, ``{Positivity from Cosmological
  Correlators},'' \href{http://dx.doi.org/10.1007/JHEP04(2024)034}{{\em JHEP}
  {\bfseries 04} (2024) 034}, \href{http://arxiv.org/abs/2310.02490}{{\ttfamily
  arXiv:2310.02490 [hep-th]}}.

\bibitem{DuasoPueyo:2023kyh}
C.~Duaso~Pueyo and E.~Pajer, ``{A cosmological bootstrap for resonant
  non-Gaussianity},'' \href{http://dx.doi.org/10.1007/JHEP03(2024)098}{{\em
  JHEP} {\bfseries 03} (2024) 098},
  \href{http://arxiv.org/abs/2311.01395}{{\ttfamily arXiv:2311.01395
  [hep-th]}}.

\bibitem{Arkani-Hamed:2023bsv}
N.~Arkani-Hamed, D.~Baumann, A.~Hillman, A.~Joyce, H.~Lee, and G.~L. Pimentel,
  ``{Kinematic Flow and the Emergence of Time},''
  \href{http://arxiv.org/abs/2312.05300}{{\ttfamily arXiv:2312.05300
  [hep-th]}}.

\bibitem{Arkani-Hamed:2023kig}
N.~Arkani-Hamed, D.~Baumann, A.~Hillman, A.~Joyce, H.~Lee, and G.~L. Pimentel,
  ``{Differential Equations for Cosmological Correlators},''
  \href{http://arxiv.org/abs/2312.05303}{{\ttfamily arXiv:2312.05303
  [hep-th]}}.

\bibitem{Chen:2023iix}
J.~Chen and B.~Feng, ``{Towards Systematic Evaluation of de Sitter Correlators
  via Generalized Integration-By-Parts Relations},''
  \href{http://arxiv.org/abs/2401.00129}{{\ttfamily arXiv:2401.00129
  [hep-th]}}.

\bibitem{Benincasa:2024leu}
P.~Benincasa and G.~Dian, ``{The Geometry of Cosmological Correlators},''
  \href{http://arxiv.org/abs/2401.05207}{{\ttfamily arXiv:2401.05207
  [hep-th]}}.

\bibitem{Benincasa:2024lxe}
P.~Benincasa and F.~Vaz\~ao, ``{The Asymptotic Structure of Cosmological
  Integrals},'' \href{http://arxiv.org/abs/2402.06558}{{\ttfamily
  arXiv:2402.06558 [hep-th]}}.

\bibitem{Werth:2024aui}
D.~Werth, L.~Pinol, and S.~Renaux-Petel, ``{CosmoFlow: Python Package for
  Cosmological Correlators},''
  \href{http://arxiv.org/abs/2402.03693}{{\ttfamily arXiv:2402.03693
  [astro-ph.CO]}}.

\bibitem{Donath:2024utn}
Y.~Donath and E.~Pajer, ``{The In-Out Formalism for In-In Correlators},''
  \href{http://arxiv.org/abs/2402.05999}{{\ttfamily arXiv:2402.05999
  [hep-th]}}.

\bibitem{Du:2024hol}
Z.-Z. Du and D.~Stefanyszyn, ``{Soft theorems for boostless amplitudes},''
  \href{http://dx.doi.org/10.1007/JHEP07(2024)011}{{\em JHEP} {\bfseries 07}
  (2024) 011}, \href{http://arxiv.org/abs/2403.05459}{{\ttfamily
  arXiv:2403.05459 [hep-th]}}.

\bibitem{Fan:2024iek}
B.~Fan and Z.-Z. Xianyu, ``{Cosmological Amplitudes in Power-Law FRW
  Universe},'' \href{http://arxiv.org/abs/2403.07050}{{\ttfamily
  arXiv:2403.07050 [hep-th]}}.

\bibitem{Grimm:2024mbw}
T.~W. Grimm, A.~Hoefnagels, and M.~van Vliet, ``{Structure and Complexity of
  Cosmological Correlators},''
  \href{http://arxiv.org/abs/2404.03716}{{\ttfamily arXiv:2404.03716
  [hep-th]}}.

\bibitem{Melville:2024ove}
S.~Melville and G.~L. Pimentel, ``{A de Sitter S-matrix from amputated
  cosmological correlators},''
  \href{http://arxiv.org/abs/2404.05712}{{\ttfamily arXiv:2404.05712
  [hep-th]}}.

\bibitem{Cohen:2024anu}
T.~Cohen, D.~Green, and Y.~Huang, ``{Operator Origin of Anomalous Dimensions in
  de Sitter Space},'' \href{http://arxiv.org/abs/2407.08581}{{\ttfamily
  arXiv:2407.08581 [hep-th]}}.

\bibitem{Stefanyszyn:2024msm}
D.~Stefanyszyn, X.~Tong, and Y.~Zhu, ``{There and Back Again: Mapping and
  Factorising Cosmological Observables},''
  \href{http://arxiv.org/abs/2406.00099}{{\ttfamily arXiv:2406.00099
  [hep-th]}}.

\bibitem{Chen:2009we}
X.~Chen and Y.~Wang, ``{Large non-Gaussianities with Intermediate Shapes from
  Quasi-Single Field Inflation},''
  \href{http://dx.doi.org/10.1103/PhysRevD.81.063511}{{\em Phys. Rev. D}
  {\bfseries 81} (2010) 063511},
  \href{http://arxiv.org/abs/0909.0496}{{\ttfamily arXiv:0909.0496
  [astro-ph.CO]}}.

\bibitem{Chen:2009zp}
X.~Chen and Y.~Wang, ``{Quasi-Single Field Inflation and Non-Gaussianities},''
  \href{http://dx.doi.org/10.1088/1475-7516/2010/04/027}{{\em JCAP} {\bfseries
  1004} (2010) 027},
\href{http://arxiv.org/abs/0911.3380}{{\ttfamily arXiv:0911.3380 [hep-th]}}.

\bibitem{Baumann:2011nk}
D.~Baumann and D.~Green, ``{Signatures of Supersymmetry from the Early
  Universe},'' \href{http://dx.doi.org/10.1103/PhysRevD.85.103520}{{\em Phys.
  Rev.} {\bfseries D85} (2012) 103520},
\href{http://arxiv.org/abs/1109.0292}{{\ttfamily arXiv:1109.0292 [hep-th]}}.

\bibitem{Chen:2012ge}
X.~Chen and Y.~Wang, ``{Quasi-Single Field Inflation with Large Mass},''
  \href{http://dx.doi.org/10.1088/1475-7516/2012/09/021}{{\em JCAP} {\bfseries
  1209} (2012) 021},
\href{http://arxiv.org/abs/1205.0160}{{\ttfamily arXiv:1205.0160 [hep-th]}}.

\bibitem{Pi:2012gf}
S.~Pi and M.~Sasaki, ``{Curvature Perturbation Spectrum in Two-field Inflation
  with a Turning Trajectory},''
  \href{http://dx.doi.org/10.1088/1475-7516/2012/10/051}{{\em JCAP} {\bfseries
  10} (2012) 051}, \href{http://arxiv.org/abs/1205.0161}{{\ttfamily
  arXiv:1205.0161 [hep-th]}}.

\bibitem{Noumi:2012vr}
T.~Noumi, M.~Yamaguchi, and D.~Yokoyama, ``{Effective field theory approach to
  quasi-single field inflation and effects of heavy fields},''
  \href{http://dx.doi.org/10.1007/JHEP06(2013)051}{{\em JHEP} {\bfseries 06}
  (2013) 051}, \href{http://arxiv.org/abs/1211.1624}{{\ttfamily arXiv:1211.1624
  [hep-th]}}.

\bibitem{Gong:2013sma}
J.-O. Gong, S.~Pi, and M.~Sasaki, ``{Equilateral non-Gaussianity from heavy
  fields},'' \href{http://dx.doi.org/10.1088/1475-7516/2013/11/043}{{\em JCAP}
  {\bfseries 11} (2013) 043}, \href{http://arxiv.org/abs/1306.3691}{{\ttfamily
  arXiv:1306.3691 [hep-th]}}.

\bibitem{Arkani-Hamed:2015bza}
N.~Arkani-Hamed and J.~Maldacena, ``{Cosmological Collider Physics},''
\href{http://arxiv.org/abs/1503.08043}{{\ttfamily arXiv:1503.08043 [hep-th]}}.

\bibitem{Chen:2015lza}
X.~Chen, M.~H. Namjoo, and Y.~Wang, ``{Quantum Primordial Standard Clocks},''
  \href{http://dx.doi.org/10.1088/1475-7516/2016/02/013}{{\em JCAP} {\bfseries
  1602} no.~02, (2016) 013},
\href{http://arxiv.org/abs/1509.03930}{{\ttfamily arXiv:1509.03930
  [astro-ph.CO]}}.

\bibitem{Chen:2016nrs}
X.~Chen, Y.~Wang, and Z.-Z. Xianyu, ``{Loop Corrections to Standard Model
  Fields in Inflation},'' \href{http://dx.doi.org/10.1007/JHEP08(2016)051}{{\em
  JHEP} {\bfseries 08} (2016) 051},
\href{http://arxiv.org/abs/1604.07841}{{\ttfamily arXiv:1604.07841 [hep-th]}}.

\bibitem{Chen:2016uwp}
X.~Chen, Y.~Wang, and Z.-Z. Xianyu, ``{Standard Model Background of the
  Cosmological Collider},''
  \href{http://dx.doi.org/10.1103/PhysRevLett.118.261302}{{\em Phys. Rev.
  Lett.} {\bfseries 118} no.~26, (2017) 261302},
\href{http://arxiv.org/abs/1610.06597}{{\ttfamily arXiv:1610.06597 [hep-th]}}.

\bibitem{Chen:2016hrz}
X.~Chen, Y.~Wang, and Z.-Z. Xianyu, ``{Standard Model Mass Spectrum in
  Inflationary Universe},''
  \href{http://dx.doi.org/10.1007/JHEP04(2017)058}{{\em JHEP} {\bfseries 04}
  (2017) 058},
\href{http://arxiv.org/abs/1612.08122}{{\ttfamily arXiv:1612.08122 [hep-th]}}.

\bibitem{Lee:2016vti}
H.~Lee, D.~Baumann, and G.~L. Pimentel, ``{Non-Gaussianity as a Particle
  Detector},'' \href{http://dx.doi.org/10.1007/JHEP12(2016)040}{{\em JHEP}
  {\bfseries 12} (2016) 040},
\href{http://arxiv.org/abs/1607.03735}{{\ttfamily arXiv:1607.03735 [hep-th]}}.

\bibitem{Chen:2017ryl}
X.~Chen, Y.~Wang, and Z.-Z. Xianyu, ``{Schwinger-Keldysh Diagrammatics for
  Primordial Perturbations},''
  \href{http://dx.doi.org/10.1088/1475-7516/2017/12/006}{{\em JCAP} {\bfseries
  1712} no.~12, (2017) 006},
\href{http://arxiv.org/abs/1703.10166}{{\ttfamily arXiv:1703.10166 [hep-th]}}.

\bibitem{An:2017hlx}
H.~An, M.~McAneny, A.~K. Ridgway, and M.~B. Wise, ``{Quasi Single Field
  Inflation in the non-perturbative regime},''
  \href{http://dx.doi.org/10.1007/JHEP06(2018)105}{{\em JHEP} {\bfseries 06}
  (2018) 105},
\href{http://arxiv.org/abs/1706.09971}{{\ttfamily arXiv:1706.09971 [hep-ph]}}.

\bibitem{An:2017rwo}
H.~An, M.~McAneny, A.~K. Ridgway, and M.~B. Wise, ``{Non-Gaussian Enhancements
  of Galactic Halo Correlations in Quasi-Single Field Inflation},''
  \href{http://dx.doi.org/10.1103/PhysRevD.97.123528}{{\em Phys. Rev. D}
  {\bfseries 97} no.~12, (2018) 123528},
  \href{http://arxiv.org/abs/1711.02667}{{\ttfamily arXiv:1711.02667
  [hep-ph]}}.

\bibitem{Iyer:2017qzw}
A.~V. Iyer, S.~Pi, Y.~Wang, Z.~Wang, and S.~Zhou, ``{Strongly Coupled
  Quasi-Single Field Inflation},''
  \href{http://dx.doi.org/10.1088/1475-7516/2018/01/041}{{\em JCAP} {\bfseries
  1801} no.~01, (2018) 041},
\href{http://arxiv.org/abs/1710.03054}{{\ttfamily arXiv:1710.03054 [hep-th]}}.

\bibitem{Kumar:2017ecc}
S.~Kumar and R.~Sundrum, ``{Heavy-Lifting of Gauge Theories By Cosmic
  Inflation},'' \href{http://dx.doi.org/10.1007/JHEP05(2018)011}{{\em JHEP}
  {\bfseries 05} (2018) 011},
\href{http://arxiv.org/abs/1711.03988}{{\ttfamily arXiv:1711.03988 [hep-ph]}}.

\bibitem{Tong:2018tqf}
X.~Tong, Y.~Wang, and S.~Zhou, ``{Unsuppressed primordial standard clocks in
  warm quasi-single field inflation},''
  \href{http://dx.doi.org/10.1088/1475-7516/2018/06/013}{{\em JCAP} {\bfseries
  1806} no.~06, (2018) 013},
\href{http://arxiv.org/abs/1801.05688}{{\ttfamily arXiv:1801.05688 [hep-th]}}.

\bibitem{Chen:2018sce}
X.~Chen, W.~Z. Chua, Y.~Guo, Y.~Wang, Z.-Z. Xianyu, and T.~Xie, ``{Quantum
  Standard Clocks in the Primordial Trispectrum},''
  \href{http://dx.doi.org/10.1088/1475-7516/2018/05/049}{{\em JCAP} {\bfseries
  1805} no.~05, (2018) 049},
\href{http://arxiv.org/abs/1803.04412}{{\ttfamily arXiv:1803.04412 [hep-th]}}.

\bibitem{Saito:2018omt}
R.~Saito and T.~Kubota, ``{Heavy Particle Signatures in Cosmological
  Correlation Functions with Tensor Modes},''
  \href{http://dx.doi.org/10.1088/1475-7516/2018/06/009}{{\em JCAP} {\bfseries
  06} (2018) 009}, \href{http://arxiv.org/abs/1804.06974}{{\ttfamily
  arXiv:1804.06974 [hep-th]}}.

\bibitem{Chen:2018xck}
X.~Chen, Y.~Wang, and Z.-Z. Xianyu, ``{Neutrino Signatures in Primordial
  Non-Gaussianities},'' \href{http://dx.doi.org/10.1007/JHEP09(2018)022}{{\em
  JHEP} {\bfseries 09} (2018) 022},
\href{http://arxiv.org/abs/1805.02656}{{\ttfamily arXiv:1805.02656 [hep-ph]}}.

\bibitem{Chen:2018cgg}
X.~Chen, A.~Loeb, and Z.-Z. Xianyu, ``{Unique Fingerprints of Alternatives to
  Inflation in the Primordial Power Spectrum},''
  \href{http://dx.doi.org/10.1103/PhysRevLett.122.121301}{{\em Phys. Rev.
  Lett.} {\bfseries 122} no.~12, (2019) 121301},
  \href{http://arxiv.org/abs/1809.02603}{{\ttfamily arXiv:1809.02603
  [astro-ph.CO]}}.

\bibitem{Chua:2018dqh}
W.~Z. Chua, Q.~Ding, Y.~Wang, and S.~Zhou, ``{Imprints of Schwinger Effect on
  Primordial Spectra},'' \href{http://dx.doi.org/10.1007/JHEP04(2019)066}{{\em
  JHEP} {\bfseries 04} (2019) 066},
\href{http://arxiv.org/abs/1810.09815}{{\ttfamily arXiv:1810.09815 [hep-th]}}.

\bibitem{Kumar:2018jxz}
S.~Kumar and R.~Sundrum, ``{Seeing Higher-Dimensional Grand Unification In
  Primordial Non-Gaussianities},''
  \href{http://dx.doi.org/10.1007/JHEP04(2019)120}{{\em JHEP} {\bfseries 04}
  (2019) 120},
\href{http://arxiv.org/abs/1811.11200}{{\ttfamily arXiv:1811.11200 [hep-ph]}}.

\bibitem{Wu:2018lmx}
Y.-P. Wu, ``{Higgs as heavy-lifted physics during inflation},''
  \href{http://dx.doi.org/10.1007/JHEP04(2019)125}{{\em JHEP} {\bfseries 04}
  (2019) 125},
\href{http://arxiv.org/abs/1812.10654}{{\ttfamily arXiv:1812.10654 [hep-ph]}}.

\bibitem{Li:2019ves}
L.~Li, T.~Nakama, C.~M. Sou, Y.~Wang, and S.~Zhou, ``{Gravitational Production
  of Superheavy Dark Matter and Associated Cosmological Signatures},''
  \href{http://dx.doi.org/10.1007/JHEP07(2019)067}{{\em JHEP} {\bfseries 07}
  (2019) 067}, \href{http://arxiv.org/abs/1903.08842}{{\ttfamily
  arXiv:1903.08842 [astro-ph.CO]}}.

\bibitem{Alexander:2019vtb}
S.~Alexander, S.~J. Gates, L.~Jenks, K.~Koutrolikos, and E.~McDonough,
  ``{Higher Spin Supersymmetry at the Cosmological Collider: Sculpting SUSY
  Rilles in the CMB},'' \href{http://dx.doi.org/10.1007/JHEP10(2019)156}{{\em
  JHEP} {\bfseries 10} (2019) 156},
  \href{http://arxiv.org/abs/1907.05829}{{\ttfamily arXiv:1907.05829
  [hep-th]}}.

\bibitem{Lu:2019tjj}
S.~Lu, Y.~Wang, and Z.-Z. Xianyu, ``{A Cosmological Higgs Collider},''
  \href{http://dx.doi.org/10.1007/JHEP02(2020)011}{{\em JHEP} {\bfseries 02}
  (2020) 011}, \href{http://arxiv.org/abs/1907.07390}{{\ttfamily
  arXiv:1907.07390 [hep-th]}}.

\bibitem{Hook:2019zxa}
A.~Hook, J.~Huang, and D.~Racco, ``{Searches for other vacua. Part II. A new
  Higgstory at the cosmological collider},''
  \href{http://dx.doi.org/10.1007/JHEP01(2020)105}{{\em JHEP} {\bfseries 01}
  (2020) 105}, \href{http://arxiv.org/abs/1907.10624}{{\ttfamily
  arXiv:1907.10624 [hep-ph]}}.

\bibitem{Hook:2019vcn}
A.~Hook, J.~Huang, and D.~Racco, ``{Minimal signatures of the Standard Model in
  non-Gaussianities},''
  \href{http://dx.doi.org/10.1103/PhysRevD.101.023519}{{\em Phys. Rev. D}
  {\bfseries 101} no.~2, (2020) 023519},
  \href{http://arxiv.org/abs/1908.00019}{{\ttfamily arXiv:1908.00019
  [hep-ph]}}.

\bibitem{Kumar:2019ebj}
S.~Kumar and R.~Sundrum, ``{Cosmological Collider Physics and the Curvaton},''
  \href{http://dx.doi.org/10.1007/JHEP04(2020)077}{{\em JHEP} {\bfseries 04}
  (2020) 077}, \href{http://arxiv.org/abs/1908.11378}{{\ttfamily
  arXiv:1908.11378 [hep-ph]}}.

\bibitem{Liu:2019fag}
T.~Liu, X.~Tong, Y.~Wang, and Z.-Z. Xianyu, ``{Probing P and CP Violations on
  the Cosmological Collider},''
  \href{http://dx.doi.org/10.1007/JHEP04(2020)189}{{\em JHEP} {\bfseries 04}
  (2020) 189}, \href{http://arxiv.org/abs/1909.01819}{{\ttfamily
  arXiv:1909.01819 [hep-ph]}}.

\bibitem{Wang:2019gbi}
L.-T. Wang and Z.-Z. Xianyu, ``{In Search of Large Signals at the Cosmological
  Collider},'' \href{http://dx.doi.org/10.1007/JHEP02(2020)044}{{\em JHEP}
  {\bfseries 02} (2020) 044},
\href{http://arxiv.org/abs/1910.12876}{{\ttfamily arXiv:1910.12876 [hep-ph]}}.

\bibitem{Wang:2019gok}
D.-G. Wang, ``{On the inflationary massive field with a curved field
  manifold},'' \href{http://dx.doi.org/10.1088/1475-7516/2020/01/046}{{\em
  JCAP} {\bfseries 01} (2020) 046},
  \href{http://arxiv.org/abs/1911.04459}{{\ttfamily arXiv:1911.04459
  [astro-ph.CO]}}.

\bibitem{Wang:2020uic}
Y.~Wang and Y.~Zhu, ``{Cosmological Collider Signatures of Massive Vectors from
  Non-Gaussian Gravitational Waves},''
  \href{http://dx.doi.org/10.1088/1475-7516/2020/04/049}{{\em JCAP} {\bfseries
  04} (2020) 049}, \href{http://arxiv.org/abs/2001.03879}{{\ttfamily
  arXiv:2001.03879 [astro-ph.CO]}}.

\bibitem{Li:2020xwr}
L.~Li, S.~Lu, Y.~Wang, and S.~Zhou, ``{Cosmological Signatures of Superheavy
  Dark Matter},'' \href{http://dx.doi.org/10.1007/JHEP07(2020)231}{{\em JHEP}
  {\bfseries 07} (2020) 231}, \href{http://arxiv.org/abs/2002.01131}{{\ttfamily
  arXiv:2002.01131 [hep-ph]}}.

\bibitem{Wang:2020ioa}
L.-T. Wang and Z.-Z. Xianyu, ``{Gauge Boson Signals at the Cosmological
  Collider},'' \href{http://dx.doi.org/10.1007/JHEP11(2020)082}{{\em JHEP}
  {\bfseries 11} (2020) 082}, \href{http://arxiv.org/abs/2004.02887}{{\ttfamily
  arXiv:2004.02887 [hep-ph]}}.

\bibitem{Fan:2020xgh}
J.~Fan and Z.-Z. Xianyu, ``{A Cosmic Microscope for the Preheating Era},''
  \href{http://dx.doi.org/10.1007/JHEP01(2021)021}{{\em JHEP} {\bfseries 01}
  (2021) 021}, \href{http://arxiv.org/abs/2005.12278}{{\ttfamily
  arXiv:2005.12278 [hep-ph]}}.

\bibitem{Bodas:2020yho}
A.~Bodas, S.~Kumar, and R.~Sundrum, ``{The Scalar Chemical Potential in
  Cosmological Collider Physics},''
  \href{http://dx.doi.org/10.1007/JHEP02(2021)079}{{\em JHEP} {\bfseries 02}
  (2021) 079}, \href{http://arxiv.org/abs/2010.04727}{{\ttfamily
  arXiv:2010.04727 [hep-ph]}}.

\bibitem{Aoki:2020zbj}
S.~Aoki and M.~Yamaguchi, ``{Disentangling mass spectra of multiple fields in
  cosmological collider},''
  \href{http://dx.doi.org/10.1007/JHEP04(2021)127}{{\em JHEP} {\bfseries 04}
  (2021) 127}, \href{http://arxiv.org/abs/2012.13667}{{\ttfamily
  arXiv:2012.13667 [hep-th]}}.

\bibitem{Maru:2021ezc}
N.~Maru and A.~Okawa, ``{Non-Gaussianity from $X, Y$ gauge bosons in
  Cosmological Collider Physics},''
  \href{http://arxiv.org/abs/2101.10634}{{\ttfamily arXiv:2101.10634
  [hep-ph]}}.

\bibitem{Lu:2021gso}
S.~Lu, ``{Axion isocurvature collider},''
  \href{http://dx.doi.org/10.1007/JHEP04(2022)157}{{\em JHEP} {\bfseries 04}
  (2022) 157}, \href{http://arxiv.org/abs/2103.05958}{{\ttfamily
  arXiv:2103.05958 [hep-th]}}.

\bibitem{Sou:2021juh}
C.~M. Sou, X.~Tong, and Y.~Wang, ``{Chemical-potential-assisted particle
  production in FRW spacetimes},''
  \href{http://dx.doi.org/10.1007/JHEP06(2021)129}{{\em JHEP} {\bfseries 06}
  (2021) 129}, \href{http://arxiv.org/abs/2104.08772}{{\ttfamily
  arXiv:2104.08772 [hep-th]}}.

\bibitem{Lu:2021wxu}
Q.~Lu, M.~Reece, and Z.-Z. Xianyu, ``{Missing scalars at the cosmological
  collider},'' \href{http://dx.doi.org/10.1007/JHEP12(2021)098}{{\em JHEP}
  {\bfseries 12} (2021) 098}, \href{http://arxiv.org/abs/2108.11385}{{\ttfamily
  arXiv:2108.11385 [hep-ph]}}.

\bibitem{Pinol:2021aun}
L.~Pinol, S.~Aoki, S.~Renaux-Petel, and M.~Yamaguchi, ``{Inflationary flavor
  oscillations and the cosmic spectroscopy},''
  \href{http://dx.doi.org/10.1103/PhysRevD.107.L021301}{{\em Phys. Rev. D}
  {\bfseries 107} no.~2, (2023) L021301},
  \href{http://arxiv.org/abs/2112.05710}{{\ttfamily arXiv:2112.05710
  [hep-th]}}.

\bibitem{Cui:2021iie}
Y.~Cui and Z.-Z. Xianyu, ``{Probing Leptogenesis with the Cosmological
  Collider},'' \href{http://dx.doi.org/10.1103/PhysRevLett.129.111301}{{\em
  Phys. Rev. Lett.} {\bfseries 129} no.~11, (2022) 111301},
  \href{http://arxiv.org/abs/2112.10793}{{\ttfamily arXiv:2112.10793
  [hep-ph]}}.

\bibitem{Tong:2022cdz}
X.~Tong and Z.-Z. Xianyu, ``{Large spin-2 signals at the cosmological
  collider},'' \href{http://dx.doi.org/10.1007/JHEP10(2022)194}{{\em JHEP}
  {\bfseries 10} (2022) 194}, \href{http://arxiv.org/abs/2203.06349}{{\ttfamily
  arXiv:2203.06349 [hep-ph]}}.

\bibitem{Reece:2022soh}
M.~Reece, L.-T. Wang, and Z.-Z. Xianyu, ``{Large-field inflation and the
  cosmological collider},''
  \href{http://dx.doi.org/10.1103/PhysRevD.107.L101304}{{\em Phys. Rev. D}
  {\bfseries 107} no.~10, (2023) L101304},
  \href{http://arxiv.org/abs/2204.11869}{{\ttfamily arXiv:2204.11869
  [hep-ph]}}.

\bibitem{Chen:2022vzh}
X.~Chen, R.~Ebadi, and S.~Kumar, ``{Classical cosmological collider physics and
  primordial features},''
  \href{http://dx.doi.org/10.1088/1475-7516/2022/08/083}{{\em JCAP} {\bfseries
  08} (2022) 083}, \href{http://arxiv.org/abs/2205.01107}{{\ttfamily
  arXiv:2205.01107 [hep-ph]}}.

\bibitem{Niu:2022quw}
X.~Niu, M.~H. Rahat, K.~Srinivasan, and W.~Xue, ``{Gravitational wave probes of
  massive gauge bosons at the cosmological collider},''
  \href{http://dx.doi.org/10.1088/1475-7516/2023/02/013}{{\em JCAP} {\bfseries
  02} (2023) 013}, \href{http://arxiv.org/abs/2211.14331}{{\ttfamily
  arXiv:2211.14331 [hep-ph]}}.

\bibitem{Niu:2022fki}
X.~Niu, M.~H. Rahat, K.~Srinivasan, and W.~Xue, ``{Parity-odd and even
  trispectrum from axion inflation},''
  \href{http://dx.doi.org/10.1088/1475-7516/2023/05/018}{{\em JCAP} {\bfseries
  05} (2023) 018}, \href{http://arxiv.org/abs/2211.14324}{{\ttfamily
  arXiv:2211.14324 [hep-ph]}}.

\bibitem{Aoki:2023tjm}
S.~Aoki, ``{Continuous spectrum on cosmological collider},''
  \href{http://dx.doi.org/10.1088/1475-7516/2023/04/002}{{\em JCAP} {\bfseries
  04} (2023) 002}, \href{http://arxiv.org/abs/2301.07920}{{\ttfamily
  arXiv:2301.07920 [hep-th]}}.

\bibitem{Chen:2023txq}
X.~Chen, J.~Fan, and L.~Li, ``{New inflationary probes of axion dark matter},''
  \href{http://dx.doi.org/10.1007/JHEP12(2023)197}{{\em JHEP} {\bfseries 12}
  (2023) 197}, \href{http://arxiv.org/abs/2303.03406}{{\ttfamily
  arXiv:2303.03406 [hep-ph]}}.

\bibitem{Tong:2023krn}
X.~Tong, Y.~Wang, C.~Zhang, and Y.~Zhu, ``{BCS in the sky: signatures of
  inflationary fermion condensation},''
  \href{http://dx.doi.org/10.1088/1475-7516/2024/04/022}{{\em JCAP} {\bfseries
  04} (2024) 022}, \href{http://arxiv.org/abs/2304.09428}{{\ttfamily
  arXiv:2304.09428 [hep-th]}}.

\bibitem{Jazayeri:2023xcj}
S.~Jazayeri, S.~Renaux-Petel, and D.~Werth, ``{Shapes of the cosmological
  low-speed collider},''
  \href{http://dx.doi.org/10.1088/1475-7516/2023/12/035}{{\em JCAP} {\bfseries
  12} (2023) 035}, \href{http://arxiv.org/abs/2307.01751}{{\ttfamily
  arXiv:2307.01751 [hep-th]}}.

\bibitem{Jazayeri:2023kji}
S.~Jazayeri, S.~Renaux-Petel, X.~Tong, D.~Werth, and Y.~Zhu, ``{Parity
  violation from emergent nonlocality during inflation},''
  \href{http://dx.doi.org/10.1103/PhysRevD.108.123523}{{\em Phys. Rev. D}
  {\bfseries 108} no.~12, (2023) 123523},
  \href{http://arxiv.org/abs/2308.11315}{{\ttfamily arXiv:2308.11315
  [hep-th]}}.

\bibitem{Chakraborty:2023qbp}
P.~Chakraborty and J.~Stout, ``{Light scalars at the cosmological collider},''
  \href{http://dx.doi.org/10.1007/JHEP02(2024)021}{{\em JHEP} {\bfseries 02}
  (2024) 021}, \href{http://arxiv.org/abs/2310.01494}{{\ttfamily
  arXiv:2310.01494 [hep-th]}}.

\bibitem{Chakraborty:2023eoq}
P.~Chakraborty and J.~Stout, ``{Compact scalars at the cosmological
  collider},'' \href{http://dx.doi.org/10.1007/JHEP03(2024)149}{{\em JHEP}
  {\bfseries 03} (2024) 149}, \href{http://arxiv.org/abs/2311.09219}{{\ttfamily
  arXiv:2311.09219 [hep-th]}}.

\bibitem{Aoki:2023wdc}
S.~Aoki, T.~Noumi, F.~Sano, and M.~Yamaguchi, ``{Analytic formulae for
  inflationary correlators with dynamical mass},''
  \href{http://dx.doi.org/10.1007/JHEP03(2024)073}{{\em JHEP} {\bfseries 03}
  (2024) 073}, \href{http://arxiv.org/abs/2312.09642}{{\ttfamily
  arXiv:2312.09642 [hep-th]}}.

\bibitem{Craig:2024qgy}
N.~Craig, S.~Kumar, and A.~McCune, ``{An Effective Cosmological Collider},''
  \href{http://arxiv.org/abs/2401.10976}{{\ttfamily arXiv:2401.10976
  [hep-ph]}}.

\bibitem{McCulloch:2024hiz}
C.~McCulloch, E.~Pajer, and X.~Tong, ``{A cosmological tachyon collider:
  enhancing the long-short scale coupling},''
  \href{http://dx.doi.org/10.1007/JHEP05(2024)262}{{\em JHEP} {\bfseries 05}
  (2024) 262}, \href{http://arxiv.org/abs/2401.11009}{{\ttfamily
  arXiv:2401.11009 [hep-th]}}.

\bibitem{Wu:2024wti}
Y.-P. Wu, ``{The cosmological collider in $R^2$ inflation},''
  \href{http://arxiv.org/abs/2404.05031}{{\ttfamily arXiv:2404.05031
  [astro-ph.CO]}}.

\bibitem{Aoki:2024uyi}
S.~Aoki, L.~Pinol, F.~Sano, M.~Yamaguchi, and Y.~Zhu, ``{Cosmological
  Correlators with Double Massive Exchanges: Bootstrap Equation and
  Phenomenology},'' \href{http://arxiv.org/abs/2404.09547}{{\ttfamily
  arXiv:2404.09547 [hep-th]}}.

\bibitem{Chen:2014joa}
X.~Chen and M.~H. Namjoo, ``{Standard Clock in Primordial Density Perturbations
  and Cosmic Microwave Background},''
  \href{http://dx.doi.org/10.1016/j.physletb.2014.11.002}{{\em Phys. Lett. B}
  {\bfseries 739} (2014) 285--292},
  \href{http://arxiv.org/abs/1404.1536}{{\ttfamily arXiv:1404.1536
  [astro-ph.CO]}}.

\bibitem{Quintin:2024boj}
J.~Quintin, X.~Chen, and R.~Ebadi, ``{Fingerprints of a Non-Inflationary
  Universe from Massive Fields},''
  \href{http://arxiv.org/abs/2405.11016}{{\ttfamily arXiv:2405.11016
  [astro-ph.CO]}}.

\bibitem{Meerburg:2016zdz}
P.~D. Meerburg, M.~M{\"u}nchmeyer, J.~B. Mu{\~n}oz, and X.~Chen, ``{Prospects
  for Cosmological Collider Physics},''
  \href{http://dx.doi.org/10.1088/1475-7516/2017/03/050}{{\em JCAP} {\bfseries
  1703} no.~03, (2017) 050},
\href{http://arxiv.org/abs/1610.06559}{{\ttfamily arXiv:1610.06559
  [astro-ph.CO]}}.

\bibitem{MoradinezhadDizgah:2017szk}
A.~Moradinezhad~Dizgah and C.~Dvorkin, ``{Scale-Dependent Galaxy Bias from
  Massive Particles with Spin during Inflation},''
  \href{http://dx.doi.org/10.1088/1475-7516/2018/01/010}{{\em JCAP} {\bfseries
  01} (2018) 010}, \href{http://arxiv.org/abs/1708.06473}{{\ttfamily
  arXiv:1708.06473 [astro-ph.CO]}}.

\bibitem{MoradinezhadDizgah:2018ssw}
A.~Moradinezhad~Dizgah, H.~Lee, J.~B. Mu{\~n}oz, and C.~Dvorkin, ``{Galaxy
  Bispectrum from Massive Spinning Particles},''
  \href{http://dx.doi.org/10.1088/1475-7516/2018/05/013}{{\em JCAP} {\bfseries
  1805} no.~05, (2018) 013},
\href{http://arxiv.org/abs/1801.07265}{{\ttfamily arXiv:1801.07265
  [astro-ph.CO]}}.

\bibitem{Kogai:2020vzz}
K.~Kogai, K.~Akitsu, F.~Schmidt, and Y.~Urakawa, ``{Galaxy imaging surveys as
  spin-sensitive detector for cosmological colliders},''
  \href{http://dx.doi.org/10.1088/1475-7516/2021/03/060}{{\em JCAP} {\bfseries
  03} (2021) 060}, \href{http://arxiv.org/abs/2009.05517}{{\ttfamily
  arXiv:2009.05517 [astro-ph.CO]}}.

\bibitem{Green:2023uyz}
D.~Green, Y.~Guo, J.~Han, and B.~Wallisch, ``{Light fields during inflation
  from BOSS and future galaxy surveys},''
  \href{http://dx.doi.org/10.1088/1475-7516/2024/05/090}{{\em JCAP} {\bfseries
  05} (2024) 090}, \href{http://arxiv.org/abs/2311.04882}{{\ttfamily
  arXiv:2311.04882 [astro-ph.CO]}}.

\bibitem{Cabass:2024wob}
G.~Cabass, O.~H.~E. Philcox, M.~M. Ivanov, K.~Akitsu, S.-F. Chen,
  M.~Simonovi\'c, and M.~Zaldarriaga, ``{BOSS Constraints on Massive Particles
  during Inflation: The Cosmological Collider in Action},''
  \href{http://arxiv.org/abs/2404.01894}{{\ttfamily arXiv:2404.01894
  [astro-ph.CO]}}.

\bibitem{Sohn:2024xzd}
W.~Sohn, D.-G. Wang, J.~R. Fergusson, and E.~P.~S. Shellard, ``{Searching for
  Cosmological Collider in the Planck CMB Data},''
  \href{http://arxiv.org/abs/2404.07203}{{\ttfamily arXiv:2404.07203
  [astro-ph.CO]}}.

\bibitem{Philcox:2024jpd}
O.~H.~E. Philcox, S.~Kumar, and J.~C. Hill, ``{Too Hot to Handle: Searching for
  Inflationary Particle Production in Planck Data},''
  \href{http://arxiv.org/abs/2405.03738}{{\ttfamily arXiv:2405.03738
  [astro-ph.CO]}}.

\bibitem{Goldstein:2024bky}
S.~Goldstein, O.~H.~E. Philcox, J.~C. Hill, and L.~Hui, ``{Massive-ish
  Particles from Small-ish Scales: Non-Perturbative Techniques for Cosmological
  Collider Physics from Large-Scale Structure Surveys},''
  \href{http://arxiv.org/abs/2407.08731}{{\ttfamily arXiv:2407.08731
  [astro-ph.CO]}}.

\bibitem{Eden:1966dnq}
R.~J. Eden, P.~V. Landshoff, D.~I. Olive, and J.~C. Polkinghorne, {\em {The
  analytic S-matrix}}.
\newblock Cambridge Univ. Press, Cambridge, 1966.

\bibitem{Mandelstam:1958xc}
S.~Mandelstam, ``{Determination of the pion - nucleon scattering amplitude from
  dispersion relations and unitarity. General theory},''
  \href{http://dx.doi.org/10.1103/PhysRev.112.1344}{{\em Phys. Rev.} {\bfseries
  112} (1958) 1344--1360}.

\bibitem{Toll:1956cya}
J.~S. Toll, ``{Causality and the Dispersion Relation: Logical Foundations},''
  \href{http://dx.doi.org/10.1103/PhysRev.104.1760}{{\em Phys. Rev.} {\bfseries
  104} (1956) 1760--1770}.

\bibitem{Paulos:2016but}
M.~F. Paulos, J.~Penedones, J.~Toledo, B.~C. van Rees, and P.~Vieira, ``{The
  S-matrix bootstrap II: two dimensional amplitudes},''
  \href{http://dx.doi.org/10.1007/JHEP11(2017)143}{{\em JHEP} {\bfseries 11}
  (2017) 143}, \href{http://arxiv.org/abs/1607.06110}{{\ttfamily
  arXiv:1607.06110 [hep-th]}}.

\bibitem{Danilkin:2014cra}
I.~V. Danilkin, C.~Fern\'andez-Ram\'\i{}rez, P.~Guo, V.~Mathieu, D.~Schott,
  M.~Shi, and A.~P. Szczepaniak, ``{Dispersive analysis of
  \ensuremath{\omega}/\ensuremath{\phi}\textrightarrow{}3\ensuremath{\pi},\ensuremath{\pi}\ensuremath{\gamma}*},''
  \href{http://dx.doi.org/10.1103/PhysRevD.91.094029}{{\em Phys. Rev. D}
  {\bfseries 91} no.~9, (2015) 094029},
  \href{http://arxiv.org/abs/1409.7708}{{\ttfamily arXiv:1409.7708 [hep-ph]}}.

\bibitem{Hoferichter:2018kwz}
M.~Hoferichter, B.-L. Hoid, B.~Kubis, S.~Leupold, and S.~P. Schneider,
  ``{Dispersion relation for hadronic light-by-light scattering: pion pole},''
  \href{http://dx.doi.org/10.1007/JHEP10(2018)141}{{\em JHEP} {\bfseries 10}
  (2018) 141}, \href{http://arxiv.org/abs/1808.04823}{{\ttfamily
  arXiv:1808.04823 [hep-ph]}}.

\bibitem{Garcia-Martin:2011nna}
R.~Garcia-Martin, R.~Kaminski, J.~R. Pelaez, and J.~Ruiz~de Elvira, ``{Precise
  determination of the f0(600) and f0(980) pole parameters from a dispersive
  data analysis},''
  \href{http://dx.doi.org/10.1103/PhysRevLett.107.072001}{{\em Phys. Rev.
  Lett.} {\bfseries 107} (2011) 072001},
  \href{http://arxiv.org/abs/1107.1635}{{\ttfamily arXiv:1107.1635 [hep-ph]}}.

\bibitem{Bjorken:1965zz}
J.~D. Bjorken and S.~D. Drell, ``{Relativistic quantum fields},''.

\bibitem{Goldberger:1955zza}
M.~L. Goldberger, H.~Miyazawa, and R.~Oehme, ``{Application of Dispersion
  Relations to Pion-Nucleon Scattering},''
  \href{http://dx.doi.org/10.1103/PhysRev.99.986}{{\em Phys. Rev.} {\bfseries
  99} (1955) 986--988}.

\bibitem{Goldberger:1955zz}
M.~L. Goldberger, ``{Causality Conditions and Dispersion Relations. 1. Boson
  Fields},'' \href{http://dx.doi.org/10.1103/PhysRev.99.979}{{\em Phys. Rev.}
  {\bfseries 99} (1955) 979--985}.

\bibitem{Caron-Huot:2017vep}
S.~Caron-Huot, ``{Analyticity in Spin in Conformal Theories},''
  \href{http://dx.doi.org/10.1007/JHEP09(2017)078}{{\em JHEP} {\bfseries 09}
  (2017) 078}, \href{http://arxiv.org/abs/1703.00278}{{\ttfamily
  arXiv:1703.00278 [hep-th]}}.

\bibitem{Alday:2017vkk}
L.~F. Alday and S.~Caron-Huot, ``{Gravitational S-matrix from CFT dispersion
  relations},'' \href{http://dx.doi.org/10.1007/JHEP12(2018)017}{{\em JHEP}
  {\bfseries 12} (2018) 017}, \href{http://arxiv.org/abs/1711.02031}{{\ttfamily
  arXiv:1711.02031 [hep-th]}}.

\bibitem{Carmi:2019cub}
D.~Carmi and S.~Caron-Huot, ``{A Conformal Dispersion Relation: Correlations
  from Absorption},'' \href{http://dx.doi.org/10.1007/JHEP09(2020)009}{{\em
  JHEP} {\bfseries 09} (2020) 009},
  \href{http://arxiv.org/abs/1910.12123}{{\ttfamily arXiv:1910.12123
  [hep-th]}}.

\bibitem{Penedones:2019tng}
J.~Penedones, J.~A. Silva, and A.~Zhiboedov, ``{Nonperturbative Mellin
  Amplitudes: Existence, Properties, Applications},''
  \href{http://dx.doi.org/10.1007/JHEP08(2020)031}{{\em JHEP} {\bfseries 08}
  (2020) 031}, \href{http://arxiv.org/abs/1912.11100}{{\ttfamily
  arXiv:1912.11100 [hep-th]}}.

\bibitem{Bissi:2022mrs}
A.~Bissi, A.~Sinha, and X.~Zhou, ``{Selected topics in analytic conformal
  bootstrap: A guided journey},''
  \href{http://dx.doi.org/10.1016/j.physrep.2022.09.004}{{\em Phys. Rept.}
  {\bfseries 991} (2022) 1--89},
  \href{http://arxiv.org/abs/2202.08475}{{\ttfamily arXiv:2202.08475
  [hep-th]}}.

\bibitem{Cespedes:2020xqq}
S.~C\'espedes, A.-C. Davis, and S.~Melville, ``{On the time evolution of
  cosmological correlators},''
  \href{http://dx.doi.org/10.1007/JHEP02(2021)012}{{\em JHEP} {\bfseries 02}
  (2021) 012}, \href{http://arxiv.org/abs/2009.07874}{{\ttfamily
  arXiv:2009.07874 [hep-th]}}.

\bibitem{kramers1927diffusion}
H.~A. Kramers, ``La diffusion de la lumiere par les atomes,'' in {\em Atti
  Cong. Intern. Fisica (Transactions of Volta Centenary Congress) Como},
  vol.~2, pp.~545--557.
\newblock 1927.

\bibitem{deL.Kronig:26}
R.~de~L.~Kronig, ``On the theory of dispersion of x-rays,''
  \href{http://dx.doi.org/10.1364/JOSA.12.000547}{{\em J. Opt. Soc. Am.}
  {\bfseries 12} no.~6, (Jun, 1926) 547--557}.
  \url{https://opg.optica.org/abstract.cfm?URI=josa-12-6-547}.

\bibitem{Adams:2006sv}
A.~Adams, N.~Arkani-Hamed, S.~Dubovsky, A.~Nicolis, and R.~Rattazzi,
  ``{Causality, analyticity and an IR obstruction to UV completion},''
  \href{http://dx.doi.org/10.1088/1126-6708/2006/10/014}{{\em JHEP} {\bfseries
  10} (2006) 014}, \href{http://arxiv.org/abs/hep-th/0602178}{{\ttfamily
  arXiv:hep-th/0602178}}.

\bibitem{Nicolis:2009qm}
A.~Nicolis, R.~Rattazzi, and E.~Trincherini, ``{Energy's and amplitudes'
  positivity},'' \href{http://dx.doi.org/10.1007/JHEP05(2010)095}{{\em JHEP}
  {\bfseries 05} (2010) 095}, \href{http://arxiv.org/abs/0912.4258}{{\ttfamily
  arXiv:0912.4258 [hep-th]}}. [Erratum: JHEP 11, 128 (2011)].

\bibitem{deRham:2017avq}
C.~de~Rham, S.~Melville, A.~J. Tolley, and S.-Y. Zhou, ``{Positivity bounds for
  scalar field theories},''
  \href{http://dx.doi.org/10.1103/PhysRevD.96.081702}{{\em Phys. Rev. D}
  {\bfseries 96} no.~8, (2017) 081702},
  \href{http://arxiv.org/abs/1702.06134}{{\ttfamily arXiv:1702.06134
  [hep-th]}}.

\bibitem{deRham:2017zjm}
C.~de~Rham, S.~Melville, A.~J. Tolley, and S.-Y. Zhou, ``{UV complete me:
  Positivity Bounds for Particles with Spin},''
  \href{http://dx.doi.org/10.1007/JHEP03(2018)011}{{\em JHEP} {\bfseries 03}
  (2018) 011}, \href{http://arxiv.org/abs/1706.02712}{{\ttfamily
  arXiv:1706.02712 [hep-th]}}.

\bibitem{Marolf:2010zp}
D.~Marolf and I.~A. Morrison, ``{The IR stability of de Sitter: Loop
  corrections to scalar propagators},''
  \href{http://dx.doi.org/10.1103/PhysRevD.82.105032}{{\em Phys. Rev. D}
  {\bfseries 82} (2010) 105032},
  \href{http://arxiv.org/abs/1006.0035}{{\ttfamily arXiv:1006.0035 [gr-qc]}}.

\bibitem{nist:dlmf}
``{NIST Digital Library of Mathematical Functions}.''
\newblock \url{https://dlmf.nist.gov}.

\bibitem{wolfram}
``{The Mathematical Functions Site}.''
\newblock \url{https://functions.wolfram.com}.

\bibitem{Slater:1966}
L.~J. Slater, {\em Generalized Hypergeometric Functions}.
\newblock Cambridge University Press, 1966.

\end{thebibliography}\endgroup
\bibliographystyle{utphys}

\end{document}